\pdfoutput=1
\documentclass[letterpaper,12pt]{report}
\usepackage[T1]{fontenc}
\usepackage[utf8]{inputenc}
\usepackage[english,brazil,spanish]{babel}
\usepackage{graphics,graphicx,epstopdf}
\usepackage{amsmath,amssymb,amsfonts}
\usepackage{float}
\usepackage[sort&compress,square,numbers]{natbib}
\usepackage{color}
\usepackage{indentfirst}             
\usepackage{tabularx}
\usepackage{makeidx}
\usepackage{enumerate}
\usepackage{pifont}
\usepackage{pdfpages}
\usepackage{hyperref}
\usepackage{cleveref}
\usepackage{hhline}
\usepackage[titletoc]{appendix}
\usepackage[a4paper,top=2.4cm, bottom=2.4cm, left=3.0cm, right=1.5cm]{geometry}
\makeindex
\newcommand*\colourcheck[1]{%
  \expandafter\newcommand\csname #1check\endcsname{\textcolor{#1}{\ding{52}}}%
}
\newcommand*\xcolourcheck[1]{%
  \expandafter\newcommand\csname #1xcheck\endcsname{\textcolor{#1}{\ding{55}}}%
}
\colourcheck{green}
\colourcheck{red}
\xcolourcheck{green}
\xcolourcheck{red}
\newcommand*{\rom}[1]{\expandafter\@slowromancap\romannumeral #1@}

\def\re#1{Re(#1)}
\def\im#1{Im(#1)}

\normalsize

\newcommand{\captionfonts}{\small}
\makeatletter  
\long\def\@makecaption#1#2{%
  \vskip\abovecaptionskip
  \sbox\@tempboxa{{\captionfonts #1: #2}}%
  \ifdim \wd\@tempboxa >\hsize
    {\captionfonts #1: #2\par}
  \else
    \hbox to\hsize{\hfil\box\@tempboxa\hfil}%
  \fi
  \vskip\belowcaptionskip}
\makeatother   

\newcolumntype{M}[1]{>{\arraybackslash}m{#1}}
\newcolumntype{N}{@{}m{0pt}@{}}

\begin{document}
\selectlanguage{brazil}
\thispagestyle{empty}
\begin{center}
    \vspace*{-1.0cm}

\textbf{\includegraphics[scale=0.25]{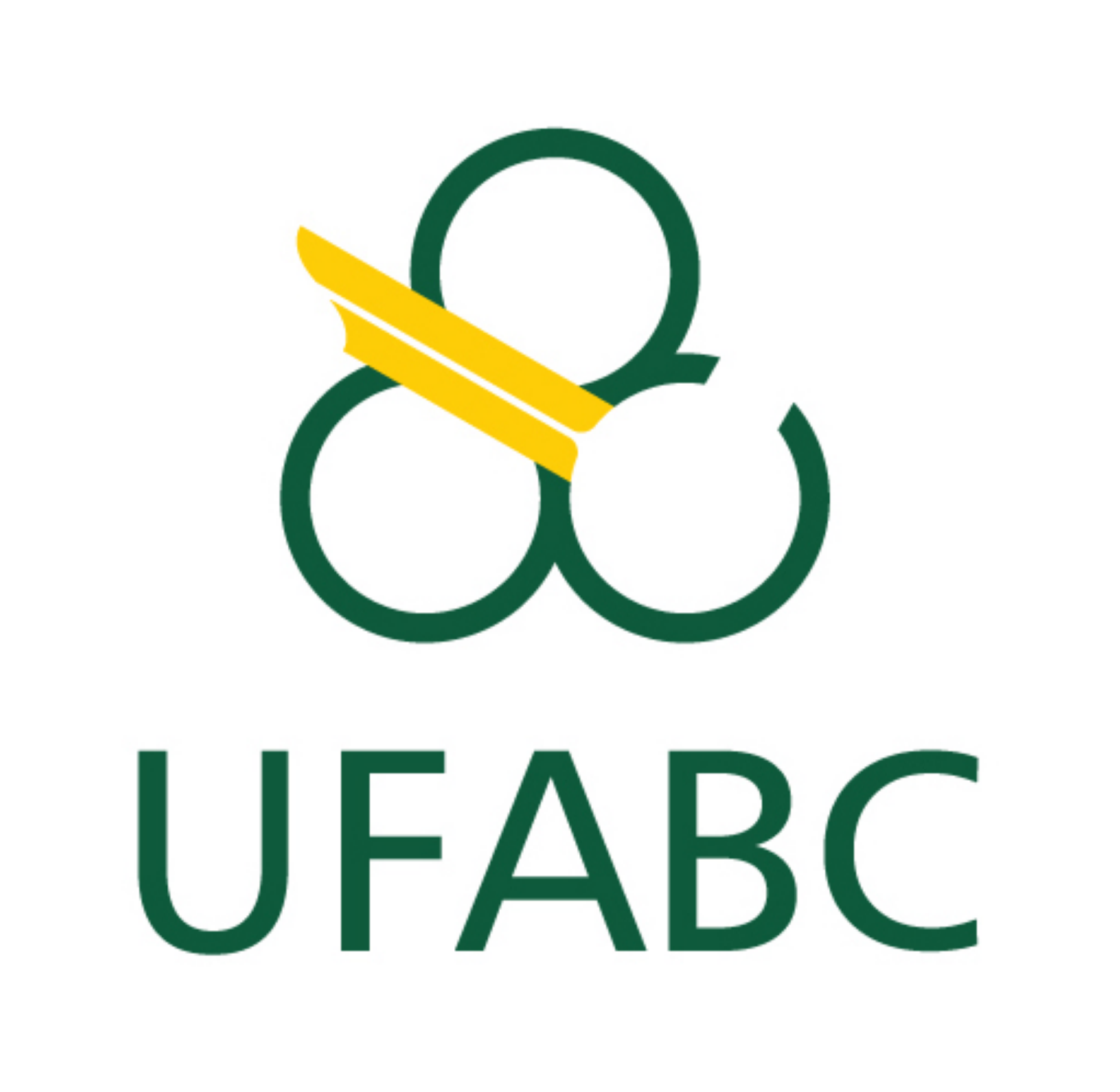} \\ \large{\bf\large
    UNIVERSIDADE FEDERAL DO ABC\\
    PROGRAMA DE PÓS-GRADUAÇÃO EM FÍSICA \vspace{3cm}\\
   BLACK HOLES AND WORMHOLES IN HIGHER-CURVATURE CORRECTED THEORIES OF GRAVITY}}\\

    \vspace{2.5cm} 
    \large{\bf TESE DE DOUTORADO}\\
   \vspace{2.5cm}
    \large{\bf Marco Antonio Cuyubamba Espinoza}\\
    \vspace{2.5cm}
 \large{\bf Orientador: Olexandr Zhydenko}\\
 \large{\bf Co-Orientador: Roman Konoplya}\\ \vspace{1.5cm}
 \large{\bf Santo André, SP, Brasil\\\vspace{-0.2 cm}
2019.}
\end{center}

\pagebreak
%

\thispagestyle{empty}
\begin{center}
\large{\bf\large
BLACK HOLES AND WORMHOLES IN HIGHER-CURVATURE CORRECTED THEORIES OF GRAVITY}\\   \vspace{3.0cm}

    \large{\bf por}\\\vspace{3.0cm}
    \large{\bf Marco Antonio Cuyubamba Espinoza}\\ \vspace{3.0cm}
\begin{flushright}
   \parbox{.45\textwidth}{\small{Tese apresentada ao Curso de doutorado do Programa de Pós-graduação em Física, da Universidade Federal do ABC (UFABC), como requisito parcial à obtenção do título de Doutor em Física}}
\end{flushright}
\vspace{3.0cm}
 \large{\bf Orientador: Olexandr Zhydenko}\\
  \large{\bf Co-Orientador: Roman Konoplya}\\ \vspace{2.0cm}
 \large{\bf Santo André, SP, Brasil\\\vspace{-0.2 cm}
     2019}
\end{center}





\pagebreak



   \vspace{1.0cm}





\includepdf[page={1}]{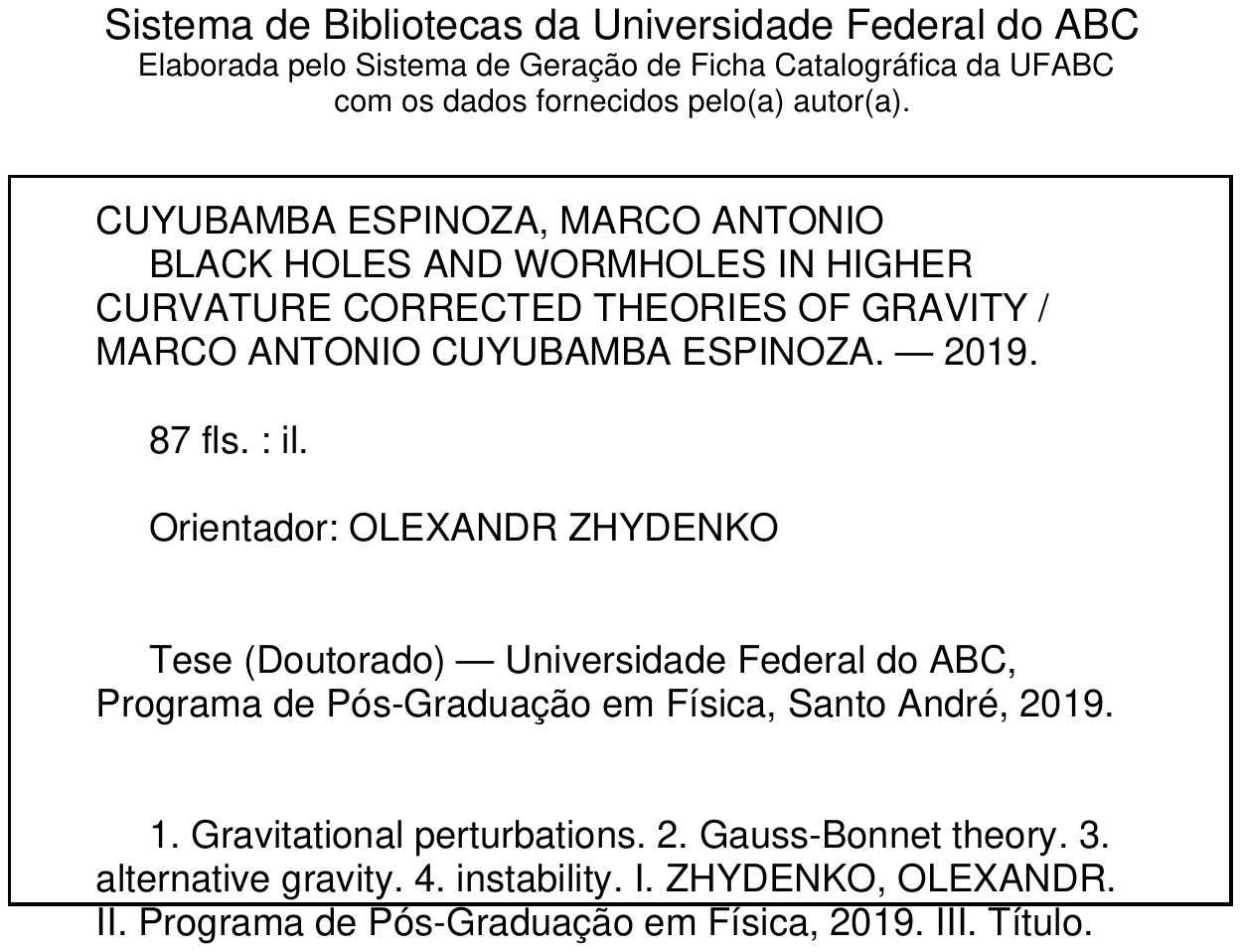}
\includepdf[page={1}]{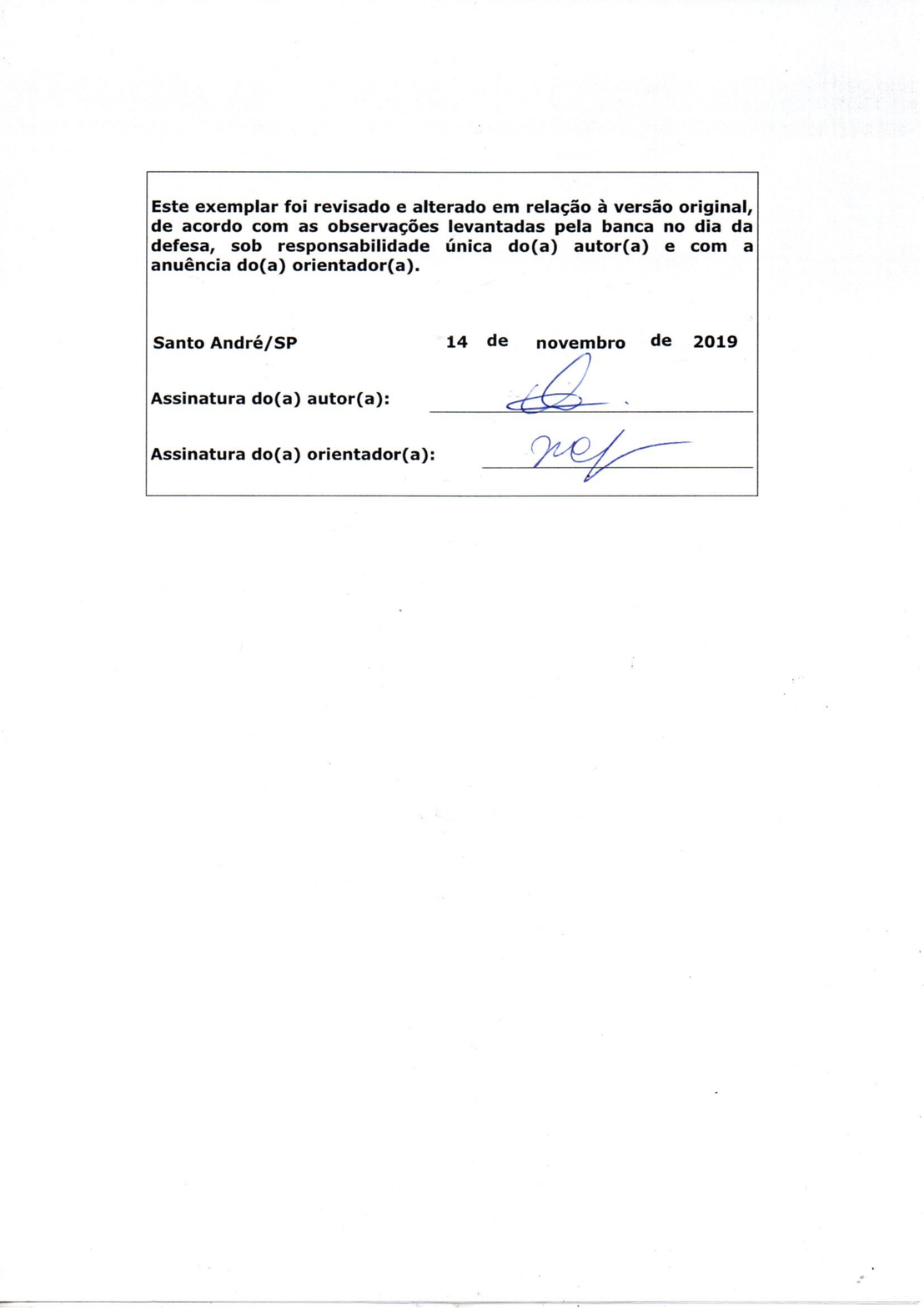}
\includepdf[page={1}]{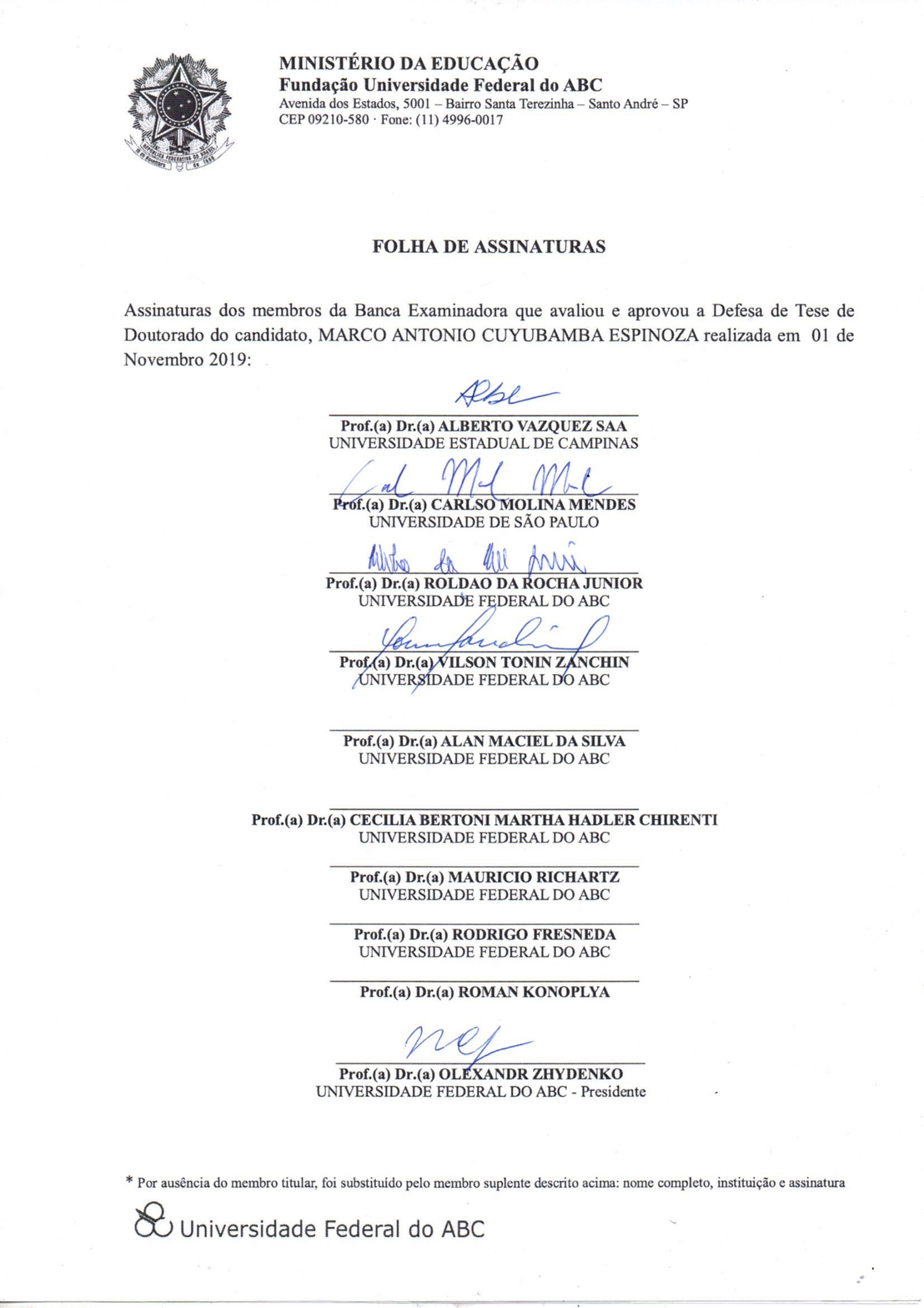}

\selectlanguage{english}

\pagenumbering{roman}
\renewcommand{\thepage}{\roman{page}}
\pagestyle{plain}
\begin{otherlanguage}{spanish}
\chapter*{Agradecimientos}
Siendo esta tesis, uno de los mayores logros en mi vida académica, deseo y quiero concentrar los agradecimientos  a una persona en especial: Mi Mamá.

Desde que tengo uso de razón, recuerdo siempre las palabras de mi mamá que decía ``{\it Estudia mucho para que seas alguien en la vida}'', probablemente palabras que toda madre diría y que para mí marcaron mucho. Mi madre al ser una persona de gran corazón siempre encontró la forma de hacer entender lo que era importante para mi. En el 2006, con 18 años recién cumplidos, comencé mi etapa académica rumbo a la vida profesional en la Universidad Nacional de Ingeniería (UNI), cuyo ingreso es considerado difícil para el estándar de esa época. Recuerdo que vi los resultados de mi ingreso junto con ella en una cabina de internet y la felicidad que mi mamá desbordaba era casi infinita. En mis cinco años de carrera, mi mamá siempre me motivaba a seguir avanzando y siempre contaba con su apoyo en la búsqueda de conseguir aún mayor conocimiento. Prácticamente me apoyaba en todo lo que yo deseaba en mi entorno académico y vivo con un feliz recuerdo de todos esos momentos.

En el 2011 acabé mi periodo de formación e ingresé a la vida laboral, donde mi mamá siempre deseó lo mejor para mí. La felicidad por mis logros y las muestras de orgullo hacia mi persona siempre me hicieron muy feliz. Luego de un corto periodo laboral, haciendo actividades un tanto fuera de la línea de mi carrera, decidí viajar a Brasil a seguir con mi etapa formativa y logrando conseguir una bolsa de estudios en la Universidade Federal do ABC (la cual tendrá también un espacio en estos agradecimientos). Con lágrimas en los ojos, mi madre se despidió de mí y deseandome siempre lo mejor. Mi vida académica nos separó pero felizmente la tecnología no, así que siempre podía escuchar sus consejos y muestras de amor de parte de ella. Luego de un arduo periodo de estudio, conseguí mi grado de Magister y claro.. mi mamá era la más feliz del mundo.

En el 2014 tomé la decisión de continuar con mis estudios de doctorado en física en la misma universidad y nuevamente mi hermosa madre tenía una mezcla de felicidad y tristeza por no poder estar a su lado. Muchos sucesos ocurrieron en esta larga etapa de estudio, algunos buenos y otros no tan buenos. Para la fecha de la sustentación de esta tesis, tengo el enorme dolor de saber que mi madre no me podrá expresar su gran amor y orgullo característico de ella. A pesar de saber que mi hermosa madre no estará a mi lado de ahora en adelante, considero que gracias a ella soy lo que soy y estaré eternamente agradecido.

Con el respeto de todos y con todo lo expuesto anteriormente, esta tesis se la dedico enteramente a mi hermosa madre \emph{Lourdes Carmen Espinoza Rojas}.

Obviamente hubieron muchos más agentes que hicieron posible mi desarrollo académico. Quiero agradecer en segundo lugar a mi linda compañera y amor de mi vida Carmen Lino, que sin su ayuda y apoyo no estaría defendiendo esta tesis. Agradezco a mi hermana Gisselle a mi cuñado Cesar, papá, tíos, primos, amigos y familia en general que siempre me empujaron a que culmine mis estudios de doctorado.

Agradezco también a mi orientador Olexandr Zhydenko y a mi co-orientador Roman Konoplya, que gracias a su paciencia y apoyo sigo en mi vida académica. Agradezco a la Universidad Federal do ABC (UFABC) por permitirme crecer académicamente y a CAPES por el apoyo financiero.

This study was financed in part by the Coordenação de Aperfeiçoamento de Pessoal de Nível Superior - Brasil (CAPES) - Finance Code 001.

Espero estar a la altura de retribuir a todos.
\end{otherlanguage} 
\newpage

\chapter*{Abstract}\label{Abstract}
The presented thesis is devoted to the study of instabilities of compact objects within the Einstein-Gauss-Bonnet theory. This theory includes higher-order corrections in curvature, which are inspired by the low energy limit of string theory. We study linear instability of higher-dimensional black holes in the de Sitter universe. The time-domain picture allows us to obtain the parametric region of stability for the gravitational perturbations in all three channels, i.e., for scalar-type, vector-type, and tensor-type perturbations. We observed that while the scalar and tensor channels show instability for some choice of the parameters, the vector-type perturbations are always stable. Furthermore, we show that the quasinormal frequencies of the scalar type of gravitational perturbations do not obey Hod's inequality, however, the other two channels, vector and tensor, have lower-lying modes that confirm Hod's conjecture.

We also studied stability of the wormholes in the four-dimensional Einstein-dilaton-Gauss-Bonnet gravity proposed by P. Kanti, B. Kleihaus, J. Kunz in [Phys. Rev. Lett. 107, 271101 (2011)]. These wormholes were claimed to be stable against linear radial perturbations. However, our time-domain analysis allowed us to prove that to prove that such wormholes are linearly unstable against general radial perturbations for any values of their parameters. We observed that the exponential growth appears after a long period of damped oscillations. This behaviour is qualitatively similar to the instability profile of the higher-dimensional black holes in the Einstein-Gauss-Bonnet theory.
\newline\newline
\noindent
Keywords: Gravitational perturbations, Gauss-Bonnet theory, alternative gravity, instability 

\begin{otherlanguage}{brazil}
\chapter*{Resumo}\label{Resumo}
Esta tese é dedicada ao estudo de instabilidades de objetos compactos na teoria de Einstein-Gauss-Bonnet. Essa teoria inclui correções de ordem superior na curvatura, que foram inspiradas pelo limite de baixa energia da teoria das cordas. Estudamos a instabilidade linear de buracos negros de dimensão superior no universo de Sitter. O perfil no domínio temporal permite-nos obter a região paramétrica de estabilidade para as perturbações gravitacionais em todos os três canais, ou seja, para perturbações do tipo escalar, vetorial e tensorial. Observamos que, embora que os canais escalar e tensorial mostram instabilidade para certos parâmetros, as perturbações do tipo vetorial são sempre estáveis. Além disso, mostramos que as freqüências quase-normais do tipo escalar de perturbações gravitacionais não obedecem a desigualdade de Hod, no entanto, os outros dois canais, vetor e tensor, têm os modos baixos que obedecem a conjetura de Hod.

Também estudamos a estabilidade dos buracos de minhoca na gravidade quadridimensional de Einstein-dilaton-Gauss-Bonnet proposta por P. Kanti, B. Kleihaus, J. Kunz em [Phys. Rev. Lett. 107, 271101 (2011)].  Estes buracos de minhoca foram declarados estáveis em relação às perturbações radiais lineares. Entretanto, nossa análise no domínio temporal permitiu-nos provar que aqueles buracos de minhoca são linearmente instáveis em relação às perturbações radiais gerais para quaisquer valores dos parâmetros. Observamos que o crescimento exponencial aparece após um longo período de oscilações amortecidas. Este comportamento é qualitativamente semelhante ao perfil de instabilidade dos buracos negros de dimensão superior na teoria de Einstein-Gauss-Bonnet.
\newline\newline
\noindent
Palavras-chave: Perturbações gravitacionais, teoría de Gauss-Bonnet, gravidade alternativa, instabilidade
\end{otherlanguage} 

\tableofcontents
\listoffigures
\listoftables
\newpage

\pagenumbering{arabic}
\renewcommand{\thepage}{\arabic{page}}
\pagestyle{plain}

\chapter{Introduction}\label{Introduction}
Recent years, theories of gravity with higher-curvature correction to the Einstein-Hilbert term of General Relativity (GR) become an interesting alternative to Einstein gravity since higher-order correction to the Einstein-Hilbert action appears in quantum gravity and string theory. In higher than four-dimensional spacetimes, the low-energy limit of the heterotic string theory predicts a second order in curvature correction, called the Gauss-Bonnet termo the Einstein action. In particular, Lovelock gravity, one of these higher curvature theories, represents the most natural generalization of Einstein action when the number of dimensions is larger than four. Recently it was found that the Einstein-Gauss-Bonnet (EGB) theory has some interesting features:
\begin{enumerate}
 \item  the anti-de Sitter spacetime, which is unstable against nonlinear perturbations in the Einstein gravity \cite{Bizon:2011gg}, restores stability in the Gauss-Bonnet theory \cite{PhysRevLett.114.071102};
 \item  unlike Schwarzschild solution, small black holes in the Einstein-Gauss-Bonnet theory are unstable against small perturbations of space-time \cite{Dotti:2004sh,Gleiser:2005ra,Konoplya:2008ix};
 \item Hawking radiation of even softly Gauss-Bonnet-corrected black holes occurs at the evaporation rate which is many orders slower than that of the black hole in the higher-dimensional Einstein theory \cite{Konoplya:2010vz}.
\end{enumerate}

Black-hole stability against gravitational perturbations is an important and crucial criterium for the viability of the particular
black hole model. Since analytical analysis of the stability is a difficult task even for relatively simple black hole solutions \cite{Konoplya:2011qq,Ishibashi:2011ws}, for the most of cases numerical treatment comes into play through consideration of the black-hole oscillations (\emph{quasinormal} spectrum) \cite{Kokkotas:1999bd}.

The quasinormal discrete spectrum of black holes have been extensively studied by theorists researchers during the past years. Nowadays, there is a strong indication that quasinormal modes have been experimentally observed in the gravitational-wave signal from a merger of binary black-hole system \cite{Abbott:2016blz,TheLIGOScientific:2016src}. It was shown that the gravitational-wave signal is consistent with the classical Einstein model of gravity. At the same time, LIGO and VIRGO collaborations simulated the evolution of the binary black hole within the Einstein theory, leaving the possible consistency of alternative theories to further investigation \cite{TheLIGOScientific:2016src}. It is shown in \cite{Konoplya:2016pmh} that simple estimations of orders lead to rather large indeterminacy of the black hole's parameters, leaving the window for alternative theories of gravity open.

In \cite{Hod:2006jw} S. Hod proposed that there exists a universal bound on the relaxation time rate. According to Hod's proposal, in the quasinormal spectrum of any black hole there is always a mode whose rate of decay (given by the absolute value of the imaginary part of the quasinormal frequency $|\im{\omega}|$) is not higher than $\pi T_{BH}$, where $T_{BH}$ is the Hawking temperature of the black hole. The Hod's inequality is satisfied for the wide range of black-hole solutions in various four- and higher-dimensional theories. Therefore, when studying the quasinormal spectrum of black holes, the natural question would be, whether this inequality is satisfied also for the considered class of black holes, i. e. whether the Hod's conjecture has counterexamples.

In the framework of higher curvature gravity, the Einstein-dilaton-Gauss-Bonnet action corresponds to a kind of minimal effective theory of the full action considered in the heterotic string theory. Therefore, there is a great interest in understanding physics of various compact objects, first of all, such as black holes and wormholes, in the Einstein-dilaton-Gauss-Bonnet theory. The negative energy density, which is necessary for existence of a wormhole, is provided by the Gauss-Bonnet term. Thus, the solution suggested in \cite{Kanti:2011jz} is an important example of a wormhole without any exotic matter, which is stipulated by the fundamental physics.

Stability against small spacetime perturbations, could be an essential criterium related to theoretical possibility of existence of wormholes. Stability of wormholes was studied in a number of works \cite{Lobo:2005yv,Eiroa:2003wp,Dzhunushaliev:2017syc}. Stability of thin-shell wormholes was investigated only against most-stable purely radial perturbations \cite{Lobo:2005yv,Eiroa:2003wp}, while (also radial) stability of wormholes in general relativity reported in \cite{Bronnikov:2013coa} requires a rather odd equation of state for the surrounding matter \cite{Konoplya:2016hmd}. Various wormholes with ghost scalar field are known to be unstable \cite{Gonzalez:2008wd,Bronnikov:2012ch}. Thus, to the best of our knowledge there is no example of a four-dimensional asymptotically flat wormhole solution whose stability would be well established. Therefore, the stability of such a wormhole solution against small spacetime perturbations is necessary to prove the viability of the wormhole model.

In  \cite{Kanti:2011jz} it was concluded that the wormhole is stable against spherical perturbations. The boundary conditions used in \cite{Kanti:2011jz} imply fixing the size of the wormhole throat, which looks nonphysical.  This was motivated in \cite{Kanti:2011jz} by the requirement that small perturbation of the dilaton $\delta \phi$ remains finite at the wormhole throat, and the Dirichlet boundary condition was imposed there. This effectively disconnected the two regions to the left and right from the throat.

This thesis consists of in six chapters. Chapter \ref{Higher-curvature} is dedicated to the complete description of black holes in a higher-curvature theory of gravity. First, we describe the general Lovelock theory in dimensions higher than four, then introduce the static, spherically symmetric black hole solution for vacuum spacetimes. We focus on the theories of the second-order corrections in curvature: by adding the Gauss-Bonnet term to the Einstein action one can obtain the higher-order corrections to the Einstein equations and find the corresponding analytical static black-hole solutions. In order to study black-hole perturbations, we use the formalism, developed by Ishibashi and Kodama \cite{Kodama:2003kk}, which allows us to obtain independent linear equations for the three types of gravitational perturbations.

In chapter \ref{NM} we describe the numerical methods used in this thesis. In order to obtain temporal profiles of the black hole evolution in the linear approximation, we use the discretization method proposed by Gundlach, Price, and Pullin \cite{Gundlach:1993tp}. Moreover, in order to find the dominant modes in the quasinormal spectrum of a black-hole we employed the Pronny's method of fitting the time-domain profile. In chapter \ref{Asymptotically} we present all the results obtained with in the numerical analysis of  in Einsten-Gauss-Bonnet gravity in the asymptotically de Sitter spacetime. Also we determine parametric regions of instability of gravitational perturbation for higher-dimensional black holes and, in the last part, discuss the obtained data for quasinormal modes and the validity of the Hod's proposal. Chapter \ref{WHG} is dedicated to the complete description of the Kanti-Kleihaus-Kunz wormhole solution in Einstein-dilaton-Gauss-Bonnet gravity in which the corrected second order term is nonvanishing in four dimensional spacetime because of the coupled dilaton field. In chapter \ref{LPKKK}, we solve the regularized wave equations under adequate boundary conditions revealing the instability of these traversable wormholes at whatever small values of the coupling constants. An explanation of the reasons for the discrepancy between our results and \cite{Kanti:2011jz} will be presented. Finally, in chapter \ref{conclusions} we give a general review of all the results presented in this thesis and the direction of future research.
\newpage
Results presented in chapters \ref{Asymptotically} and \ref{LPKKK} were published in a high impact factor international journal,
\begin{itemize}
  \item M. A. Cuyubamba, R. A. Konoplya, and A. Zhidenko. Quasinormal modes and a new instability of Einstein-Gauss-Bonnet black holes in the de Sitter world. Phys. Rev., D\textbf{93}, 104053 (2016) \cite{Cuyubamba:2016cug};
  \item M. A. Cuyubamba, R. A. Konoplya, A. Zhidenko, No stable wormholes in Einstein-dilaton-Gauss-Bonnet theory, Phys. Rev., D\textbf{98}, 044040 (2018) \cite{Cuyubamba:2018jdl}.
\end{itemize} 

\chapter{Higher-curvature corrected gravity}\label{Higher-curvature}
In 1915, Albert Einstein \cite{EinsteinGR} published one of the most important scientific accomplishment of the last century, the general theory of relativity. The principle of equivalence and the special theory of relativity were the two ideas behind this extraordinary mathematical construction. The principle of equivalence, in the strong version, states that the laws of physics are invariant under any chosen coordinate and in small enough neighborhood, the laws of physic are described in the same way as in special relativity without gravity. In other words, it is impossible to distinguish between an experiment in the presence of gravitational force and the same experiment in an accelerated system. In general relativity, the concepts of absolute space and time are eliminated. Instead, space and time are related to the matter distribution in the universe, which is opposed to Newton's description where space is a 3-dimensional region where a body can move and time has an important role as being an independent parameter measuring the evolution of a system. In the Einstein theory the spacetime is described as a four-dimensional manifold endowed with a metric, which is a mathematical mechanism to measure ``distances'' between points in four-dimensional spacetime. This metric is not the usual locally Euclidean one (is not a four-dimensional Euclidean space), but rather of Lorentzian signature $(-,+,+,+)$ on a manifold. The line element in coordinates $x^{\mu}$, $\mu=0\ldots 3$ can be written as

\begin{equation}
  ds^2=g_{\mu\nu}(x)dx^\mu dx^\nu\label{HC:Lovelock:-1}.
\end{equation}
Moreover, in the description of Einstein, in order to describe the source of the gravitational field, which is the matter content in the spacetime, the stress-energy tensor $T_{\mu\nu}$ is used. The final step in order to complete the description of the classical theory is the choice of the appropriate action ($S$). The action that describes the interaction dynamics is called the Einstein-Hilbert action,

\begin{equation}\label{HC:Lovelock:01}
  S=\frac{1}{16\pi\,G_4}\int d^4x\sqrt{-g}\,\mathcal{L}_{EH}+S_{matter},
\end{equation}
where
\begin{equation}\label{HC:Lovelock:02}
  \mathcal{L}_{EH}=R-2\Lambda
\end{equation}
is the Einstein-Hilbert lagrangian with cosmological constant $\Lambda$, which was first introduced by Einstein in order to describe a static universe. Later, the Hubble redshift observations indicated that the universe is actually in the state of accelerating expansion. This is mostly thought to be due to the presence of some fluid known as Dark Energy. The Einstein-Hilbert action (\ref{HC:Lovelock:01}) leads to the famous Einstein equation
\begin{equation}
  G_{\mu\nu}\equiv R_{\mu\nu}-\frac{1}{2}g_{\mu\nu}R+\Lambda\, g_{\mu\nu}=\frac{8\pi\,G_4}{c^4}\,T_{\mu\nu},\label{HC:Lovelock:0}
\end{equation}
where the first term in the left hand side is called the Einstein tensor and is expressed in terms of the Ricci tensor $R_{\mu\nu}$. It is derived from the Riemann curvature tensor
$R^{\alpha}_{\phantom{\alpha}\beta\gamma\sigma}$ by contraction of first and third indices and $R=R^\mu_\mu$ the scalar curvature. In order to approach the Newtonian limit, the constant of proportionality on the right-hand side is chosen as $G_4$ the four-dimensional Newton gravitational constant. Vermeil \cite{Vermeil:1917au} in 1917, then Weyl and Cartan showed that the tensor $G_{\mu\nu}$ is the only symmetric and conserved (divergence free $G^{\mu\nu}_{\phantom{\mu\nu};\nu}=0$) tensor depending only on the metric, its first derivative and depending linearly on second derivatives of the metric. The Einstein tensor contains the geometrical information of the spacetime, whilst the other side carries all the field sources encoded in the stress-energy tensor.

Solid formulation for Einstein general relativity has been provided by the consideration of the post-Newtonian corrections to the Newton theory in the Solar system limit, the indirect indications of existence of gravitational waves from binary pulsars \cite{Weisberg:1981mt}, and the recent observations of binary black holes from LIGO and Virgo detectors \cite{Abbott:2016blz,TheLIGOScientific:2016src}. Further, the Event Horizon Telescope (EHT) obtained the first image of the supermassive black hole \cite{Akiyama:2019cqa} located in the center of the giant elliptical galaxy $M87$, it is one of the most amazing things of this year (2019) for astrophysics and black hole researchers since it provides a direct visual evidence of the existence of black holes in our universe.

In addition to the trivial solution of the Einstein equation without matter, which is a flat spacetime described by the Minkowski metric, Karl Schwarzschild \cite{Schwarzschild:1916uq} in 1916 was the first to find an analytical solution of the general relativity theory with a singular character in the solution which, first, seemed to be a mathematical curiosity of no physical relevance. Such an object was later called a black hole. At present, black holes are formed due to the collapse of matter \cite{PhysRevLett.14.57} such as that of a dying star. If a star is sufficiently large, at the last stage of its evolution it undergoes a gravitational collapse and becomes a black hole.

The black holes are self-gravitating objects so dense that nothing, not even light, can escape from their gravitational attraction (hence the name 'black' hole). Nowadays, black holes are thought to be objects in the universe which are present at the center of the galaxies like Sgr $A^*$, a supermassive black hole at the center of the milky way with mass $M=4\times 10^6$ times the solar mass ($M_{\odot}$).

Being a vacuum solution to the Einstein equation (\ref{HC:Lovelock:0}), the Schwarzschild metric, given by the line element
\begin{equation}\label{HC:Black:03}
  ds^2=-\left(1-\frac{r_s}{r}\right)dt^2+\frac{dr^2}{\left(1-\frac{r_s}{r}\right)}+r^2\left(d\theta^2+\sin^2\theta\right)d\varphi^2,
\end{equation}
describes not only a black hole, but also the outside region of any spherically symmetric body with total mass $M$, where $t$ and $r$ are the time and radial coordinate respectively, and the angular coordinates are $\theta$ and $\varphi$ on a two-sphere. The parameter $r_s$ is called the Schwarzschild radius ($r_s=2M$ in geometric units) and this particular position coincides with the well-known event horizon. In order to connect it to gravitational physics in Newtonian gravity, in asymptotically approximation $r\rightarrow\infty$, the Schwarzschild metric approaches to flat Minkowski spacetime $g_{\mu\nu}\sim$ diag$\left(-1,1,1,1\right)$. The event horizon also plays an important role in the geometrical description, since it actually divides the spacetime in two regions in a sense disconnected. Nothing happening in the interior region can influence the dynamics of the outside of the event horizon. In the center of the black hole $r=0$, the curvature of the spacetime becomes infinity, which corresponds to a physical singularity.

A couple of years after Schwarzschild published his work, a metric for black holes with electric and magnetic charges, called the Reissner-Nordström black hole \cite{ReissnerQ,NordstromQ}, was found by solving the Einstein equation with the electromagnetic source term. However, it took almost a half century before Kerr found a vacuum solution for an axisymmetric spacetime \cite{PhysRevLett.11.237} which corresponds to a rotating black hole. Its maximal analytical extension was derived by Boyer and Lindquist \cite{BoyerLinquist}. The Kerr-Newman black hole \cite{NewmanQJ} generalizes the Kerr solution and includes electric and magnetic charges (see Table~\ref{tab:blackholes}).
\begin{table}[ht!]
  \centering
  \begin{tabular}{|c|c|c|c|}
    \hhline{|====|}
    {\bf Spacetime} & $M$ & $Q$ & $J$\\
    \hline\hline
    Schwarzschild \cite{Schwarzschild:1916uq} & \greencheck & \redxcheck & \redxcheck\\
    \hline
    Reissner-Nordstr\"om \cite{ReissnerQ,NordstromQ} & \greencheck & \greencheck & \redxcheck\\
    \hline
    Kerr \cite{PhysRevLett.11.237,BoyerLinquist} & \greencheck & \redxcheck & \greencheck\\
    \hline
    Kerr-Newmann \cite{NewmanQJ} & \greencheck & \greencheck & \greencheck\\
    \hline
    \end{tabular}
  \caption{Black hole spacetime specified by three parameter: the mass $M$, the total charge $Q$ and the angular momentum $J$}
  \label{tab:blackholes}
\end{table}

Furthermore, there is a general consensus that general relativity would get modified at some scale of energy and above that quantum corrections becomes non-trivial, in this sense, alternative theories with better quantum behavior which are also consistent with the experimental tests of GR are taken as serious candidates. Generalized theory of gravity which solves the incompatibility connection between the general relativity and the quantum mechanics has been of physical importance in the last years. String theory is currently the most important candidate that is assumed to solve the problem and encompasses GR as a classical limit. Such a generalized theory as an extension of the GR of Einstein was first studied by D. Lovelock which was pointed out later by B. Zwiebach \cite{Zwiebach:1985uq}.

\section{Lovelock gravity}
One of the natural generalization of the Einstein theory of gravity to higher-dimensional spacetimes ($D>4$) was carried out by David Lovelock \cite{Lovelock:1971} in 1971, providing a general form for a tensor, which is conserved, symmetric, quasi-linear in the second derivatives of the metric, and does not have any higher derivatives. The corresponding Lagrangian is higher-order in curvature can be written as
\begin{equation}\label{HC:Lovelock:1}
\mathcal{L}=\sum_{m=0}^{k} c_m\,\mathcal{L}_m,
\end{equation}
where $c_m$ are arbitrary constants of the theory and $\mathcal{L}_m$ corresponds to the $m$-order Lovelock term
\begin{equation}\label{HC:Lovelock:2}
\mathcal{L}_m=\frac{1}{2^m}\delta_{\lambda_1 \rho_1\ldots\lambda_m\rho_m}^{\mu_1\nu_1\ldots\mu_m\nu_m}R_{\mu_1\nu_1}^{\phantom{\mu_1\nu_1}\lambda_1\rho_1}\ldots R_{\mu_m\nu_m}^{\phantom{\mu_m\nu_m}\lambda_m\rho_m},
\end{equation}
and $R_{\mu\nu}^{\phantom{\mu\nu}\lambda\sigma}$ and $\delta_{\lambda_1 \rho_1\ldots\lambda_m \rho_m}^{\mu_1\nu_1\ldots\mu_m\nu_m}$ are the $D$-dimensional Riemann tensor and the generalized Kronecker delta respectively. The latter is defined by
\begin{equation}\label{HC:Lovelock:3}
\delta_{\nu_1 \nu_2\ldots\nu_p}^{\mu_1\mu_2\ldots\mu_p}=\left|
\begin{matrix} 
\delta_{\nu_1}^{\mu_1} & \delta_{\nu_2}^{\mu_1} & \ldots & \delta_{\nu_p}^{\mu_1}\\ \delta_{\nu_1}^{\mu_2} & \delta_{\nu_2}^{\mu_2} & \ldots & \delta_{\nu_p}^{\mu_2}\\ \vdots & \vdots &
\ddots & \vdots\\ \delta_{\nu_1}^{\mu_p} & \delta_{\nu_2}^{\mu_p} & \ldots & \delta_{\nu_p}^{\mu_p} 
\end{matrix}
\right|=\varepsilon_{\nu_1\ldots\nu_p}\,\varepsilon^{\mu_1\ldots\mu_p},
\end{equation}
where $\varepsilon_{\nu_1\ldots\nu_p}$ is the Levi-Civita symbol. As we can see, $\mathcal{L}_1$ is the Einstein-Hilbert term whereas $\mathcal{L}_2$ is the Gauss-Bonnet combination. By construction, the Lovelock terms vanish for $2m>D$, so that, the Lagrangian in $D$-dimensional space-times (\ref{HC:Lovelock:1}) contains $k\equiv [(D-1)/2]$ terms.

Thus, the four-dimensional Lovelock gravity, with $c_0=-2\Lambda$ and $c_1=1$, corresponds to the Einstein-Hilbert Lagrangian (\ref{HC:Lovelock:01}), being
\begin{eqnarray}
  \mathcal{L}_0 &=& 1 \nonumber,\\
  \mathcal{L}_1 &=& R \nonumber.
\end{eqnarray}

In five or six dimensions we obtain also the Gauss-Bonnet term ($k=2$), which yields a lagrangian quadratic in the Riemann tensor
\begin{equation}\label{GBterm}
    \mathcal{L}_2 = R_{\mu\nu\lambda\sigma}R^{\mu\nu\lambda\sigma}-4R_{\mu\nu}R^{\mu\nu}+R^2.
\end{equation}
Therefore, the second-order Lovelock theory is also called the Einstein-Gauss-Bonnet theory.
Higher than six dimensional Lovelock theories can also contain higher than second order curvature corrections, such as
\begin{eqnarray} &\mathcal{L}_3 = 24 R_{\mu\nu}^{\phantom{\mu\nu}\rho\sigma}R_{\rho}^{\phantom{\rho}\nu}R_{\sigma}^{\phantom{\sigma}\nu}-24R_{\mu\nu}^{\phantom{\mu\nu}\rho\sigma}
R_{\rho\eta}^{\phantom{\rho\eta}\mu\nu}R_{\sigma}^{\phantom{\sigma}\eta} + 3R R_{\mu\nu}^{\phantom{\mu\nu}\rho\sigma}R_{\sigma\rho}^{\phantom{\sigma\rho}\mu\nu}\nonumber\\ &+
8R_{\mu\nu}^{\phantom{\mu\nu}\rho\sigma}R_{\rho\eta}^{\phantom{\rho\eta}\mu\kappa}R_{\kappa\sigma}^{\phantom{\kappa\sigma}\nu\eta} +
2R_{\mu\nu}^{\phantom{\mu\nu}\rho\sigma}R_{\rho\sigma}^{\phantom{\rho\sigma}\eta\kappa}R_{\eta\kappa}^{\phantom{\eta\kappa}\mu\nu} +
16R_{\mu}^{\phantom{\mu}\nu}R_{\nu}^{\phantom{\nu}\rho}R_{\sigma}^{\phantom{\sigma}\mu}\nonumber\\ &- 12R R_{\mu}^{\phantom{\mu}\nu}R_{\nu}^{\phantom{\nu}\mu} + R^3 \nonumber\end{eqnarray}
and so on, including higher Lovelock terms. Thus, the Einstein-Gauss-Bonnet theory is the simplest one, when only $c_2$ is nonvanishing, being of special interest in this thesis.

String theory predicts quantum corrections to classical General Relativity and the first relevant correction corresponds to the Gauss-Bonnet correction. In higher than four dimensional spacetimes, the low-energy limit of the heterotic string theory predicts the Gauss-Bonnet correction to the
Einstein action. The Einstein-Gauss-Bonnet action takes the form
\begin{equation}\label{HC:EGB:0}
  S_{EGB}=\frac{1}{16\pi G_D}\int d^Dx\,\mathcal{L}_{EGB},
\end{equation}
where
\begin{equation}\label{HC:EGB:1}
  \mathcal{L}_{EGB}=R-2\Lambda+\alpha\left(R_{\mu\nu\lambda\sigma}R^{\mu\nu\lambda\sigma}-4R_{\mu\nu}R^{\mu\nu}+R^2\right)
\end{equation}
 and $\alpha=1/{2\pi\,\ell^2_s}=2c_2$ is a coupling constant. The variation of action (\ref{HC:EGB:0}) allows one to find the Einstein-Gauss-Bonnet equation. All the geometrical
 information is now contained in the Einstein-Gauss-Bonnet tensor
\begin{equation}\label{HC:EGB:6}
  G_{\mu}^{\phantom{\mu}\nu}=\Lambda\,G_{(0)\mu}^{\phantom{(0)\mu}\nu}+ G_{(1)\mu}^{\phantom{(1)\mu}\nu}+\alpha\, \ G_{(2)\mu}^{\phantom{(2)\mu}\nu},
\end{equation}
where
\begin{eqnarray}
  G_{(0)\mu}^{\phantom{(0)\mu}\nu} &=& \delta_\mu^\nu,\label{BHP:3} \\
  G_{(1)\mu}^{\phantom{(1)\mu}\nu} &=& R_\mu^{\phantom{\mu}\nu}-\frac{1}{2}\delta_\mu^\nu\,R\label{BHP:4}
\end{eqnarray}
and
\begin{eqnarray}
  G_{(2)\mu}^{\phantom{(2)\mu}\nu} &=& R_{\lambda\mu}^{\phantom{\lambda\mu}\sigma\rho}R_{\sigma\rho}^{\phantom{\sigma\rho}\lambda\nu}-
  2R_\sigma^{\phantom{\sigma}\lambda}R_{\lambda\mu}^{\phantom{\lambda\mu}\sigma\nu} -2R_\mu^{\phantom{\mu}\lambda}R_\lambda^{\phantom{\lambda}\nu}+R\,R_\mu^{\phantom{\mu}\nu}- \nonumber\\
  & & \frac{1}{4}\delta_\mu^\nu\left(R_{\lambda\sigma}^{\phantom{\lambda\sigma}\rho \delta}R_{\rho\delta}^{\phantom{\rho\delta}\lambda\sigma}
  -4R_\lambda^{\phantom{\lambda}\sigma}R_\sigma^{\phantom{\sigma}\lambda}+ R^2\right)\label{BHP:5}
\end{eqnarray}
is the contribution to the Einstein equations up to the second order in curvature.

Some time ago higher-dimensional black holes attracted much interest, because it was expected that the Large Hadron Collider (LHC) might produce the miniature black holes living in higher than four dimensional spacetime.
Another motivation is due to the AdS/CFT correspondence which implies that the dynamics of $D-$dimensional field theory at finite temperature is equivalent to a $(D+1)-$dimensional black hole. In spite of everything said, the original and most important motivation for the study of higher-dimensional gravity is the String Theory where a higher number of dimensions is needed for the consistency. Higher-dimensional black holes have brought forward many new features, which could not have been studied in four dimensions.

\section{Black-hole solutions}\label{HC:Black}
We consider here static spherically symmetric black holes in $D$ dimensions. Thus, the metric coefficients do not depend on the time coordinate associated with the time-like Killing vector, orthogonal to the spatial hypersurfaces, and on the $n=D-2$ angular coordinates, being the functions of the radial coordinate only. The corresponding line element can be written in the form
\begin{equation}\label{HC:Black:1}
  ds^2=-f(r)dt^2+\frac{dr^2}{f(r)}+r^2\bar{\gamma}_{ij}dx^i\,dx^j,
\end{equation}
where $\bar{\gamma}_{ij}$ is the metric of the $n$-dimensional manifold. The vacuum solution of the Lovelock gravity is given by
\begin{equation}\label{HC:Black:2}
  f(r)=\kappa-r^2\,\psi(r),
\end{equation}
with the constant curvature $\kappa=-1,\,0$ or $1$ (the latter case, $\kappa=1$, corresponds to the compact spherical black hole), where $\psi(r)$ satisfies
\begin{equation}\label{HC:Black:3}
  W[\psi]=\sum_{m=2}^{k}\left[\frac{n\,c_m}{2}\left(\prod_{p=1}^{2m-2}(n-p)\right)\psi^m\right]+\frac{n}{2}\psi -\frac{\Lambda}{n+1}=\frac{\mu}{r^{n+1}}
\end{equation}
and $\mu$ is a constant of integration which is related to the ADM mass of the black hole
\begin{equation}\label{HC:Black:4}
  M=\frac{n\,\mu\,\pi^{(n+1)/2}}{\Gamma\left((n+1)/2\right)}.
\end{equation}

Now, let us fix the horizon radius of the black hole at $r_H$ which corresponds to one positive root of the equation $f(r)=0$. For $\Lambda\leq0$, $r_H$ is the greatest root. For $\Lambda>0$ the greatest one corresponds to the cosmological horizon $r_C$, so the horizon radius is the second largest root. From (\ref{HC:Black:2}), we find $\psi_H\equiv\psi(r_H)=\kappa/r_H^2$. Using this relation and (\ref{HC:Black:3}), we obtain the algebraic expression for the mass parameter in terms of the horizon radius
\begin{eqnarray}
  W[\psi_H] &=& \sum_{m=2}^{k}\left[\frac{n\,c_m}{2}\left(\prod_{p=1}^{2m-2}(n-p)\right)\psi_H^m\right]+\frac{n}{2}\psi_H -\frac{\Lambda}{n+1}=\frac{\mu}{r_H^{n+1}}\nonumber \\
  \mu &=& \frac{n\,r_H^{n-1}}{4}\left(2\kappa+ \kappa^m\,\sum_{m=2}^{k}\frac{2c_m(n-1)!}{(n-2m+1)!\,r_H^{2m-2}}- \frac{4\,\Lambda}{n(n+1)}r_H^2\right).\label{HC:Black:5}
\end{eqnarray}

In order to examine the singularity in the solutions, we calculate the Kretschmann scalar
\begin{equation}\label{HC:Black:6}
  R_{\mu\nu\lambda\rho}R^{\mu\nu\lambda\rho}=f''+\frac{2n\,(f')^2}{r^2}+2n(n-1)\frac{(\kappa-f)^2}{r^4},
\end{equation}
where primes denote derivatives with respect to the radial coordinate $r$.

Hence, a singularity appears either at $r=0$ or in a point where $f'$ diverges. From (\ref{HC:Black:2}), we find $$f'=-2r\,\psi+r^2\,\psi',$$ and conclude that $f'$ diverges only at the point where $\psi'$ diverges. Taking the derivative of (\ref{HC:Black:3}), we have
\begin{equation}\label{HC:Black:7}
  \psi'=-\frac{(n+1)\mu}{r^{(n+2)}\partial_\psi\,W[\psi]}.
\end{equation}
Thus, singularities appear either in the points where $\partial_\psi\,W=0$ or at $r=0$.

An exact static vacuum solution of Einstein-Gauss-Bonnet equations can be written as (\ref{HC:Black:1}) and, (\ref{HC:Black:3}) takes the form
\begin{equation}\label{HC:EGB:3}
  W[\psi]=\frac{\alpha\,n(n-1)(n-2)}{4}\psi^2+\frac{n}{2}\psi-\frac{\Lambda}{n+1}=\frac{\mu}{r^{n+1}}.
\end{equation}
For $\kappa=1$ we have a compact (spherical) black hole with the event-horizon radius $r_H$ which corresponds to the greatest positive root of the equation $f(r)=0$ as we already mentioned above. Being measured in units of length for any value of $D$, this quantity is convenient for parametrization of the black-hole mass, which can be expressed from (\ref{HC:Black:5}) as
\begin{equation}\label{HC:EGB:3.5}
  \mu=\frac{n\,r_H^{n-1}}{4}\left(2+\frac{\alpha(n-2)(n-1)}{r_H^2}-\frac{4\Lambda\,r_H^2}{n(n+1)}\right).
\end{equation}
where $\alpha$ is a coupling constant, which needs in general be a real quantity, and may to be negative in the context of holography within the AdS/CFT correspondence. In the context of higher dimensional gravity, the Gauss-Bonnet coupling constant $\alpha$ is positive, since it is inversely proportional to the square of the string length.
The quadratic equation (\ref{HC:EGB:3}) has two solutions, given by
\begin{equation}\label{HC:EGB:4}
  \psi(r)=\frac{1}{\alpha(n-1)(n-2)}\left(\varepsilon\sqrt{1+\frac{4\alpha(n-1)(n-2)}{n}\left(\frac{\mu}{r^{n+1}}+ \frac{\Lambda}{n+1}\right)}-1\right),
\end{equation}
where $\varepsilon=\pm 1$. If the cosmological constant vanishes, the two branches correspond to asymptotically flat ($\varepsilon=1$, $f(r)\sim 1$ as $r\to\,+\infty$) and
asymptotically anti-de Sitter ($\varepsilon=-1$, $f(r)\sim 1+\Lambda_{-}r^2$ as $r\to\,+\infty$) solutions.

\section{Black holes in EGB-dS space-times}\label{dS:Black}
Recent observations show that the Universe expands with acceleration and, within General Relativity, it can be modeled by a de Sitter space, which corresponds to a positive cosmological constant ($\Lambda>0$). In higher-dimensional spacetimes, de Sitter metric takes the form (\ref{HC:Black:1})-(\ref{HC:Black:2}), where the function $\psi(r)$ take the form of (\ref{HC:EGB:4}) for $\varepsilon=1$.
Moreover, the cosmological constant $\Lambda$ can be parametrized as
\begin{equation}\label{dS:Black:2}
  \Lambda=\frac{n(n+1)}{2}\left(\frac{r_C^{n-1}-r_H^{n-1}}{r_C^{n+1}-r_H^{n+1}} +\frac{\alpha(n-1)(n-2)}{2}\frac{r_C^{n-3}-r_H^{n-3}}{r_C^{n+1}-r_H^{n+1}}\right),
\end{equation}
where $r_C>r_H$ is the cosmological horizon.

In the limit $r_C\rightarrow r_H$ we obtain the extremal value of the cosmological constant, which is given as follow
\begin{equation}\label{dS:Black:3}
  \Lambda_{extr}=\frac{n(n-1)}{2r_H^4}\left(r_H^2+\frac{(n-2)(n-3)}{2}\alpha\right).
\end{equation}
Limit $r_C\rightarrow +\infty$ corresponds to the asymptotically flat spacetime ($\Lambda=0$). Hereafter we measure all the quantities in units of the event horizon, i.e., we introduce dimensionless parameters, $0\leq\frac{r_H}{r_C}<1$ and $\frac{\alpha}{r_H^2}\geq 0$, while frequencies are measured in units of the inverse horizon radius $r_H^{-1}$.

\begin{figure}[H]
\centering
\resizebox{0.8\linewidth}{!}{\includegraphics*{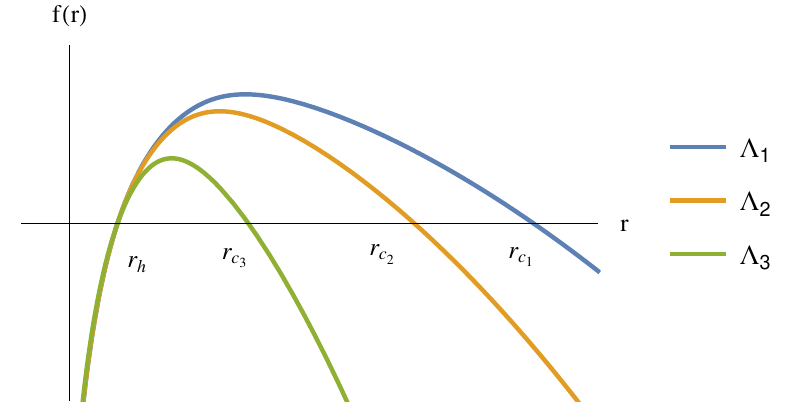}}
\caption{Event horizons and cosmological horizons ($\Lambda_1<\Lambda_2<\Lambda_3$) are two roots of $f(r)=0$.}\label{profile-03}
\end{figure}

For an observer placed between $r_C$ and $r_H$ (horizon radius), static black-hole solutions (\ref{HC:Black:2}) and (\ref{HC:EGB:4}) in an asymptotically de Sitter space are limited by the event horizon and the cosmological horizon $r_C\geq r_H$ in the spatial coordinate, where the cosmological horizon approaches to the event horizon for higher values of the cosmological constant (as we can see in the figure \ref{profile-03}).

Furthermore, if $\varepsilon=1$, the limit $\alpha\rightarrow 0$ of the Einstein-Gauss-Bonnet gravity leads to a higher-dimensional solution of the Einstein theory -- Schwarzschild-Tangherlini -de Sitter spacetime
\begin{eqnarray}
  f(r) &=& \kappa-\frac{2r^2}{n}\left(\frac{\mu}{r^{n+1}}+\frac{\Lambda}{n+1}\right),\label{HC:EGB:5} \\
  \Lambda &=& \frac{n(n+1)}{2r_C^2}.\label{HC:EGB:501}
\end{eqnarray}

\section{Black-hole thermodynamics}\label{HC:Hawking}
In 1970, Demetrios Christodoulou \cite{PhysRevLett.25.1596} introduced the term of irreducible mass $m_{ir}$, as a nondecreasing quantity for a particular process of the stationary axisymmetric solution of Einstein equation (Kerr black hole). It turns out that $m_{ir}$ is proportional to the square root of the black hole area $m_{ir}^2=A_{H}/{16\pi}$, thus, it implies that the area always increases. This result is currently known as the second law of the thermodynamics or Hawking area theorem since in 1971 Stephen Hawking \cite{PhysRevLett.26.1344} gave a general proof of this statement within classical general relativity. As the mass increases because of the matter falling into the black hole we observe the growth of the black-hole size. In 1973, Bekenstein \cite{PhysRevD.7.2333} argued that black holes should have a finite entropy proportional to the area of the event-horizon surface, though this could contradict the fact that they have zero temperature and the second law of the thermodynamics (which states that entropy of any closed system can never decrease) might be violated. Bekenstein indicated that actually the second law is not violated, considering that the entropy of the universe $S$ including that of the black hole $S_H$ cannot decrease whatever happens. Some months later, J.M. Barden, B. Carter and S.W. Hawking \cite{Bardeen1973} proposed the four laws which were very similar to the laws of thermodynamics.
\begin{itemize}
  \item Zeroth law : The surface gravity $\kappa_H$ is constant over the event horizon.
  \item First law : The variation of the black-hole mass $M$ (energy) is related to the variations of the angular momentum $J$ and charge $Q$, in such a way that its area plays a role of the entropy
      \begin{equation}\label{HC:Hawking:0.5}
        \delta M=\frac{\hbar\kappa_H}{2\pi\,G_N}\delta A_H+\Omega_H\,\delta J+\Phi_H\,\delta Q.
      \end{equation}
  \item Second law : In any classical process, the area of the black hole event horizon never decreases.
  \item Third law : In any physical process it is not possible to reduce the surface gravity $\kappa_H$ to zero.
\end{itemize}
The zeroth law would be the analogue of the thermal equilibrium ($T$ constant) of a thermodynamical system. Since $T_H\propto\kappa_H$ and the surface gravity $\kappa_H$ is constant on the horizon, thus the temperature is also constant on the event horizon. The first law and considering the proportional relation between the area of stationary black hole and the entropy
\begin{equation}\label{HC:Hawking:0}
  S_H=\frac{A_H}{4 l_s^2},\qquad\qquad l_s^2=\hbar G/c^3,
\end{equation}
where $l_s$ is the Planck length, show that any variations of the black hole mass are expressed as contributions of variations in the angular momentum, charge or entropy. Once quantum mechanics comes into play we need to modify the second law, because Hawking radiation implies reducing the surface area of the black hole. Thus in terms of the black hole entropy $S_H$, the correct statement of the second law would be, the entropy of the universe $S=S_H+S_{ext}$ cannot decrease, where $S_{ext}$ is the entropy of the world excluding the black hole. Furthermore, the third law discards the formation of extremal configurations in any physical process, like the extremal charged and extremal Kerr black holes, since they violate the statement.

Yet, physical temperature of a \emph{classical} black hole is considered as absolute zero, so no physical relation between $\kappa_H$ and $T_H$ could be established and, consequently, it also would be inconsistent to assume a physical relation between $S_H$ and $A_H$. Some years later, Hawking solved this apparent paradox, by using quantum theory. In the presence of a quantum field near the horizon Hawking proved that a black hole behaves like a black body, having a characteristic thermal spectrum given by the effective temperature
\begin{equation}\label{HC:Hawking:1}
  T_H=\frac{\hbar\kappa_H}{2\pi c k}.
\end{equation}

This temperature has not been directly observed yet. For an astrophysical black hole, Hawking's temperature is some orders of magnitude lower than the cosmic microwave background, which makes it practically impossible to measure. Furthermore, the thermodynamics of the Lovelock theory has been studied during the recent years. The zeroth law has been shown to hold in \cite{Sarkar:2012wy} when the matter satisfies the dominant energy condition. The first law was also verified and used in \cite{PhysRevD.38.2434,Jacobson:1993xs} in order to obtain the expression for the entropy of Lovelock black holes. An extended version of the first law has also been proposed in \cite{Kastor:2010gq}, where the Lovelock couplings play the role of extra thermodynamic variables. It has been shown in \cite{Akbar:2008vz,Sarkar:2010xp,Kolekar:2012tq} that the second law holds for general Lovelock theories. Some studies of the third law can be also found in \cite{Torii_2006}. For static spherically symmetric black holes (\ref{HC:Black:1}), the Hawking temperature is
\begin{equation}\label{HC:Hawking:2}
  T_H=\frac{f'(r_+)}{4\,\pi}.
\end{equation}

\section{Gravitational perturbations}\label{BHP:Gravitational}
On February 11th 2016 there was the first announcement of gravitational wave observations by LIGO and Virgo collaborations \cite{Abbott:2016blz,TheLIGOScientific:2016src} from a pair of merging binary black holes. This observation gave birth to a new era of gravitational waves astronomy of astrophysical compact objects. There are, up today, 11 confirmed gravitational waves detections being 10 from binary black holes (BBH) mergers and one from a binary neutron star (BNS) spiral GW$170817$. Three BBH reported in its first observing run ($O_1$) \cite{TheLIGOScientific:2016src,PhysRevLett.116.241103} from 12th September 2015 to 19th January 2016, others three BBH and one BNS were reported in the second run $O_2$ from 30th November 2016 to 25th August 2017. After refining of estimates, 4 news detections (initially considered as false alarms) were renamed as gravitational wave (GW) signals \cite{LIGOScientific:2018mvr}.

Theoretical treatment of gravitational perturbations in the linear approximation was widely studied in $4$-dimensional and higher spacetimes. When adding a small perturbation $h_{\mu\nu}$ to the unperturbed metric (\ref{HC:Black:1}) labeled by $g^0_{\mu\nu}$,
\begin{equation}\label{BHP:1}
  g_{\mu\nu}\rightarrow g^0_{\mu\nu}+h_{\mu\nu},\qquad\quad |h_{\mu\nu}|\ll|g^0_{\mu\nu}|,
\end{equation}
the linearized equation for $h_{\mu\nu}$ is derived using the new perturbed metric in the Einstein-Gauss-Bonnet equation. The variation of (\ref{HC:EGB:6}), for vacuum spacetimes
\begin{equation}\label{BHP:2}
  \delta G_{\mu}^{\phantom{\mu}\nu}=\Lambda\,\delta G_{(0)\mu}^{\phantom{(0)\mu}\nu}+\delta G_{(1)\mu}^{\phantom{(1)\mu}\nu}+\alpha\, \delta G_{(2)\mu}^{\phantom{(2)\mu}\nu}=0,
\end{equation}
leads to a second order differential equation for $h_{\mu\nu}$. It is convenient to represent the tensor components of $h_{\mu\nu}$ with respect to the transformation law under rotations
on the $(D-2)$-sphere. The linear perturbations, then, can be classified into \emph{tensor}, \emph{vector}, and \emph{scalar} types denoted by $s$, $v$, and $t$, respectively,
\begin{equation}
h_{\mu\nu}=
\begin{pmatrix}
& s & s & v & \cdots & v\\
& s & s & v & \cdots & v\\
& v & v & t & \cdots & t\\
& \vdots & \vdots & \vdots & \ddots & \vdots\\
& v & v & t & \cdots & t
\end{pmatrix},
\label{BHP:6}
\end{equation}
each of which can be treated independently from the others. After a lot of algebra and separation of variables, the gravitational perturbation equations within the EGB de Sitter theory can be reduced to a second-order master differential equations with some effective potentials,
\begin{equation}\label{ci0}
  \left(\frac{\partial^2}{\partial\,t^2}-\frac{\partial^2}{\partial\,r_*^2}+V_i(r(r_*))\right)\Psi(t,r(r_*))=0,
\end{equation}
where $i$ stands for $t$ (tensor), $v$ (vector), and $s$ (scalar) perturbations and $r_*$ is the tortoise coordinate, which satisfies $dr_*=dr/f(r)$. The explicit form of the effective potential is given in the following subsections. We will present a brief summary of this approach to the perturbation analysis applied to the Einstein-Gauss-Bonnet gravity \cite{Takahashi:2010ye}.

\subsection{Tensor-type perturbations}\label{BHP:Gravitational:Tensor}
Under transverse ($h_{\mu\nu;\lambda}=0$) and traceless ($h_\mu^{\phantom{\mu}\mu}=0$) conditions, tensor-type perturbations take the form
\begin{equation}\label{BHP:7}
  h_{\mu\nu}=\left(
  \begin{array}{cc}
    \begin{array}{cc}
      0 & 0 \\
      0 & 0
    \end{array} & \begin{array}{ccc}
                    0 & \cdots & 0 \\
                    0 & \cdots & 0
                  \end{array} \\
    \begin{array}{cc}
      0 & 0 \\
      \vdots & \vdots \\
      0 & 0
    \end{array} & {\Huge\left(\boldsymbol{\bar{h}_{ij}}\right)}
  \end{array}
  \right)\,.
\end{equation}
Since, $\bar{h}_{ij}$ is a tensor on the unit surface of constant curvature, $S_n^\kappa$, it can be expanded in tensor harmonics $\mathbb{T}_{ij}$, as $\bar{h}_{ij}=r^2\phi(t,r)\mathbb{T}_{ij}$, satisfying
\begin{equation}\label{BHP:8}
  \left(\bar{\nabla}^k\bar{\nabla}_k + \kappa_t^2 \right)\mathbb{T}_{ij}= 0,\qquad\mathbb{T}_{ij;k}=0,\qquad\mathbb{T}_i^{\phantom{i}i}=0,
\end{equation}
where $\kappa_t^2$ are discrete eigenvalues $\kappa_t^2=\ell(\ell+n-1)-2$, ($\ell=2,3,\ldots$) for $\kappa=1$, and real positive values for $\kappa=-1,0$. In order to find a master perturbation equation, we substitute (\ref{BHP:8}) into (\ref{BHP:2}). After some algebra the resultant equation is the master second-order differential equations given by (\ref{ci0}) and the effective potential for tensor-type perturbation is
\begin{equation}
V_t(r)=\dfrac{\ell(\ell+n-1)f(r)T^{\prime\prime}(r)}{(n-2)r\,T^{\prime}(r)}+ \dfrac{f(r)}{R(r)}\dfrac{d}{dr}\left(f(r)\dfrac{d\,R(r)}{dr}\right),\label{BHP:11}
\end{equation}
where
\begin{eqnarray}
  T(r)&=&r^{n-1}\left(1+r^{-2}\alpha(n-1)(n-2)(\kappa-f(r))\right),\label{BHP:9}\\
  R(r)&=&r\sqrt{T^\prime(r)}.\label{BHP:9.5}
\end{eqnarray}

\subsection{Vector-type perturbations}\label{BHP:Gravitational:Vector}
Taking the Regge-Wheeler gauge, the vector-type perturbations are given by
\begin{equation}\label{BHP:12}
  h_{\mu\nu}=\left(
  \begin{array}{cc}
    \begin{array}{cc}
      0 & 0 \\
      0 & 0
    \end{array} & {\large\left(\boldsymbol{r\,f_a\,\mathbb{V}_i}\right)} \\
   {\Large{\bf Sym}} &
   \begin{array}{ccc}
      0 & \cdots & 0 \\
      \vdots & \ddots & \vdots\\
      0 & \cdots & 0
   \end{array}
  \end{array}
  \right)
\end{equation}
where $f_a$ is a function of $(t,r)$, ``{\bf Sym}'' represents the symmetric part of $h_{\mu\nu}$, $\mathbb{V}_i$ are vector harmonics, defined on $S_n^\kappa$, and satisfying the divergence-free condition
\begin{equation}\label{BHP:13}
  \left(\bar{\nabla}^k\bar{\nabla}_k + \kappa_v^2 \right)\mathbb{V}_i= 0,\qquad\bar{\nabla}_ i\mathbb{V}^i=0,
\end{equation}
where $\kappa_v^2$ are discrete eigenvalues $\kappa_v^2=\ell(\ell+n-1)-1$, ($\ell=2,3,\ldots$) for $\kappa=1$, and real non-negative numbers for $\kappa=0,-1$. In the same way as for the tensor-type perturbations, we reduce the linear Lovelock equations (\ref{BHP:2}) to the master equation (\ref{ci0}). Therefore the explicit form of the effective potential is given by
\begin{equation}\label{BHP:14}
  V_v(r)=\frac{(\ell-1)(\ell+n)f(r)T^\prime(r)}{(n-1)r\,T(r)}+ f(r)R(r)\dfrac{d}{dr}\left(f(r)\dfrac{d}{dr}\left(\dfrac{1}{R(r)}\right)\right).
\end{equation}
Note that we have not assumed any particular form for $f(r)$, so that the Schr\"odinger-type master equations govern linear perturbations of any given static spherically symmetric spacetime within the Lovelock theory.

\subsection{Scalar-type perturbations}\label{BHP:Gravitational:Scalar}
For the scalar-type perturbations of static spherically symmetric black holes (\ref{HC:Black:1}) we use the Zerilli gauge,
\begin{equation}\label{BHP:15}
  h_{\mu\nu}=\left(
  \begin{array}{cc}
   {\large\left(\boldsymbol{\mathcal{F}_{ab}\mathbb{S}}\right)}  & {\large\left(\boldsymbol{r\,\mathcal{F}_a\,\mathbb{S}_i}\right)} \\
   {\Large{\bf Sym}} & {\large\left(\boldsymbol{2r^2\,H_L\,\bar{\gamma}_{ij}\,\mathbb{S}}\right)}
  \end{array}
  \right).
\end{equation}
where $\mathcal{F}_a$, $\mathcal{F}_{ab}$, $H_L$ are functions of $(t,r)$, ``{\bf Sym}'' represents the symmetric part of $h_{\mu\nu}$, and $\mathbb{S}$ is a scalar harmonic
\begin{equation}\label{BHP:16}
  \left(\bar{\Delta} + \kappa_s^2 \right)\mathbb{S}= 0,
\end{equation}
and $\mathbb{S}_i$ is the scalar-type vector harmonic
\begin{equation}\label{BHP:17}
  \mathbb{S}_i=-\frac{1}{\kappa_s}\bar{\nabla}_i\,\mathbb{S},
\end{equation}
which satisfies
\begin{eqnarray}
  &\left(\bar{\Delta}+\kappa_s^2-(n-1)\kappa\right)\mathbb{S}_i=0, \label{BHP:18}\\
  &\bar{\nabla}_i\,\mathbb{S}^i=\kappa_s\,\mathbb{S}.\label{BHP:19}
\end{eqnarray}
where $\kappa_s^2$ are discrete eigenvalues $\kappa_s^2=\ell(\ell+n-1)$, ($\ell=2,3,\ldots$) for $\kappa=1$, and real non-negative numbers for $\kappa=0,-1$. It is always possible to expand the perturbation $h_{\mu\nu}$ in scalar harmonics, then, the Lovelock equation $\delta G_{\mu\nu}=0$ can be reduced again into a master equation ($\ref{ci0}$), where the effective potential is
\begin{equation}\label{BHP:20}
  V_s(r)=\frac{2\ell(\ell+n-1)f(r)}{nr^2\,B(r)}\frac{d}{dr}\left(r\,B(r)\right)+ f(r)B(r)\dfrac{d}{dr}\left(f(r)\dfrac{d}{dr}\left(\dfrac{1}{B(r)}\right)\right),
\end{equation}
and
\begin{equation}\label{BHP:21}
  B(r)=\frac{2(\ell-1)(\ell+n)-nr^3\psi^\prime}{R(r)}T(r).
\end{equation}

\section{Quasinormal modes and stability}\label{BHP:Stability}
\begin{figure}[H]
\resizebox{\linewidth}{!}{\includegraphics*{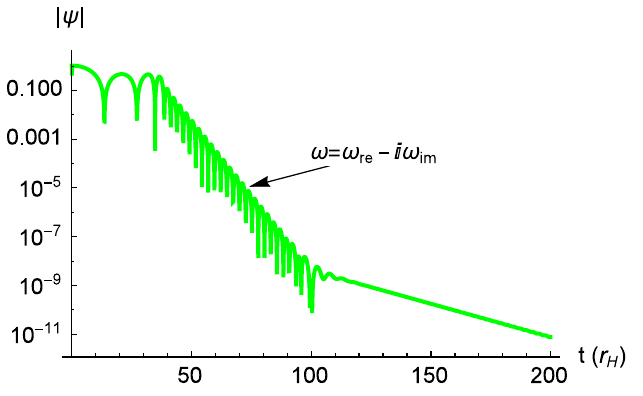}}
\caption{Oscillatory decay of black hole perturbations (quasinormal ringing).}\label{profile-01}
\end{figure}

Small perturbations of a black hole lead to emission of gravitational waves dominated by damped oscillations, which are characterized by a set of complex frequencies $\omega$. The real part $\re{\omega}$ describes the oscillation frequency of the perturbation whereas the imaginary part $\im{\omega}$ shows us the rate of decay (see Fig. \ref{profile-01}). These complex frequencies are called \emph{quasinormal modes}, since the waves lose energy at the event horizon of a black hole and at spatial infinity. Moreover, the quasinormal frequencies (QNMs) depend on the parameters of the black hole, such as mass $M$, charge $Q$, and angular momentum $J$ and do not depend on the way of the initial perturbation. In this sense, the QN frequencies are called the fingerprint of the black hole.

\begin{figure}[H]
\resizebox{\linewidth}{!}{\includegraphics*{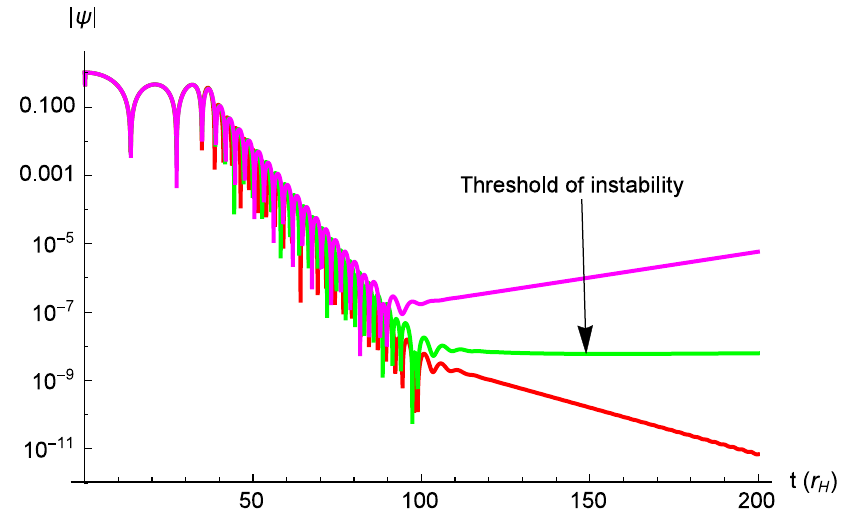}}
\caption{Evolution of perturbations near the threshold of instability.}\label{profile-02}
\end{figure}

In order to study quasi-normal spectrum of a black hole it is convenient to make the following transformation for the function in (\ref{ci0}),
\begin{equation}\label{BHP:21.5}
  \Psi(t,r)=\sum_{n=0}^{\infty}\Phi_{\omega_n}e^{-i\omega_n t},
\end{equation}
where $\omega_n$ are the quasinormal spectrum and the functions $\Phi_{\omega_n}$ satisfy the linear equation
\begin{equation}
\left(\frac{d^2}{dr_*^2}+\omega^2-V_i(r)\right)\Phi_\omega(r)=0.
\end{equation}
The Schr\"odinger wave-like equation with the appropriate boundary conditions, such as
\begin{eqnarray}
  r_*\rightarrow -\infty &:& \Phi_\omega \propto e^{-i\omega r_*}\label{BHP:22},\\
  r_*\rightarrow +\infty &:& \Phi_\omega \propto e^{+i\omega r_*}\label{BHP:23},
\end{eqnarray}
leads to the discrete spectrum of frequencies mentioned above. These boundary conditions are purely ingoing wave at the event horizon and outgoing wave at the cosmological horizon if $\Lambda>0$. For asymptotically flat solutions ($\Lambda=0$), the boundary condition at the event horizon is the same and the one at spatial infinity is purely outgoing. It is clear, that a black hole can exist only if its perturbations are of a finite amplitude. If the perturbations grow infinitely, the black hole is unstable. We observe, therefore, that the parametric threshold of instability happens when the imaginary part of the dominant mode is zero. Gravitational signal from a perturbed black hole is dominated by a nonoscillatory tail at late times (see Fig. \mbox{\ref{profile-02}}), and the threshold of the instability happens when the ``tail'' changes from decaying to the growing one.

QNMs and linear (in)stability were studied by several researchers within the context of Einstein's gravity. Moreover, Gauss-Bonnet gravity was also under consideration, as far as we are aware, first in 1989 by B.R. Iyer, S. Iyer and C.V. Vishveshwara \cite{Iyer_1989}. There, they analyzed scalar field around a neutral black hole in six dimensions within the Einstein-Gauss-Bonnet gravity. Around 15 years later, R. Konoplya \cite{Konoplya:2004xx} studied modes in five and six dimensions for the charged configuration. It gave rise to further publications in this area, for example, for black holes in the asymptotically de-Sitter (dS) and anti-de-Sitter (AdS) backgrounds \cite{Abdalla:2005hu}. Test fields analyzed by the time-domain integration \cite{Abdalla:2005hu,Zhidenko:2008fp,Gonzalez:2017gwa} showed linear stable perturbations. In addition, linear gravitational perturbations of the asymptotically flat Gauss-Bonnet black holes \cite{Dotti:2005sq,Gleiser:2005ra} found unstable solutions and similar instabilities occurs also for asymptotically A(dS) spacetimes, though an extensive study showed that gravitational perturbations become stable for lower multiple numbers and lower $\alpha$-coupling parameter \cite{Konoplya:2008ix}. The following natural step would be to study the stability of black holes in asymptotically de-Sitter spacetimes, which is the main work of the first part of this thesis, and is summarized in a work \cite{Cuyubamba:2016cug}. Moreover, there were already found instabilities for black holes in asymptotically AdS spacetime in the EGB configuration within linear perturbations in some region of parameters \cite{Konoplya:2017ymp}.

In the general Lovelock theory, there were found instabilities of gravitational perturbations for sufficiently small black holes \cite{Takahashi:2010ye,Takahashi:2010gz,Yoshida:2015vua}, then extended to the charged black hole \cite{Takahashi:2011qda,Takahashi:2012np} and to the purely Lovelock black hole \cite{Gannouji:2013eka}. We summarize the existing literature on this topic for black holes in the Lovelock theory in Table \ref{ReviewBH}.
\begin{table}
\centering
\begin{tabular}{|c|c|c|c|c|}
  \hhline{|=====|}
  Type of black hole & Test fields & Tensor-type & Vector-type & Scalar-type \\
  \hhline{|=====|}
  Gauss-Bonnet & \cite{Konoplya:2004xx,Abdalla:2005hu,Zhidenko:2008fp,Gonzalez:2017gwa} & \cite{Dotti:2005sq,Konoplya:2008ix} & \multicolumn{2}{c|}{\cite{Gleiser:2005ra,Konoplya:2008ix}}\\
    \hline
  Lovelock & - & \cite{Takahashi:2010ye,Takahashi:2010gz,Yoshida:2015vua} & \multicolumn{2}{c|}{\cite{Takahashi:2010ye,Takahashi:2010gz,Yoshida:2015vua}}\\
    \hline
  Lovelock with charge & - & \cite{Takahashi:2011qda} & \multicolumn{2}{c|}{\cite{Takahashi:2012np}}\\
    \hline
  Pure Lovelock & - & \cite{Gannouji:2013eka} & \multicolumn{2}{c|}{-} \\
    \hline
  Gauss-Bonnet-AdS & \cite{Abdalla:2005hu} & \multicolumn{3}{c|}{\cite{Konoplya:2017ymp}} \\
    \hline
  Gauss-Bonnet-dS & \cite{Abdalla:2005hu} & \multicolumn{3}{c|}{\cite{Cuyubamba:2016cug,Konoplya:2017ymp}} \\
    \hhline{|=====|}
\end{tabular}
\caption{Review of papers on quasinormal modes, (in)stability in linear perturbations of black holes in the modified Lovelock theory.}\label{ReviewBH}
\end{table} 

\chapter{Numerical Methods}\label{NM}
\section{Characteristic integration}\label{CI}
In our case, since the dynamical wave-like equation (\ref{ci0}) has a cumbersome form. The most straightforward way to prove (in)stability is to integrate the master equation numerically in the time domain which takes into consideration contributions from all the modes. We use the discretization scheme proposed by Gundlach, Price, and Pullin \cite{Gundlach:1993tp}. This method was used for calculation of quasinormal modes in a great number of works (see reference in \cite{Konoplya:2011qq}). Comparisons of the time-domain numerical data with the accurate frequency-domain calculations show excellent agreement not only in cases when a black hole is stable, but also near the onset of instability.

Rewriting (\ref{ci0}) in terms of the light-cone coordinates $du=dt-dr^*$ and $dv=dt+dr^*$, one finds that
\begin{equation}\label{ci0.5}
  4\frac{\partial^2\,\Psi}{\partial u\partial v}=-V_i(\frac{v-u}{2})\Psi.
\end{equation}

\begin{figure}[H]
\centering
\resizebox{0.8\linewidth}{!}{\includegraphics*{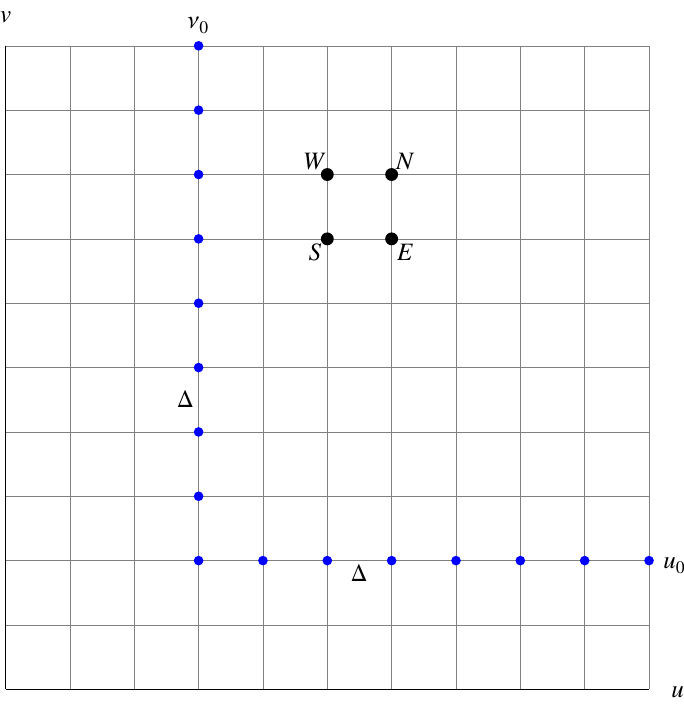}}
\caption{Grid of step $\Delta$ in light-cone coordinates plane. The blue points
represent the initial data.}\label{GridFigure-1}
\end{figure}

The discretization scheme has the following form
\begin{eqnarray}
\Psi(N)&=&\Psi(W)+\Psi(E)-\Psi(S)\nonumber\\
&&-\frac{\Delta^2}{8}V(S)\left[\Psi(W)+\Psi(E)\right]+\mathcal{O}(\Delta^4),\label{ci1}
\end{eqnarray}
where $N$, $W$, $E$ and $S$ are the points in a grid with step $\Delta$ in the discretized $u$-$v$ plane: $S=(u,v)$, $W=(u,v+\Delta)$, $E=(u+\Delta,v)$ and $N=(u+\Delta,v+\Delta)$. With the initial data specified on two null-surfaces $u = u_0$ and $v = v_0$ we are able to find values of the function $\Psi$ at each point of the grid (See figure \ref{GridFigure-1}). Since quasinormal modes and the asymptotical behavior of perturbations do not depend on initial conditions (as confirmed by several numerical simulations), we set $\Psi(u,v_0)=0$ on the $u$-axis and consider the Gaussian wave initial data on the $v$-axis.

This discretization scheme requires a number of operations which is proportional to $\Delta^{-2}$, what implies that the corresponding accumulated error is $\mathcal{O}(\Delta^{2})$. By decreasing the step $\Delta$ for the same initial data, we check the convergence of the integration scheme. Here we present the resulting profiles for the sufficiently small step $\Delta$, such that its further decreasing does not change the time-domain picture. Another source of error comes from numerical truncations in computations which we control by increasing precision of the floating-point arithmetics.

In order to determine the effective potential in each point of the grid, we need to calculate $r(r_*)=r((v-u)/2)$. We solve numerically the equation $dr_*=dr/f(r)$ using the Explicit-Runge-Kutta method implemented in the \emph{NDSolve} procedure of Wolfram Mathematica, then we invert the function using the Newton-Raphson method in order to evaluate the effective potential at each point of the grid in the $(u,v)$-plane. By applying the discretization scheme we calculate the amplitude of the wave function $\Psi$ for all the points of the grid. The diagonal elements ($u=v$) correspond to the time evolution of the perturbation at fixed coordinate $\Psi(t=n\Delta,r_*=(v_0-u_0)/2)$.

\section{Prony's method}\label{PM}
In order to find the quasinormal modes from the exponentially damping profile of the temporal evolution, we employ a direct approach known as the Prony's method considering the fitting of the profile data $\{\Psi(0),\Psi(\Delta)\cdots\Psi((N-1)\Delta)\}$ by a linear combination of damping exponents, such
\begin{equation}\label{pm-1}
  \Psi(t)=\sum_{k=1}^{p}a_ke^{-i\omega_k t},
\end{equation}
where $\omega_k$ are complex numbers corresponding to the quasinormal frequencies and $a_k$ are coefficients of the expansion. Since the data is discretized $\Psi(n\Delta)$, $n=0,\cdots,N$ and $\Delta$ is the step of the discretization, we have
\begin{eqnarray}
  \Psi(n\Delta) &=& \sum_{k=1}^{p}a_k\left(e^{-i\omega_k\Delta}\right)^n,\nonumber \\
  \Psi(n\Delta) &=& \sum_{k=1}^{p}a_k\eta_k^n.\nonumber
\end{eqnarray}
where $\eta_k=e^{-i\omega_k\Delta}$. In order to find the values of $\eta_k$, and thus $\omega_k$, we consider a polynomial equation, such that $\eta_k$ are its roots, $\prod_{k=1}^{p}(\eta-\eta_k)=0$. We rewrite the equation as
\begin{equation}\label{pm-2}
  \prod_{k=1}^{p}(\eta-\eta_k)=\sum_{m=0}^{p}c_m \eta^{p-m}\quad\Rightarrow\quad\sum_{m=0}^{p}c_m \eta_k^{p-m}=0,\,\,\,c_0=1,
\end{equation}
where $c_m$ are the unknown coefficients. Let us consider the following expression,
\begin{eqnarray}
   \sum_{m=0}^{p}c_m\Psi((n-m)\Delta) &=& \sum_{m=0}^{p}c_m\sum_{k=1}^{p}a_k\eta_k^{n-p}\eta_k^{p-m},\nonumber \\
    &=& \sum_{k=1}^{p}a_k\eta_k^{n-p}\sum_{m=0}^{p}c_m\eta_k^{p-m},\nonumber \\
    &=& 0.\nonumber
\end{eqnarray}
Since $c_0=1$ we obtain
\begin{equation}\label{pm-3}
   \sum_{m=1}^{p}c_m\Psi((n-m)\Delta) = -\Psi(n\Delta),
\end{equation}
or, explicitly, in the matrix form
\begin{equation}\label{pm-4}
   \underbrace{\left(
  \begin{array}{cc}
    \begin{array}{cc}
      \Psi((p-1)\Delta) & \Psi((p-2)\Delta) \\
      \Psi(p\Delta) & \Psi((p-1)\Delta)
    \end{array} &
    \begin{array}{cc}
      \cdots & \Psi(0) \\
      \cdots & \Psi(1)
    \end{array} \\
    \begin{array}{cc}
      \vdots & \vdots \\
      \Psi((N-1)\Delta) & \Psi((N-2)\Delta)
    \end{array} &
   \begin{array}{cc}
      \ddots & \vdots  \\
      \cdots & \Psi((N-p)\Delta)
   \end{array}
  \end{array}
  \right)}_M
  \underbrace{\left(
  \begin{array}{c}
    c_1 \\
    c_2 \\
    \vdots \\
    c_p
  \end{array}
  \right)}_C =
  -\underbrace{\left(
  \begin{array}{c}
    \Psi(p\Delta) \\
    \Psi((p+1)\Delta) \\
    \vdots \\
    \Psi(N\Delta)
  \end{array}
  \right)}_N.\nonumber
\end{equation}

The system of linear equations is overdetermined, and can be solved in the least-squares sense
 \begin{equation}\label{pm-5}
   C=-\left(M^+M\right)^{-1}M^+N,\nonumber
 \end{equation}
where, $M^+$ is the hermitian transposition of $M$. Moreover, $C$ represent the column matrix with the coefficient of the polynomial equation (\ref{pm-2}) we can determine, by numerical NDSolve procedure, the roots $\eta_k$. Once we calculate the coefficients $C$, we can solve numerically the polynomial equation (\ref{pm-2}) with respect to
\begin{equation}\label{pm-6}
  \eta_k=-e^{-i\omega_k\Delta}\quad\Rightarrow\quad\omega_k=\frac{i}{\Delta}\ln(\eta_k).
\end{equation}

Since the wave-like equation (\ref{ci0.5}) is real, for each frequency $\omega$, we obtain also its conjugate, $(-\omega^*)^*$, as a solution to (\ref{pm-2}).

Practically, the Prony's method allows us to determine at most the first three modes, because numerical error dominates over the contribution of higher overtones. For our purpose it is usually sufficient to determine the fundamental mode only. 

\chapter{(In)Stability in GB de Sitter black holes}\label{Asymptotically}
In this chapter, we will study stability regions for higher dimensional black hole which are asymptotically de Sitter ones ($\Lambda>0$). In order to determine the (in)stability regions, we use the characteristic integrations starting from the temporal profiles described in section \ref{CI}. It was observed in \cite{Konoplya:2008ix} that the accurate computation of $\psi(r)$ of (\ref{HC:EGB:4}) requires higher precision of arithmetic operations. We use its alternative equivalent form
\begin{equation}\label{ds:Black:1.5}
  \psi(r)=\frac{4\left(\frac{\mu}{r^{n+1}}+\frac{\Lambda}{n+1}\right)}{n+ \sqrt{n^2+4\alpha\,n(n-1)(n-2)\left(\frac{\mu}{r^{n+1}}+\frac{\Lambda}{n+1}\right)}},
\end{equation}
which apparently allows us to decrease the relative error. In order to check the stability of the algorithm, we compare the profiles found with the floating-point arithmetics of different precision. In particular, we observe that the double precision arithmetics (64 bits) is sufficient for our computations. We have also compared the obtained time-domain profiles for $\Lambda=0$ with those obtained with the high-precision code in \cite{Konoplya:2008ix} and found that the difference is smaller than the discretization-scheme error order.

\begin{figure}[H]
\centering
\resizebox{0.6\linewidth}{!}{\includegraphics*{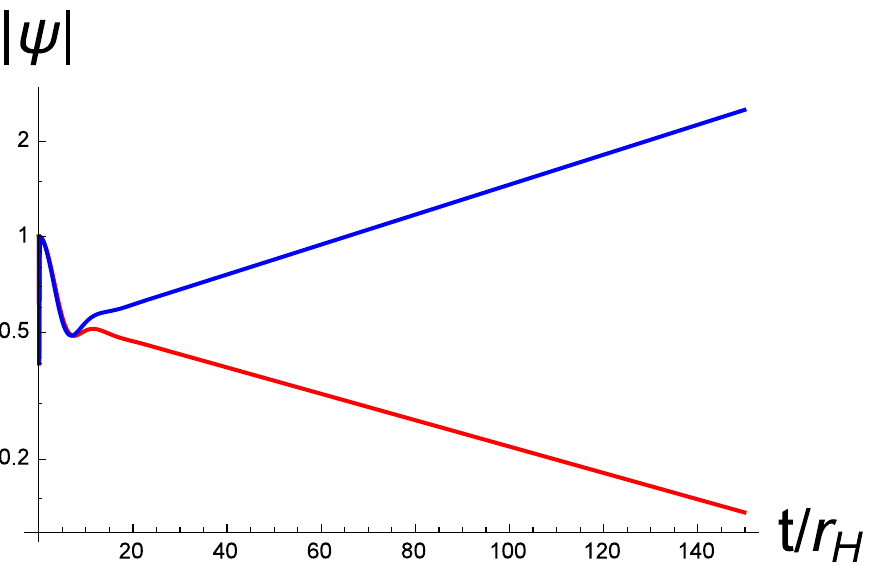}}
\caption{Nonoscillatory temporal profiles for scalar perturbations for $\alpha=0.5r_H^2$. The stable (red) and unstable (blue) profiles correspond to $\Lambda=0.8/r_H^2$ and $\Lambda=0.9/r_H^2$.}\label{profile-1}
\end{figure}

After fixing the horizon radius $r_H$ and the dimension $D$, the solution depends on two-parameters, $\Lambda$ and $\alpha$, which are the cosmological constant and the coupling parameter, respectively. Varying these parameters, the time-domain integration allows us to catch the threshold of instability. We had to obtain a great number of profiles in order to distinguish between a typical damped (stable) or a growing (unstable) time-domain profiles near the threshold of instability. Such profiles look as shown on fig.~\ref{profile-1}.

\section{Stability regions}\label{dS:Stability}
The results of our numerical time-domain integration at various values of parameters of GB-dS black holes will be show in this section. Looking for a threshold of instability, we observed instability in scalar and tensor channels of perturbations, while the vector sector showed no growing time-domain profiles.
\subsection{Instability of tensor-type perturbations}

In tensor-type perturbation, we found that the most unstable mode develops at $\ell=\infty$ which is why this type of instability is called the eikonal instability, emphasizing the fact the regime of geometrical optics. It means that the region of stability is getting smaller when going over to higher multiple numbers $\ell$ represented in the $(r_H/r_c,\alpha/r_H^2)$-plane. In order to catch the threshold limit at $\ell=\infty$, we fixed the cosmological constant $\Lambda$ ($r_H/r_c$ is fixed) and consider the critical $\alpha$ as a function of the inverse multiple number $\alpha(\ell^{-1})$. Then, after a linear extrapolation of several higher multiple numbers, the critical value of $\alpha_{crit}$ is reached. Moreover, the effective potential for the tensor-type perturbations has a negative gap outside the black hole, near its event horizon. Although intuitively, increasing of the multipole number $\ell$ should lead to a higher barrier of the effective potential, this is not the only effect that comes from increasing $\ell$. The higher $\ell$, the deeper is the negative gap, so that, quite contr-intuitively, the higher $\ell$ are more unstable.

\begin{figure}
\centering
\resizebox{0.7\linewidth}{!}{\includegraphics*{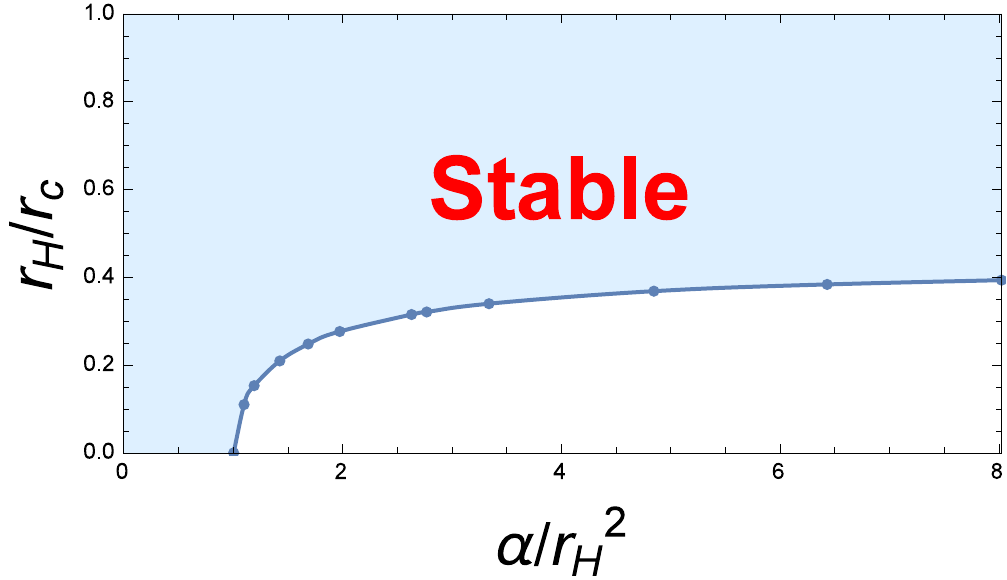}}
\caption{Stability/instability regions for tensor-type gravitational perturbations in $(D=6)$-dimensional Gauss-Bonnet-de Sitter spacetimes. This is the eikonal instability with the $\ell=\infty$ most unstable mode.}
\label{Tensor-6}
\end{figure}

The parametric region of in(stability)of black holes for tensor-type perturbations in $(D=6)$ dimensions is shown in fig.~\ref{Tensor-6}. For higher dimensions ($D>6$), as well as $D=5$, the tensor-type perturbations do not show instability at least to our numerical treatment. From fig.~\ref{Tensor-6}, we can observe that small black holes are always unstable and sufficiently large black hole are stable. Moreover, for a fixed value of $\alpha$,  the threshold of instability is reached at some critical value fraction designated by $\mu_c=r_H/r_C$ corresponding to the limit $\ell\to\infty$. Apparently, this critical value fraction $\mu_c$ approaches to some constant value for large $\alpha$. In this sense, there should exist a minimum value $\mu_C$ such that black hole with $r_H/r_C>\mu_c$ are stable against tensor perturbations in Gauss-Bonnet-de Sitter gravity for any values of coupling constant $\alpha$. It is worth mention that, our results in the limit of the asymptotically Minkowski spacetime ($r_C\rightarrow\infty$, corresponding to $\alpha/r_H^2$-axis in fig.~\ref{Tensor-6}), are in concordance with the results presented in \cite{Konoplya:2008ix}.

\subsection{Instabilities of scalar-type perturbations}

\begin{figure}
\centering
\resizebox{0.7\linewidth}{!}{\includegraphics*{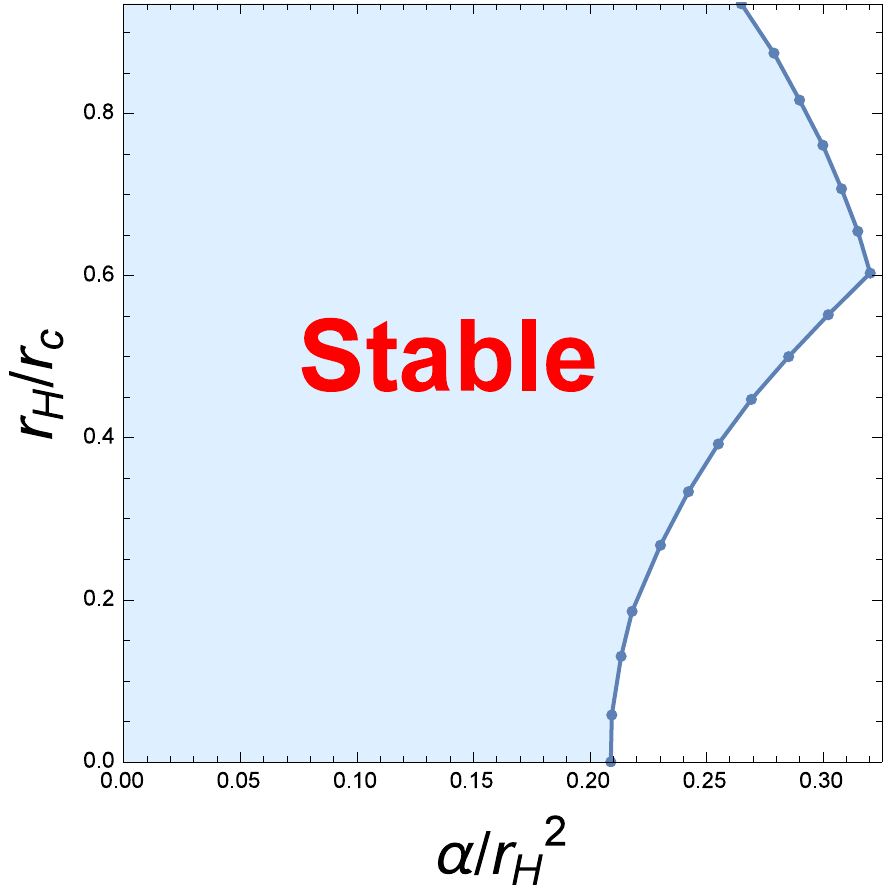}}
\caption{Stability/instability regions for scalar-type gravitational perturbations in 5 dimensions. Upper right corner corresponds to the $\Lambda$-instability, while the lower right corner -- to the eikonal instability. The overlap of regions of both types of instability produces the instability region for $D=5$ case.}\label{InstabilityS5}
\end{figure}

\begin{figure*}
\centering
\resizebox{\linewidth}{!}{\includegraphics*{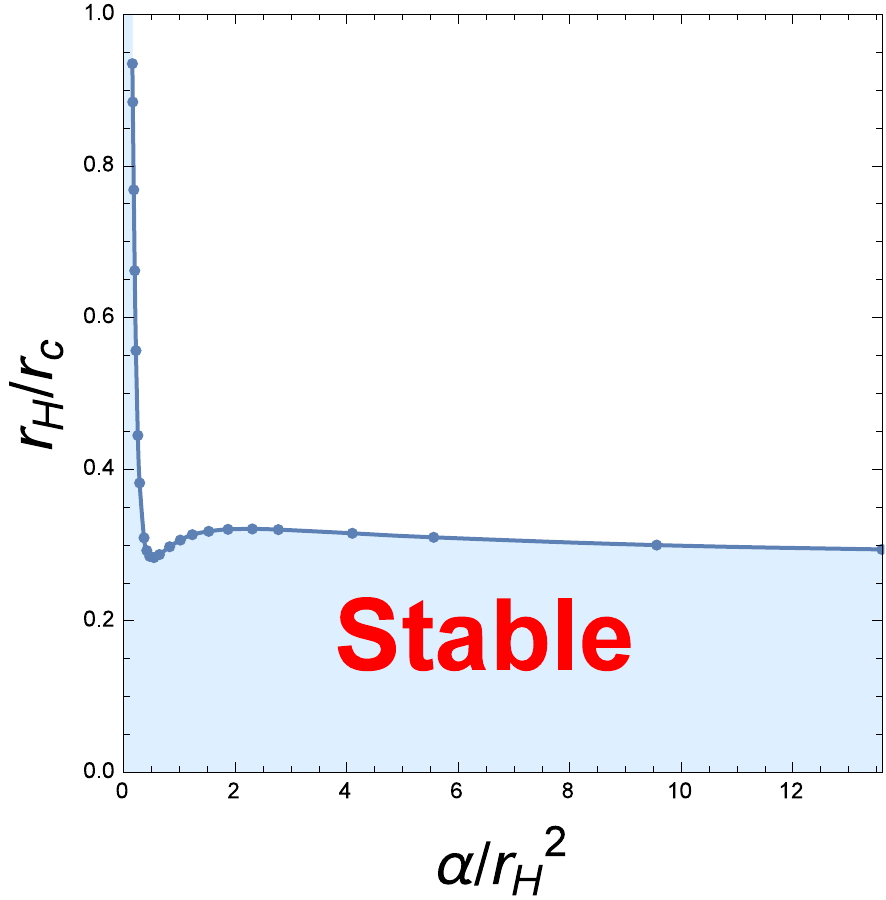}\includegraphics*{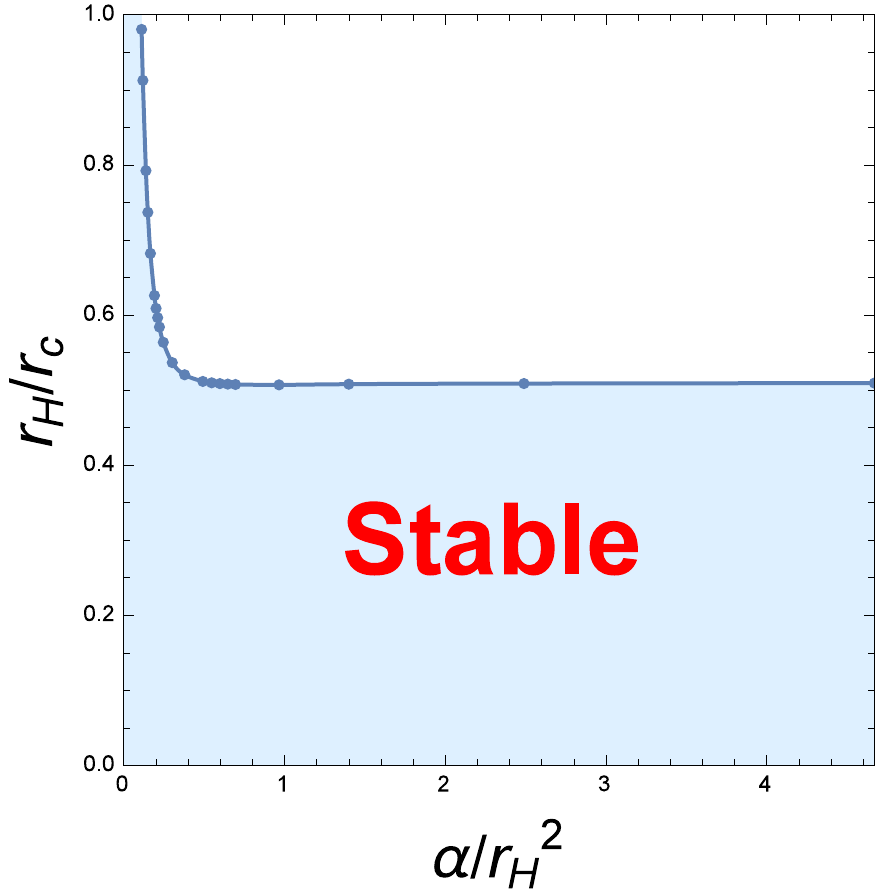}\includegraphics*{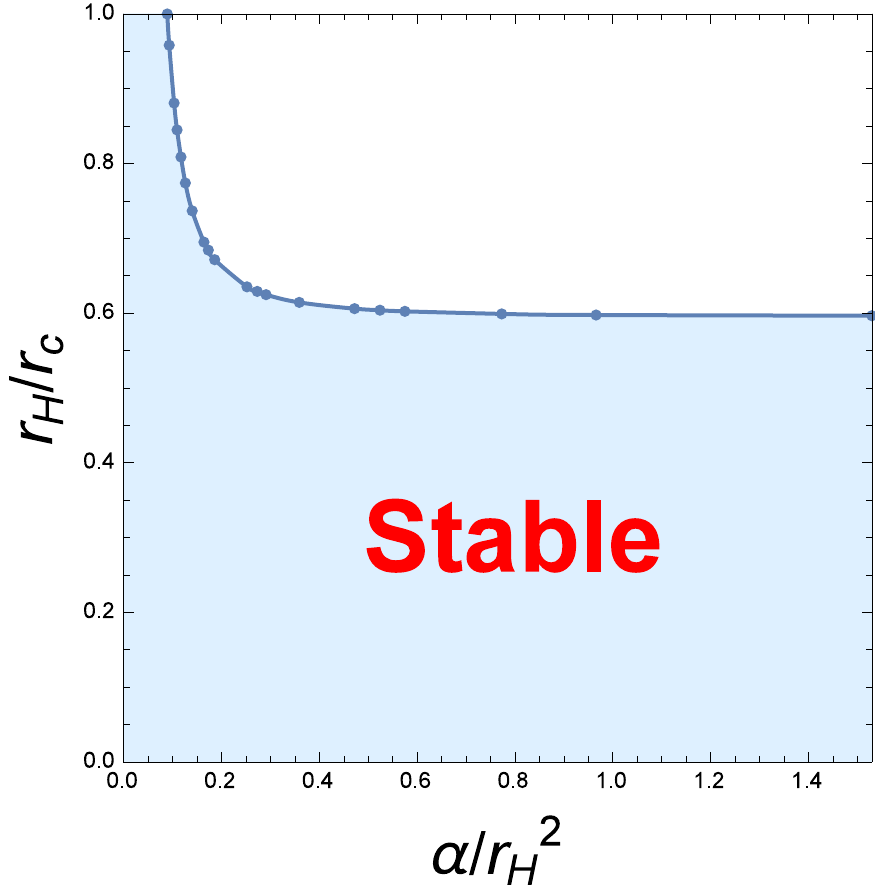}}
\caption{Stability/instability regions for scalar-type gravitational perturbations in $D=6$, $D=7$, and $D=8$ (from left to right).}\label{InstabilityS}
\end{figure*}

In the scalar channel, we can observe that temporal profiles with unbounded growth (unstable) show nonoscillatory behavior near the threshold of instability. This characteristic is in agreement with the statement proved in \cite{Konoplya:2008yy}, that \emph{unstable mode cannot be oscillatory} for spherically symmetric black hole perturbations.

Analyzing the scalar-type of gravitational perturbations for $D\geq 5$, we can observe that a new kind of instability is present for black holes with both sufficiently large values of parameters $\alpha$ and $\Lambda$, corresponding to the GB coupling parameter and cosmological constant respectively (see fig.~\ref{InstabilityS5} and fig.~\ref{InstabilityS}). This new kind of instability occurs for the lowest multiple number $\ell=2$ (unlike the eikonal instability) and, therefore, is not related to the non-hyperbolicity problem of the differential equations. We can note that this kind of instability does not appear in the asymptotically flat case \cite{Konoplya:2008yy}, therefore, we will call it \emph{$\Lambda$-instability} throughout this thesis.

For $D=5$, we can observe from fig.~\ref{InstabilityS5}, both kinds of instabilities ($\Lambda$- and eikonal instability) take place in the parametric representation ($\alpha/r_H^2, r_H/r_C$). The right peak represents the intersection of both thresholds which occurs close to $\Lambda=1.6r_H^2$($r_H=0.6r_C$), so that the upper curve corresponds to the $\Lambda$-instability and the lower one corresponds to the eikonal instability. In this sense, there is no minimal critical value $\mu=r_H/r_C$, for black holes to be stable for arbitrary values of $\alpha$-coupling parameter. From fig.~\ref{InstabilityS5}, the parametric region has two regimes of instability:\\

a) \emph{Small} $D=5$ black holes are unstable once the GB-coupling is larger than some critical value (for a given $r_H/r_C$) $\alpha=\alpha_{crit}$ The instability region is dominated by the eikonal $\ell=\infty$ regime. \\

b) \emph{Large} $D=5$ black holes are unstable for values of the coupling $\alpha$ above some critical (at a given $r_H/r_C$). This instability happens only $\ell=2$ modes.\\

\begin{figure}[H]
\centering
\resizebox{0.7\linewidth}{!}{\includegraphics*{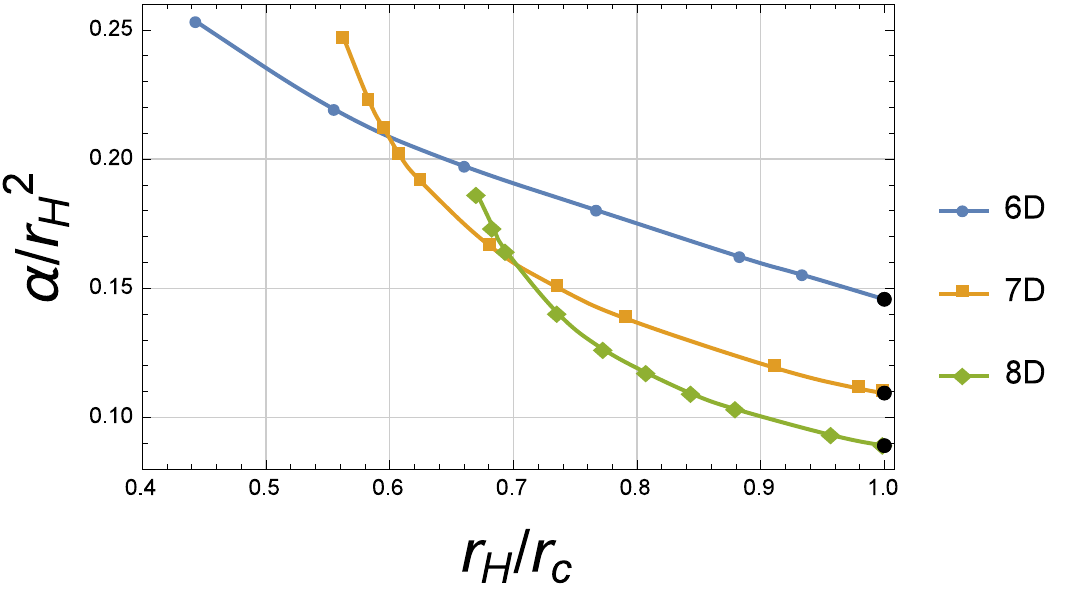}}
\caption{Scalar type of gravitational perturbations: Critical value of $\alpha$, corresponding to the threshold of instability as function of $r_H/r_C$ in dimensions $6$, $7$, and $8$.}\label{alpha-min}
\end{figure}

\begin{table}[ht!]
  \centering
  \begin{tabular}{|c|c|c|}
    \hline
    D=6 & D=7 & D=8\\
    \hline
    \begin{tabular}{c|c}
     $r_H/r_c(\Lambda\,r_H^2)$ & $\alpha/r_H^2$\\
    \hline
    0.662(4) & 0.197 \\
    0.768(5) & 0.180 \\
    0.884(6) & 0.162 \\
    0.935(6.4) & 0.155\\
    1 (6.876) & 0.146
    \end{tabular}
     &
     \begin{tabular}{c|c}
     $r_H/r_c(\Lambda\,r_H^2)$ & $\alpha/r_H^2$\\
     \hline
     0.737(9) & 0.150\\
    0.792(10) & 0.138\\
    0.912(12) & 0.119\\
     0.980(13) & 0.111\\
     1 (13.27) & 0.109
    \end{tabular}
     &
    \begin{tabular}{c|c}
     $r_H/r_c(\Lambda\,r_H^2)$ & $\alpha/r_H^2$\\
    \hline
    0.845(19) & 0.109\\
    0.881(20) & 0.103\\
    0.957(22) & 0.093\\
    0.999(23) & 0.089\\
    1 (23.01) & 0.089
    \end{tabular}\\
    \hline
  \end{tabular}
  \caption{Scalar type of gravitational perturbations: Critical values of $\alpha$ corresponding to the threshold of instability. The bottom line is an extrapolation of the numerical data to the extremal $\Lambda$}
  \label{tab:interpolateval}
\end{table}

On fig.~\ref{InstabilityS}, we show that in higher than five dimensions of spacetime the scalar sector of the gravitational perturbations is stable for small black holes for arbitrary GB-coupling parameter. For an arbitrary size of Gauss-Bonnet black hole, the threshold of instability is reached for a sufficiently large value of the cosmological constant $\Lambda$, so that, there is a maximum value of $\mu_c$ such that for $r_H<\mu_cr_C$ these black holes are stable. Moreover, Gauss-Bonnet black holes for $D\geq 5$ are stable for all $r_H/r_C$, including the extremal limit $r_H=r_C$ for the coupling parameter $\alpha$ lower than a some $\alpha_{crit}$, which depends on spacetime dimension $D$. This minimum value decreases as $D$ grows, what can be seen in fig.~\ref{alpha-min} and in the table~\ref{tab:interpolateval}.

\subsection{Overlap of instabilities of tensor- and scalar-type perturbations}

\begin{figure}[H]
\centering
\resizebox{0.7\linewidth}{!}{\includegraphics*{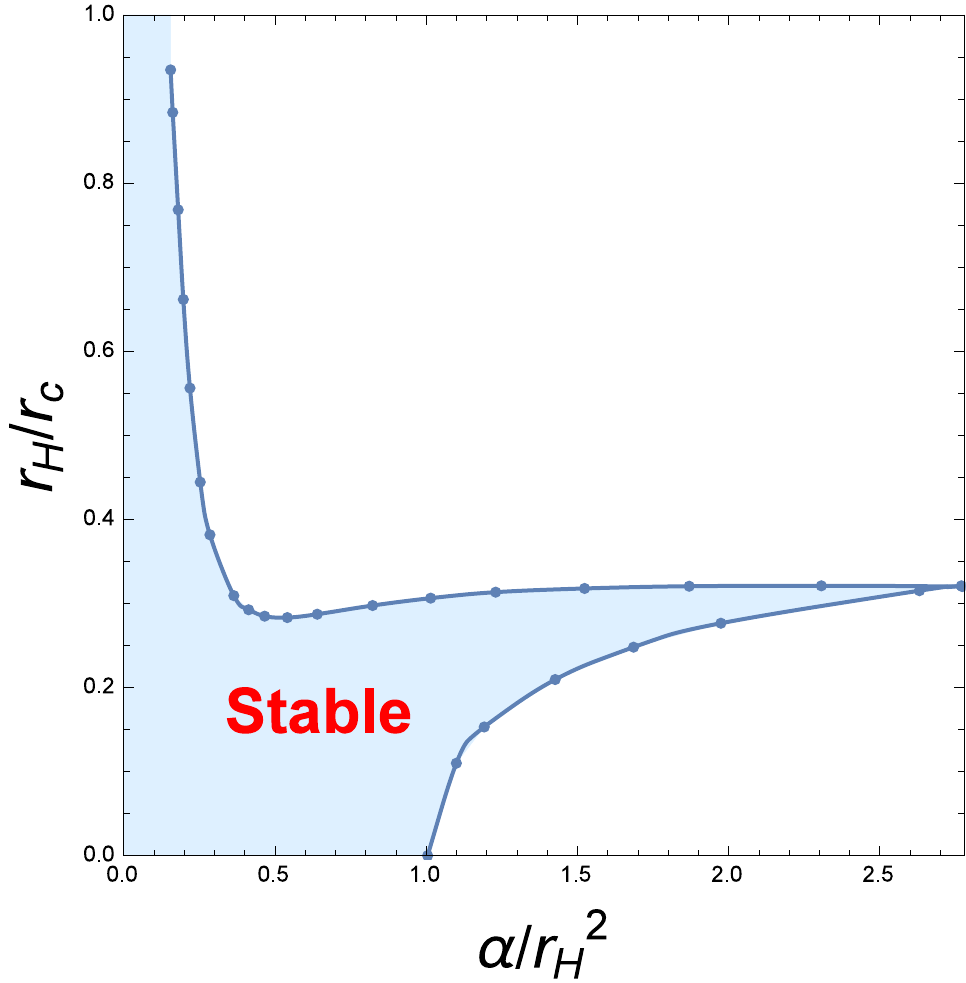}}
\caption{Stability/instability regions for gravitational perturbations in $D=6$ as overlap of the $\Lambda$-instability in the scalar channel and eikonal instability in the tensor channel.} \label{InstabilityS6}
\end{figure}

 First of all, we would like to remind that the vector-type of perturbations show dynamical stability for all $D$. Then, $D =7, 8, \dots$ black holes have only the $\Lambda$-instability in the scalar-type of gravitational perturbation. The regions of this instability are shown on figs.~\ref{InstabilityS}~and~\ref{alpha-min}. For $D=5$ and $D=6$ dimension spacetimes we obtain more complicated regions of instability, which are determined by the overlapping of $\Lambda$-instability ($\ell=2$) and eikonal instability ($\ell=\infty$): in the case of $D=5$ the $\Lambda$-instability in the scalar channel combines with the eikonal instability in the same channel (fig.~\ref{InstabilityS5}), while for $D=6$ the $\Lambda$-instability in the scalar channel combines with the eikonal instability in the tensor channel (fig.~\ref{InstabilityS6}). The complete region of instability of a black holes consists of a combination of the instability regions of all types of gravitational perturbations. Accordingly, the black-hole stability region is the overlap of the corresponding stability regions.

\section{Hod's conjecture for GB - de Sitter}\label{dS:Hod}
In 2007, Shahar Hod formulated an interesting conjecture \cite{Hod:2006jw}, stating the black holes could be considered as the most extreme thermodynamical objects in nature because relaxation times (time required for a perturbing system to return to an equilibrium state) are of the same order of magnitude as the minimum relaxation timescale $\tau_{min}=\hbar/{\pi T_H}$. Since the fundamental mode of the black-hole quasinormal spectrum gives the dynamical relaxation timescale $\tau$ for a generic perturbation, the Shahar Hod's proposal leads to an upper bound in the imaginary part which depends on the Hawking temperature of the black hole. Namely, he argued that
\begin{equation}
|\im{\omega}| \leq \pi T_{H},
\end{equation}
where $T_H$ is the Hawking temperature. For static spherically symmetric black holes, it is given by  (\ref{HC:Hawking:2}).

Using numerical and analytical results for quasinormal modes of four- and higher-dimensional Schwarzschild \cite{Konoplya:2003ii}, Schwarzschild-de Sitter \cite{Zhidenko:2003wq} and Schwarzschild-anti-de Sitter black holes \cite{Horowitz:1999jd} Hod illustrated that his inequality is fulfilled for asymptotically flat black holes as well as for non-asymptotically flat ones. The arguments were based on semi-classical consideration and thermodynamic ideas.

\begin{figure}[H]
\centering
\resizebox{\linewidth}{!}{\includegraphics*{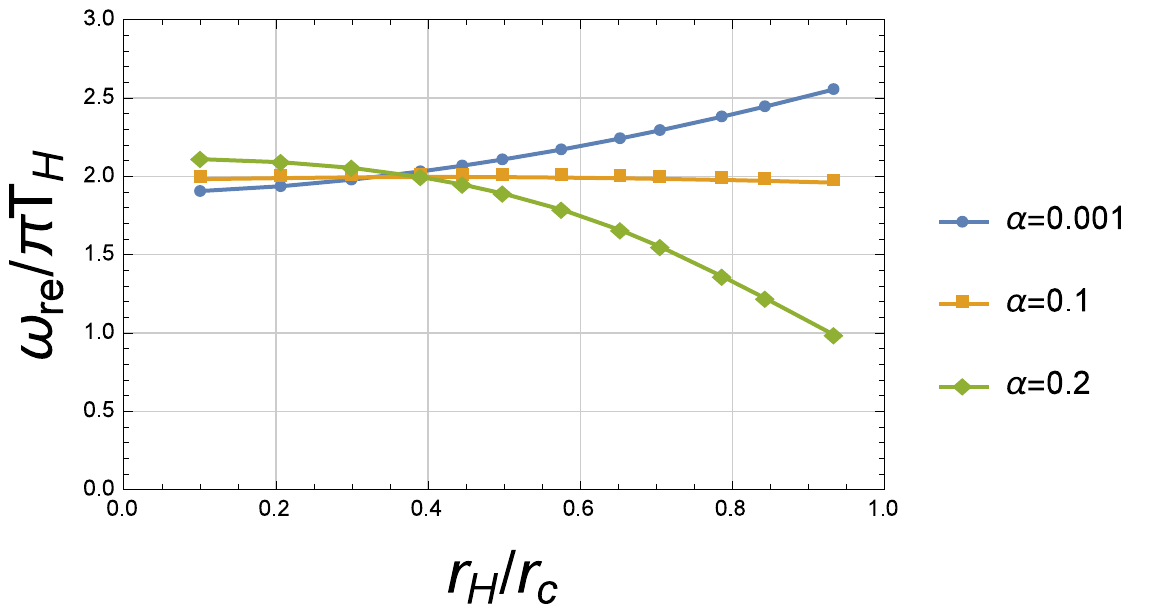}\includegraphics*{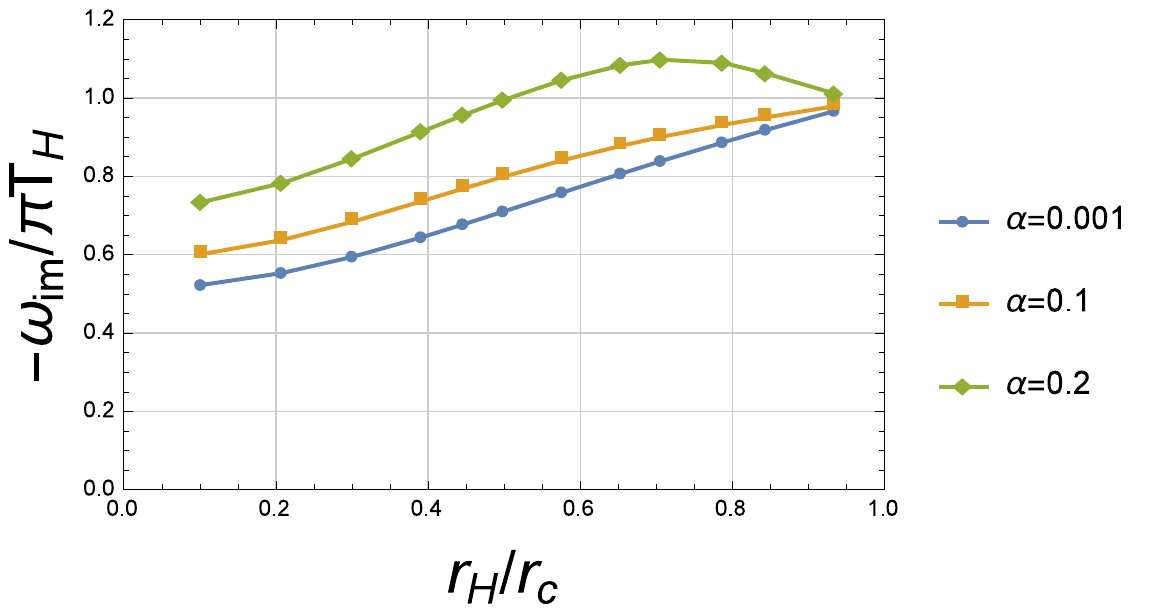}}
\caption{Variation of the real (left panel) and imaginary (right panel) parts of the dominant quasinormal mode as functions of $r_H/r_C$ for scalar-type perturbations in 5 dimensions.}\label{QNMT}
\end{figure}

\begin{figure}[H]
\centering
\resizebox{0.7\linewidth}{!}{\includegraphics*{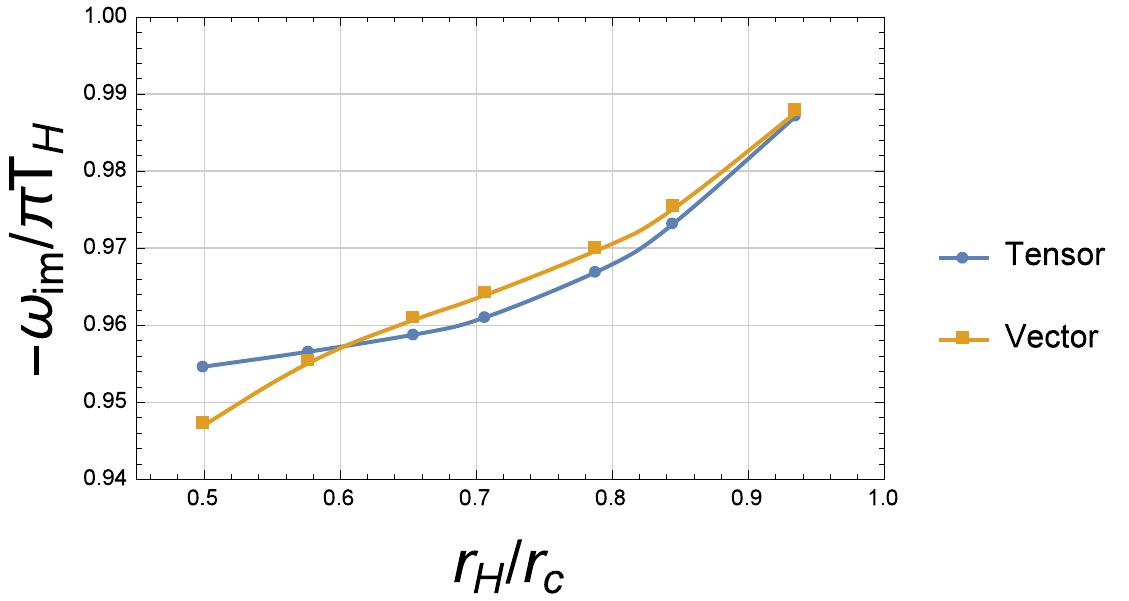}}
\caption{Variation of the imaginary part of the dominant quasinormal mode with respect to $r_H/r_C$ for vector- and tensor-type perturbations in 5 dimensions for $\alpha=0.2r_H^2$.}\label{QNMTvt}
\end{figure}

We verified whether Hod's proposal is valid for a more general Einstein-Gauss-Bonnet-de Sitter black hole solution. We present here an analysis for $D=5$ dimensional black hole for some free parameters ($\Lambda,\alpha$), which correspond to gravitationally stable regions. The break down of the inequality at the onset of instability would be a perfect proof of the proposal. However, fig.~\ref{QNMT} shows that the lowest mode of the scalar type of gravitational perturbations have the imaginary part $\im{\omega}$ for which

\begin{equation}
|\im{\omega}| > \pi T_{H},
\end{equation}
for $\alpha =0.2r_H^2$ in the range $0.5\lesssim r_H/r_C \lesssim 0.94$, that is inside the stability region according to fig.~\ref{InstabilityS5}.
Nevertheless, vector and tensor perturbations in the stable sector (for $\alpha=0.2r_H^2$), unlike scalar perturbations, do not breakdown the conjecture, as we can see in  (fig. \ref{QNMTvt}). The corresponding data of the figures above mentioned are shown in Tables ~\ref{DataQNMs} and ~\ref{DataQNMvt}, which was obtained from the Prony's method described in the section ~\ref{PM}.

\begin{table}[H]
\centering
\begin{tabular}{|c|c|c|c|c|c|c|}
  \hhline{|=======|}
  \multicolumn{7}{|c|}{Scalar-type perturbation}  \\
  \hline
   & \multicolumn{2}{c|}{$\alpha=$0.001$r_H^2$} & \multicolumn{2}{c|}{$\alpha=$0.1$r_H^2$} & \multicolumn{2}{c|}{$\alpha=$0.2$r_H^2$} \\
  \hhline{|-------|}
  $r_H/r_C$ & QNMs & $\pi\,T_H$ &  QNMs & $\pi\,T_H$ & QNMs & $\pi\,T_H$   \\
  \hhline{|=======|}
  0.103 & 0.9315-0.2553$\dot{\imath}$ & 0.4886 & 0.8087-0.2453$\dot{\imath}$ & 0.4080 & 0.7376-0.2567$\dot{\imath}$ &  0.3497\\
    \hline
  0.209 & 0.8862-0.2531$\dot{\imath}$ & 0.4574 & 0.7595-0.2435$\dot{\imath}$ & 0.3819 &0.6841-0.2563$\dot{\imath}$ &  0.3274\\
    \hline
  0.302 & 0.8231-0.2473$\dot{\imath}$ & 0.4158 & 0.6922-0.2376$\dot{\imath}$ & 0.3472 &0.6110-0.2516$\dot{\imath}$ &  0.2976\\
    \hline
  0.392 & 0.7437-0.2359$\dot{\imath}$ & 0.3659 & 0.6099-0.2251$\dot{\imath}$ & 0.3056 &0.5223-0.2395$\dot{\imath}$ &  0.2619\\
    \hline
  0.447 & 0.6884-0.2255$\dot{\imath}$ & 0.3327 & 0.5544-0.2137$\dot{\imath}$ & 0.2778 &0.4632-0.2278$\dot{\imath}$ &  0.2381\\
    \hline
  0.500 & 0.6313-0.2127$\dot{\imath}$ & 0.2994 & 0.4987-0.1999$\dot{\imath}$ & 0.2500 &0.4047-0.2133$\dot{\imath}$ &  0.2143 \\
    \hline
  0.577 & 0.5420- 0.1894$\dot{\imath}$ & 0.2495 & 0.4149-0.1752$\dot{\imath}$ & 0.2083 &0.3188-0.1868$\dot{\imath}$ &  0.1786 \\
    \hline
  0.655 & 0.4478-0.1611$\dot{\imath}$ & 0.1996 & 0.3312-0.1464$\dot{\imath}$ & 0.1667 & 0.2362-0.1548$\dot{\imath}$ &  0.1429 \\
    \hline
  0.707 & 0.3818-0.1396$\dot{\imath}$ & 0.1663 & 0.2755-0.1251$\dot{\imath}$ & 0.1389 &0.1842-0.1307$\dot{\imath}$ &  0.1190 \\
    \hline
  0.788 & 0.2774-0.1033$\dot{\imath}$ & 0.1164 & 0.1922-0.0906$\dot{\imath}$ & 0.0972 &0.1131-0.0908$\dot{\imath}$ &  0.0833 \\
    \hline
  0.845 & 0.2035-0.0764$\dot{\imath}$ & 0.0831 & 0.1369- 0.0660$\dot{\imath}$ & 0.0694 &0.0725-0.0632$\dot{\imath}$ &  0.0595  \\
    \hline
  0.935 & 0.0851-0.0322$\dot{\imath}$ & 0.0333 & 0.0544-0.0272$\dot{\imath}$ & 0.02778 &0.0234-0.0241$\dot{\imath}$ &  0.0238  \\
    \hhline{|=======|}
\end{tabular}
\caption{Quasinormal frequencies and $\pi T_H$ in the scalar channel for $\alpha=$ 0.001 , 0.1 and 0.2, obtained in order to check the Hod's conjecture. The first column corresponds to the ratio $r_H/r_C$ where the data were determined}\label{DataQNMs}
\end{table}

\begin{table}[H]
\centering
\begin{tabular}{|c|c|c|c|c|}
  \hhline{|=====|}
   & \multicolumn{2}{c|}{Vector-type} & \multicolumn{2}{c|}{Tensor type}  \\
  \hhline{|-----|}
  $r_H/r_C$ & QNMs & $\pi\,T_H$ &  QNMs & $\pi\,T_H$  \\
  \hhline{|=====|}
  0.500 & 0.6132-0.2029$\dot{\imath}$ & 0.2143 & 1.0337- 0.2046$\dot{\imath}$ & 0.2143 \\
    \hline
  0.577 & 0.5102-0.1706$\dot{\imath}$ & 0.1786 &0.8786- 0.1708$\dot{\imath}$ & 0.1786 \\
    \hline
  0.655 & 0.4088-0.1372$\dot{\imath}$ & 0.1429 &0.7199- 0.1370$\dot{\imath}$ & 0.1429 \\
    \hline
  0.707 & 0.3417- 0.1148$\dot{\imath}$ & 0.1190 &0.6112- 0.1144$\dot{\imath}$ & 0.1190 \\
    \hline
  0.788 & 0.2411- 0.0808$\dot{\imath}$ & 0.0833 & 0.4419- 0.0806$\dot{\imath}$ & 0.0833 \\
    \hline
  0.845 & 0.1735- 0.0580$\dot{\imath}$ & 0.0595 &0.3235- 0.0579$\dot{\imath}$ & 0.0595 \\
    \hline
  0.935 & 0.0705- 0.0235$\dot{\imath}$ & 0.0238 &0.1349- 0.0235$\dot{\imath}$& 0.0238 \\
    \hhline{|=====|}
\end{tabular}
\caption{Quasinormal frequencies and $\pi T_H$ in the vector and tensor channels for $\alpha=$0.2, obtained in order to check the Hod's conjecture. The first column corresponds to the ratio $r_H/r_C$ where the data were determined}\label{DataQNMvt}
\end{table}

Our results imply the following: If we admit that, a black hole could be excited only in the scalar channel, i.e. the vector and tensor sector could have zero excitation factors, the counterexample for the Hod's proposal would be provided. Even though the three types of gravitational perturbations can be analyzed independently within the linear approximation, we do not believe that it would be easy to prepare such process of ``purely scalar'' perturbation in nature.
An immediate reason for this is the fact that, although a classical black hole preserves all the geometrical symmetries given by a set of Killing vectors, tiny quantum fluctuations definitely break down the exact symmetries and within the quantum picture, the separation of a real process of perturbation into fully independent scalar, vector and tensor channels would apparently be impossible. In other words, if due to specially prepared initial perturbations only one of the channels (scalar) is excited, then the other two channels (vector and tensor) inevitably acquire some, whatever small, but nonzero, excitations. Therefore, we conclude that here we rather confirmed the Hod's proposal than disproved it.

We can also see that the higher value of the $\alpha$-coupling is, the quicker Hawking temperature grows with respect to the oscillation frequency. The decay rate of oscillation, $\im{\omega}/\pi\,T_H$ reaches a constant value when approaching the extremal state. Indeed, the imaginary part of the QNM approaches zero for extremal black holes, so that $\im{\omega}\propto T_H\approx0$ for $r_H\approx r_C$. In the limit $\alpha \rightarrow 0$ quasinormal modes of the Einstein-Gauss-Bonnet black hole go over into those of (asymptotically de Sitter) Tangherlini black holes in Einstein theory \cite{Konoplya:2003dd,Konoplya:2007jv}. 

\chapter{Wormholes in Higher-Curvature gravity}\label{WHG}
In 1916, Ludwing Flamm, when examining Schwarzschild's solution, realized that Einstein's equations allowed for another solution (currently known as white hole). In contrast to black holes, white holes eject matter and light from their event horizon. In this sense, a black hole is an ``entrance'' and a white hole is an ``exit'' of such matter and light. These two solutions could represent two different regions which would be spacetime connected by a conduit, named later a ``bridge'' by Einstein and Rosen \cite{PhysRev.48.73}, who had considered such a model in 1935 rather as a mathematical curiosity. Some decades later, Misner and Wheeler first introduced the term ``wormhole'' in \cite{MISNER1957525},
\begin{quote}
\it
This analysis forces {\ldots} where there is a net flux of lines of force through what topologists would call a handle of the multiply-connected space and what physicists might
perhaps be excused for more vividly terming a ``wormhole'' \nobreak\hfill\penalty50\hskip1em\null\nobreak\hfill\mbox{-- C. W. Misner and J. A. Wheeler}
\end{quote}
showing that the bridge (wormhole) could connect two distant points of our own universe or different universes by a spacetime shortcut. Yet, the existence of Schwarzschild wormholes was later ruled out because they are not traversable since the throat opens and closes so quickly that not even light can pass through and moreover, the white holes (exits) become unstable under small perturbations that convert them into black holes \cite{PhysRevLett.33.442}.

However, adding of a scalar field coupled to gravity allows one to prevent the wormhole's throat from closing. These new classes of solutions, found first by Ellis \cite{Ellis:1973yv} and independently by Bronnikov \cite{Bronnikov:1973fh}, are known as Ellis-Bronnikov wormholes, belonging, actually, to a more general class of wormholes. In 1988, Morris and Throne \cite{Morris:1988cz} provided some conditions for the wormholes to be traversable. These solutions are regular (posses no horizon), though an unusual type of matter must be introduced in order to keep the throat open. The energy-momentum tensor of such matter must violate the null energy condition which is needed in order to the wormhole be traversable. However, this exotic feature can be attributed to the so-called dark energy, supported by contemporary cosmological observations of the universe expanding with acceleration.

The estimations for the acceleration of the universe expansion suggest that the dark energy dominates over other types of matter, contributing about 70\% of total energy in the observable universe. Playing such an important role in cosmology, exotic matter may manifest itself also in local phenomena, for example around black holes or wormholes.
It can be modeled by a self-interacting phantom scalar field, characterized by negative values of pressure source $\omega=p/\rho < -1$. Several studies have been developed in this context: regular black-hole \cite{Bronnikov:2005gm} and wormhole \cite{Ellis:1973yv,Bolokhov:2012kn} solutions, both supported by the phantom scalar field, and testing of instability against gravitational perturbations \cite{Bronnikov:2012ch}.

Fortunately, alternative theories of gravity circumvent the use of these exotic matters giving opportunities for existence of traversable wormholes. One of such theories is the Einstein-Gauss-Bonnet gravity which is supplemented by the presence of a non-minimal coupling dilaton scalar field. It turns out that the Einstein-dilaton-Gauss-Bonnet gravity allows for a traversable wormhole solution obtained by P. Kanti et al. \cite{Kanti:2011jz}, and some generalizations to rotating black holes were further developed \cite{Kanti:2011yv,Blazquez-Salcedo:2016yka,Cunha:2016wzk}.

\section{Einstein-dilaton-Gauss-Bonnet theory}\label{EdGB}
Since string theory is one of the candidates of a quantum theory of gravity, the quadratic curvature is a natural correction at the low-energy limit. In fact, the full heterotic string theory \cite{METSAEV1987385,GROSS198741} includes the dilaton (a scalar field) along with axions, fermions and gauge fields. In this framework, the Einstein-dilaton-Gauss-Bonnet action corresponds to a kind of minimal effective theory that contains, in addition to the scalar curvature $R$ and the kinetic term of the dilaton, also a nonvanishing quadratic curvature correction in four dimensions, non-minimally coupled to the dilaton field. Thus, the effective theory in four dimensions is described by the action,

\begin{eqnarray}\label{act}
S&=&\frac{1}{16 \pi}\int d^4x \sqrt{-g} \Biggr[R - \frac{1}{2}
 \partial_\mu \phi \,\partial^\mu \phi
+ \alpha  e^{-\gamma \phi} \mathcal{L}_2	  \Biggr],
\end{eqnarray}
where $\alpha$ is a positive parameter proportional to the Regge slope, $\gamma$ is a coupling constant of the dilaton field $\phi$ and $\mathcal{L}_2$ is the Gauss-Bonnet correction given by eq. \ref{GBterm}. The variation of the action, $\delta S=0$, allows one to find the equations of motion for the dilaton and for the gravitational field,
\begin{eqnarray}
\nabla^2 \phi & = & \alpha \gamma  e^{-\gamma \phi}\mathcal{L}_2,	
\label{eqs}\\
G_{\mu\nu} & = &
\frac{1}{2}\left[\nabla_\mu \phi \nabla_\nu \phi
                 -\frac{1}{2}g_{\mu\nu}\nabla_\lambda \phi \nabla^\lambda\phi
		 \right]
\nonumber\\
& &
-\alpha e^{-\gamma \phi}
\left[	H_{\mu\nu}
  +4\left(\gamma^2\nabla^\rho \phi \nabla^\sigma \phi
           -\gamma \nabla^\rho\nabla^\sigma \phi\right)	P_{\mu\rho\nu\sigma}
		 \right],\label{eqs2}
\end{eqnarray}
where $G_{\mu\nu}=R_{\mu\nu}-\frac{1}{2}g_{\mu\nu}R$ is the Einstein tensor and
\begin{eqnarray}\nonumber
H_{\mu\nu} & = & 2\left[R R_{\mu\nu} -2 R_{\mu\rho}R^\rho_\nu
                        -2 R_{\mu\rho\nu\sigma}R^{\rho\sigma}
			+R_{\mu\rho\sigma\lambda}R_\nu^{\ \rho\sigma\lambda}
		   \right]
		   -\frac{1}{2}g_{\mu\nu}\mathcal{L}_2	\ ,
\\\nonumber
P_{\mu\nu\rho\sigma} & = & R_{\mu\nu\rho\sigma}
+2 g_{\mu\sigma} R_{\rho\nu}-2 g_{\mu\rho} R_{\sigma\nu}
+2 g_{\nu\rho} R_{\sigma\mu}-2 g_{\nu\sigma} R_{\rho\mu}
+R g_{\mu\rho} g_{\sigma\nu}-R g_{\mu\sigma} g_{\rho\nu} \ .
\label{HP}
\end{eqnarray}

Since astrophysical objects are also interesting to study in alternative theories, in addition to wormholes, it is worth mention that the four-dimensional Einstein-dilaton-Gauss-Bonnet theory allows for the black holes with a scalar hair.
Static spherically symmetric black hole solutions were found in EDGB theory \cite{Kanti:1995vq} whose gravitational background has the line element in the form
\begin{equation}
ds^2=g_{\mu\nu}dx^\mu dx^\nu = -e^{\Gamma(r)}dt^2+e^{\Lambda(r)}dr^2
+r^2\left(d\theta^2+\sin^2\theta d\varphi^2 \right) ,\nonumber
\end{equation}
where $\Lambda$ and $\Gamma$ depend on $r$ solely. A regular solution with asymptotically Minkowski geometry at infinity, $e^{\Gamma(\infty)}$,$e^{\Lambda(\infty)}\to 1$ was obtained in \cite{Kanti:1995vq,Torii:1996yi,Pani:2009wy}. Further, higher dimensional black holes have also been studied in \cite{Guo:2008hf,Guo:2008eq}, and, recently, the generalization of the Kerr black hole has been discussed as an astrophysical viable model \cite{Blazquez-Salcedo:2016yka} within the four-dimensional Einstein-dilaton-Gauss-Bonnet theory.

\section{Kanti-Kleihauss-Kunz solution}\label{KKK}
\subsection{Spherically symmetric wormhole}
The set of equations (\ref{eqs}) and (\ref{eqs2}) allow for asymptotically flat spherically symmetric wormholes that can be described by the following line element \cite{Kanti:2011jz}
\begin{eqnarray}
ds^2 &=& -e^{2\nu(l)}dt^2+f(l)dl^2
+r^2\left(d\theta^2+\sin^2\theta d\varphi^2 \right)\ ,\nonumber
\\ r^2&\equiv& l^2+r_0^2\ ,\label{metricL}
\end{eqnarray}
where $r_0$ is the radius of the throat, metric functions $f(l)$, $\nu(l)$ and the dilaton field $\phi(l)$ are written down in terms of the new coordinate $l$ which goes from $-\infty$ to $+\infty$. The condition of asymptotic flatness is fulfilled at both sides of the wormhole $l\to\pm\infty$. Substitution of the metric equation (\ref{metricL}) into the dilaton equation (\ref{eqs}) and the modified Einstein equation (\ref{eqs2}), after some algebra, yield a coupled system of ordinary differential equations \cite{Kanti:2011yv}, which reads as
\begin{eqnarray}
f' + \frac{f (r^2 f + l^2 - 2 r^2)}{l r^2}
& = &
\frac{r^2 f \phi'^2}{4 l}+2\alpha \gamma\frac{e^{-\gamma\phi}}{l r^2} \bigg\{
  2 (r^2 f - l^2)(\gamma\phi'^2-\phi'')\nonumber\\
  &&
 \left.+ \phi'\left[\frac{f'}{f}\,(r^2 f-3 l^2) +\frac{4 l r_0^2}{r^2}\right]\right\},
 \label{ode1} \\[1mm] && \nonumber\\
\nu' - \frac{r^2 f -l^2}{2 l r^2}
& = &
\frac{\phi'^2 r^2}{8 l}
+2\alpha \gamma\frac{e^{-\gamma\phi}}{l r^2 f}
\,\nu'\phi' (r^2 f- 3 l^2)\,,\label{ode2}  \\[1mm] && \nonumber
\\ \label{ode3}
 \nu''+\nu'^2 + \frac{\nu' (2 l f - r^2 f')}{2 r^2 f}
 & + &
  \frac{ 2 r_0^2 f - l  r^2 f'}{2 r^4 f} + \frac{\phi'^2}{4} = 2\alpha \gamma\frac{e^{-\gamma\phi}}{r^2 f}\times \\ \nonumber
&&
\left\{2 l \left[\nu' (\gamma\phi'^2-\phi'') -\phi' (\nu'^2+\nu'')\right]
+ \nu' \phi' \left(\frac{3lf'}{f}-\frac{2r_0^2}{r^2}\right)\right\},
\\ \label{ode4}
\phi'' + \nu'\phi' + \frac{\phi' (4 l f - r^2 f')}{2 r^2 f}
& = &
 4\alpha \gamma\frac{e^{-\gamma\phi}}{r^4 f} \times \\ \nonumber
&&
\left\{
-2 (r^2 f - l^2)(\nu'^2+\nu'')
+\nu'\left[\frac{f'}{f}\,(r^2 f-3 l^2) +\frac{4 l r_0^2}{r^2}\right]\right\},
\end{eqnarray}
where the prime designates the derivative with respect to the new radial coordinate $l$. The equations (\ref{ode1}), (\ref{ode2}) and (\ref{ode3}) were obtained from the $tt$, $ll$ and $\theta\theta$ components of the modified Einstein equation (\ref{eqs2}) respectively, and the equation (\ref{ode4}) follows from the dilaton equation (\ref{eqs}). We note that the equation (\ref{ode1}) is a first-order differential equation for $f=f(l)$ and contains terms with the second derivatives of $\phi(l)$, the equation (\ref{ode2}) is a first-order differential equation for $\nu=\nu(l)$ which has no terms with second derivative of $\phi(l)$, so that (\ref{ode4}) can be reduced to the second-order differential equation for $\phi(l)$ with no derivatives of $f(l)$ and $\nu(l)$. Equation (\ref{ode3}), involving second derivatives in $\nu$ and $\phi$, is satisfied as long as the other three equations are satisfied.

\subsection{Numerical metric solution}
Since the metric wormhole solution (\ref{metricL}) is regular, we expand the metric functions $f$, $\nu$  and the dilaton function $\phi$ in the vicinity of the throat $l=0$, in the following form
\begin{eqnarray}
  f(l) &=& f_0+f_1 l +\ldots,\label{ft1}\\
  e^{2\nu(l)} &=& e^{2\nu_0}\left(1+\nu_1 l\ldots\right), \label{nut1}\\
  \phi(l) &=& \phi_0+\phi_1 l+\ldots, \label{phit1}
\end{eqnarray}
where $f_i$, $\nu_i$ and $\phi_i$, $i=0,1\ldots$ are constant coefficients which will be determined by a recursive relation obtained from the set of coupled differential equation (\ref{ode1}), (\ref{ode2}), and (\ref{ode4}), while $f_0$, $\nu_0$ and $\phi_0$ are the free parameters at $l=0$. The following initial conditions are imposed at the throat
\begin{eqnarray}
\nonumber f(0)&=&f_0,\\
\nonumber \phi(0)&=&\phi_0,\\
\nonumber \nu(0)&=&\nu_0,\\
\phi'(0)^2 &=& \phi_1^2\, =\, \frac{f_0(f_0-1)}{2\alpha\gamma^2 e^{-\gamma \phi_0}
\left[f_0-2(f_0-1)\frac{\alpha}{r_0^2} e^{-\gamma \phi_0}\right]}\ .
\end{eqnarray}
After the diagonalization of the dilaton and Einstein equations (\ref{ode1})-(\ref{ode4}) in the limit $l\to 0$, the latter of the four conditions follows from the requirement of regularity of the metric coefficients at the throat. Since the right-hand side is positive-definite, we must impose constraint relations for the $(\alpha,f_0,r_0,\phi_0)$ parameters. Despite $\alpha$-coupling constant in general is a real quantity, we consider $\alpha>0$ which is in concordance with the interpretation of $\alpha$ in terms of Regge slope parameter. We must impose the constraint $f_0> 1$ which introduces a boundary in the wormhole solutions, then $\alpha/r_0^2 < e^{\gamma\phi_0}/2$ in order to have a positive expression in the brackets in the denominator.

At infinity we impose asymptotic flatness of the solution and vanishing dilaton field. Then at $l\to\infty$ the corresponding functions take the form
\begin{eqnarray}
  f(l) &=& 1+\frac{2M}{l}+\ldots, \nonumber\\
  \nu(l) &=& -\frac{M}{l}+\ldots, \\
  \phi(l) &=& -\frac{D}{l}+\ldots,\nonumber
\end{eqnarray}
where $M$ and $D$ are the mass and dilaton charge of the wormhole respectively.

\begin{figure*}
\resizebox{\linewidth}{!}{\includegraphics*{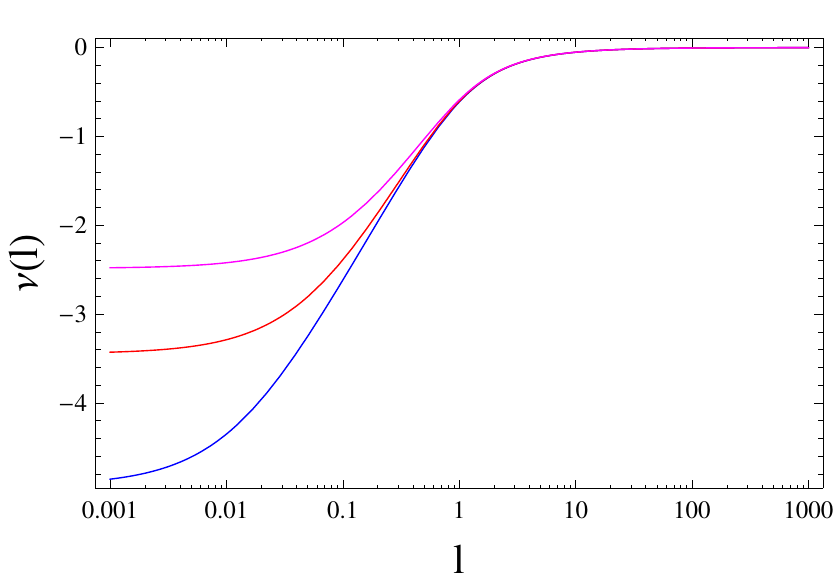}\includegraphics*{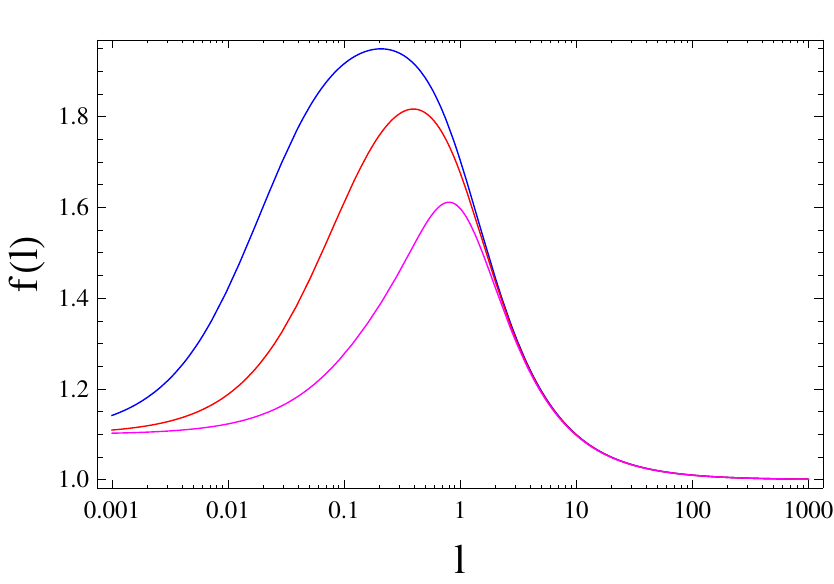}}
\caption{Metric functions, $\nu(l)$ (left panel, from bottom to top) and $f(l)$ (right panel, from top to bottom), for the wormholes $\alpha=0.005r_0^2$ (blue), $\alpha=0.02r_0^2$ (red), and $\alpha=0.05r_0^2$ (magenta) for $f_0=1.1$.}\label{mplot}
\end{figure*}

Due to the scaling symmetry (see Sec. IIC in \cite{Kanti:2011yv}), without loss of generality we take $\gamma=1$ and $r_0=1$, i.~e. we measure $\alpha$ and all dimensional quantities in the units of $r_0$. Then, for given $\alpha>0$ and $f_0>1$, we choose $\nu_0$ and $\phi_0$ such that
\begin{equation}
\lim_{l\to \infty}\nu = 0 \,, \qquad \lim_{l\to \infty}\phi = 0 \, .
\end{equation}
Once the units are chosen in this way, the wormhole solution can be found for every value of $\alpha$ for which (see fig.~\ref{mplot})
\begin{equation}
\frac{\alpha}{r_{0}^2}\lessapprox 0.13.
\end{equation}

The above setup defines two families of solutions to the equations (\ref{ode1})-(\ref{ode4}) for $l\geq0$ corresponding to two possible signs of $\phi'(0)$. All the wormhole configurations presented in \cite{Kanti:2011yv} correspond to the choice $\phi'(0)<0$. It is clear that if one requires the smoothness of the solution at $l=0$ the resulting configuration is not symmetric with respect to the throat. Indeed, if one replaces $l\to-l$, then $\phi'(0)\to-\phi'(0)$ (as well as derivatives of the metric functions). It turns out that such solutions have singularities after crossing the throat. Therefore, we will study symmetric wormholes such that $\phi'(+0)=-\phi'(-0)<0$, i.~e. with discontinuities of the first derivatives at the throat. Although such a geometry looks artificial, it apparently does not lead to problems because the observable quantities remain finite at the throat \cite{Kanti:2011yv}. The discontinuities can be attributed to the presence of matter with positive energy density and pressure at the throat \cite{Kanti:2011jz}.

\section{Traversable Wormhole}

\subsection{Embedding}
The construction of the embedding diagram allows us to imagine how the wormhole will look like in a given manifold. The corresponding geometry of a particular slice can be found once we know the numerical function $f(l)$. Since the solution is spherically symmetric, we consider an equatorial slice $\theta=\pi/2$ in order to simplify the analysis and a fixed time $dt=0$ without loss of generality, then from (\ref{metricL}) the spatial metric is
\begin{equation}
  ds^2=f(l)dl^2+(l^2+r_0^2)d\varphi^2.
\end{equation}
Now we embed this two-dimensional slice into the three-dimensional Euclidean space $\mathbb{R}^3$ written in cylindrical coordinates $(\eta,\varphi,z)$. The origin of our new coordinate system is placed at $z=0$ (throat of the wormhole) and $\varphi$ runs from $0$ to $2\pi$ (spherical symmetry) and the radial coordinate $\eta$ goes from 0 to $\infty$. The geometry in cylindrical coordinates has the form,
\begin{equation}
  dS^2=d\eta^2+dz^2+\eta^2 d\varphi^2.
\end{equation}

Then we set $ds^2\equiv dS^2$ in order to label any point in the space and regarding $z$ and $\eta$ as functions of $l$, we find
\begin{equation}\label{eta123}
  \eta(l)=\sqrt{l^2+r_o^2},\,\,\,\,\,\,\,\,\, \left(\frac{dz}{dl}\right)^2+\left(\frac{d\eta}{dl}\right)^2=f(l).
\end{equation}

\begin{figure*}
\begin{center}
  \resizebox{0.5\linewidth}{!}{\includegraphics*{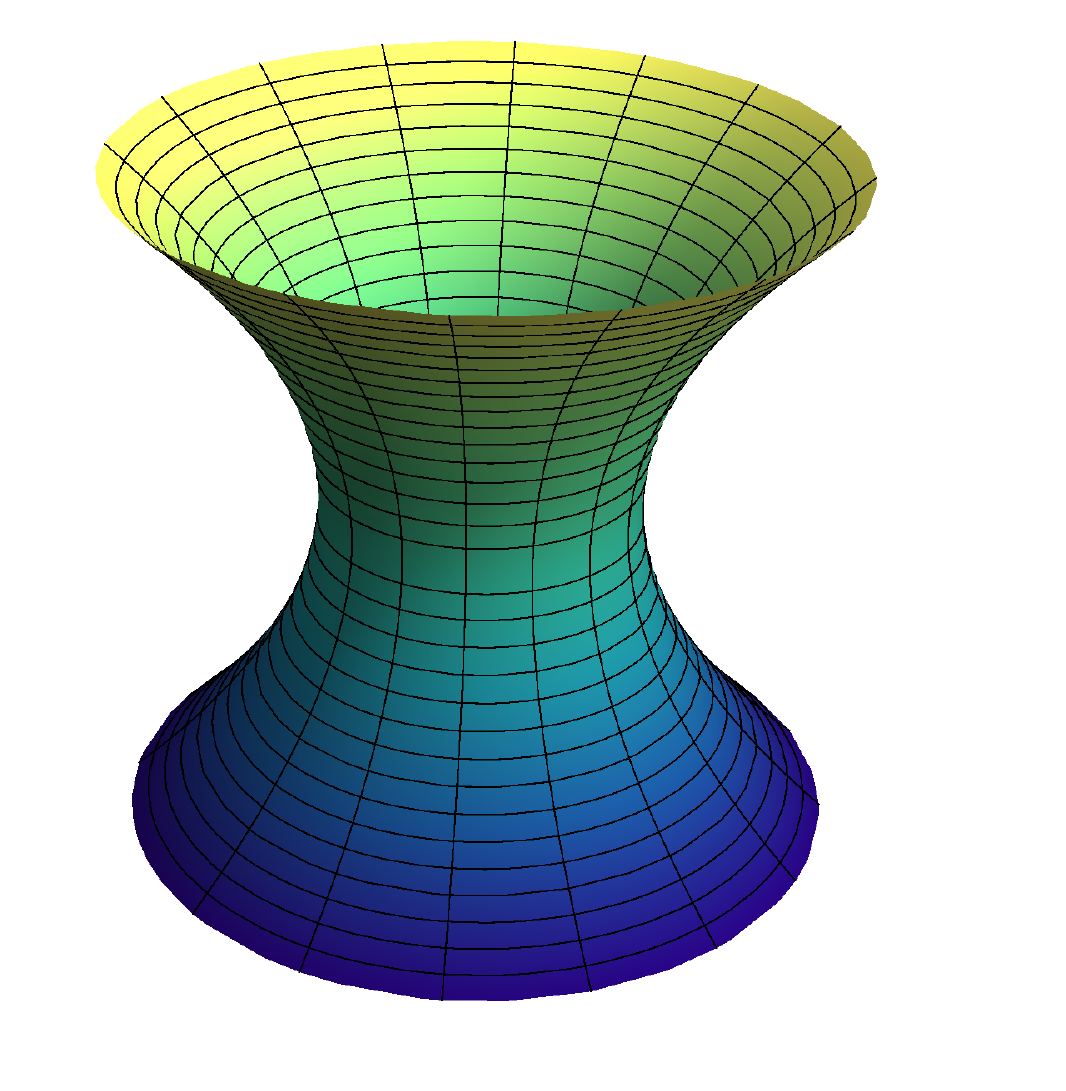}}
\caption{Embedding diagram for a wormhole solution for $\alpha/r_0^2=0.02$ and $f_0=1.1$.}\label{wormT}
\end{center}
\end{figure*}

From the last equation, we easily conclude that,
\begin{equation}\label{emb2}
  z(l)=\pm\int_{0}^{l}\sqrt{f(l')-\frac{l'^2}{l'^2+r_0^2}}dl'.
\end{equation}
Once the function $f(l)$ is obtained from the numerical integration of the equations (\ref{ode1})-(\ref{ode4}), one can find $z(l)$ numerically.
The parametric representation of the $\theta=\pi/2$- slice of the embedded three-dimensional space in cylindrical coordinates $(\eta(l),\varphi,z(l))$ is shown on fig. \ref{wormT}.

In order to have a viable wormhole solution the imbedded surface must flare out, that is the inverse of the embedding function $\eta=\eta(z)$ must have a nonvanishing minimum radius, $\eta(0)=r_0$  and $d\eta/dz\to0$ at the throat. The second derivatives must satisfy, $d^2\eta/dz^2>0$ near the throat $r_0$. If we now look at the slope
\begin{eqnarray}
  \frac{d\eta}{dz} &=& \frac{d\eta}{dl}\frac{dl}{dz};\qquad\qquad \frac{dl}{dz}=\pm\left(f(l)-\frac{l^2}{l^2+r_0^2}\right)^{-1/2}\nonumber \\
   &=& \pm\frac{l}{\sqrt{l^2+r_0^2}}\left(f(l)-\frac{l^2}{l^2+r_0^2}\right)^{-1/2}\nonumber \\
   &=& \pm\,l\left(f(l)\left(l^2+r_0^2\right)-l^2\right)^{-1/2},
\end{eqnarray}
then, at the throat $\left(\frac{d\eta}{dz}\right)_{l\to 0}=0$ and the second derivative becomes
\begin{eqnarray}
  \frac{d^2\eta}{dz^2} &=& \frac{d}{dz}\left(\frac{d\eta}{dz}\right)=\frac{d}{dl}\left(\frac{d\eta}{dz}\right)\left(\frac{dl}{dz}\right),\nonumber \\
   &=& \pm\frac{d}{dl}\left(\frac{l^2+r_o^2}{l^2}f(l)-1\right)^{-1/2}.\frac{dl}{dz},
\end{eqnarray}
being at the throat $\left.\frac{d^2\eta}{dz^2}\right|_{l\to0}=1/r_o f_0$. Despite the function $\eta(z)$ satisfies these conditions, we may define the proper distance $\xi$ from the throat as an independent coordinate
\begin{equation}\label{emb3}
  \xi(l)=\int_{0}^{l}\sqrt{f(l')}dl'.
\end{equation}
It is easily seen that the above conditions are indeed satisfied,
\begin{equation}\label{transversable12}
  \left.\frac{d\eta}{d\xi}\right|_{l=0}=0,\qquad\qquad\left.\frac{d^2\eta}{d\xi^2}\right|_{l=0}=\frac{1}{r_0f_0}>0.
\end{equation}

\subsection{Violation of the energy conditions}
One way to solve the Einstein equation in the presence of some kind of matter is to specify the sources of the fields involved in the stress-energy tensor $T_{\mu\nu}$, solve the differential equations and thus obtain the metric $g_{\mu\nu}$. The inverse process is nonetheless also possible: to consider a metric solution $g_{\mu\nu}$, and solve the equations for the stress-energy tensor components $T_{\mu\nu}$. However, despite the validity of this process, not all stress-energy tensor solutions are allowed from the point of view of the basic physical principles. Certain energy conditions must be satisfied by these tensors.

These energy conditions are invariant restrictions of the energy-momentum tensor, and in general true in almost every classical considerations. These scalar invariant quantities can be constructed contracting the stress-energy tensor with timelike and null vectors. We write this condition for a generic stress-energy tensor
\begin{equation}\label{stress-energy}
  T^{\mu\nu}=\left[
  \begin{array}{cccc}
    \rho & 0 & 0 & 0 \\
    0 & p_1 & 0 & 0 \\
    0 & 0 & p_2 & 0 \\
    0 & 0 & 0 & p_3
  \end{array}
  \right],
\end{equation}
where $\rho$ is the density of the matter and $p_i$ are the principal pressures along the three spacelike directions $e_i$. Thus for $p_1=p_2=p_3$ we have a perfect fluid. The energy conditions are summarized in the following table:

\begin{table}[ht!]
  \centering
  \begin{tabular}{@{}m{0pt}@{}|M{3.5cm}|M{3.5cm}|M{4cm}|M{3.5cm}|@{}m{0pt}@{}}
    \hhline{|======|}
    \multicolumn{6}{|c|}{Energy conditions}  \\
    \hhline{|------|}
    &\begin{center}
          Null
        \end{center}
        &
       \begin{center}
          Weak
        \end{center}
        &
        \begin{center}
          Strong
        \end{center}
        &
        \begin{center}
          Dominant
        \end{center}& \\[-15pt]
    \hhline{|======|}
   &
    -The NEC states that,\newline
    \scalebox{0.9}{$T_{\mu\nu}k^{\mu}k^{\nu} \geq 0$}\newline
    for any null vector $k^\mu$ &

    -The WEC states that,\newline
    \scalebox{0.9}{$T_{\mu\nu}u^{\mu}u^{\nu} \geq 0$}\newline
    for any timelike vector $u^\mu$ &

    -The SEC states that,
    \scalebox{0.9}{$\left(T_{\mu\nu}-\frac{1}{2}T g_{\mu\nu}\right)u^\mu u^\nu \geq 0$}\newline
    for any timelike vector $u^\mu$\newline &

    -The DEC states that,
    \scalebox{0.9}{$T_{\mu\nu}u^{\mu}u^{\nu} \geq 0$}\newline
    and \scalebox{0.9}{$T_{\mu\nu}u^{\mu}$} is not spacelike
    for any timelike vector $u^\mu$\newline & \\[50pt]
    \hline
       & -For the stress-energy tensor
         of the form (\ref{stress-energy})\newline
        \scalebox{0.9}{$\rho+p_i\geq 0,\,\,\, i=1\ldots 3$} &

        -For the stress-energy tensor
         of the form (\ref{stress-energy})\newline
        \scalebox{0.9}{$\rho\geq 0$},\newline
        \scalebox{0.9}{$\rho+p_i\geq 0,\,\,\, i=1\ldots 3$} &

        -For the stress-energy tensor
         of the form (\ref{stress-energy})\newline
        \scalebox{0.9}{$\rho+p_i\geq 0$},\newline
        \scalebox{0.9}{$\rho+\sum_{i}p_i\geq 0,\,\,\, i=1\ldots 3$} &

        -For the stress-energy tensor
         of the form (\ref{stress-energy})\newline
        \scalebox{0.9}{$\rho\geq 0$},\newline
        \scalebox{0.9}{$p_i\in[-\rho,\rho],\,\,\,i=1\ldots 3$} & \\[50pt]
    \hhline{|=====|}
    \end{tabular}
  \caption{The null, weak, strong and dominant energy conditions.}
  \label{Energy_conditions}
\end{table}
Here $T=T_{\mu\nu}g^{\mu\nu}$ is the trace of stress-energy tensor. Among various energy conditions presented above in the context of general relativity, the null energy condition (NEC) represents a special interest. The NEC is quite robust in classical theory, because almost every classical field satisfies it, with the exception of some matter fields with nonminimal coupling to gravity. However, in the realm of quantum physics, these energy conditions may no longer be true, so that the observer could measure negative energy densities. There are several examples where these apparently unphysical conditions are fulfilled, for instance, due to the Casimir effect and the Hawking evaporation of black holes.

Despite the consistency of the null energy conditions in classical physics, it is the weakest among all the energy conditions since its violation leads to the violation of the other energy conditions. Wormholes are interesting physical entities whose existence relies on the violation of the null energy condition. In this sense, wormholes violate all energy conditions above mentioned.

For the Kanti-Kleihauss-Kunz wormhole solution, we can choose the general null vectors of the form $(-e^\nu,f^{1/2},0,0)$ and $(-e^\nu,0,r,0)$, thus the null energy condition can be expressed respectively as
\begin{eqnarray}\label{EC-Wormhole}
  &&-T_0^0+T_l^l\geq 0,\qquad \text{and}\\
  &&-T_0^0+T_\theta^\theta\geq 0.
\end{eqnarray}

If one or both of the these inequalities do not hold in some region of spacetime, the null energy condition is violated. Using the expansion of the metric function and dilaton fields near the throat, (\ref{ft1}), (\ref{nut1}), and (\ref{phit1}), we find that
\begin{equation}\label{EC-Wormhole2}
  \left[-T_0^0+T_l^l\geq 0\right]_{l=0}=-\frac{2}{f_0r_0^2}<0,
\end{equation}
thus, for the wormhole solutions close to the throat the null energy condition is violated. 

\chapter{Linear wormhole perturbations}\label{LPKKK}
\section{Linearized spherically symmetric perturbation}\label{WLP}
In order to study linear perturbations of Kanti-Kleihauss-Kunz spherically symmetric wormholes, we perturb the background which corresponds to perturbations of the dilaton $\delta\phi$ and the metric functions, $\delta f$ and $\delta\nu$,
\begin{eqnarray}\nonumber
\phi\to\phi(l)+\delta\phi(t,l),\\\label{perturbations}
f\to f(l)+\delta f(t,l),\\
\nu\to\nu(l)+\delta\nu(t,l).\nonumber
\end{eqnarray}

After substitution the perturbed functions in the modified Einstein equation and the dilaton equation, these can be reduced to one dynamical equation for $\delta\phi$ \cite{Kanti:2011yv}
\begin{equation}
\frac{\partial^2\delta\phi}{\partial l^2} + q_1(l) \frac{\partial\delta\phi}{\partial l} + q_0(l) \delta\phi(t,l) - q_\sigma(l) \frac{\partial^2\delta\phi}{\partial t^2} = 0 \ ,
\label{lineq}
\end{equation}
where the functions $q_{\sigma}(l)$, $q_0(l)$ and $q_1(l)$ depend on the background solution (metric and dilaton field) and have some cumbersome expressions. Although $q_0(l)$ and $q_1(l)$ diverge at the throat ($l=0$) as $\mathcal{O}(l^{-1})$, the function $q_\sigma$ is finite everywhere. Furthermore, the perturbations of the metric functions, $\delta f$ and $\delta\nu$, can be expressed in terms of $\delta\phi$ and its derivatives, in this sense, by solving the differential equation (\ref{lineq}), we also obtain the perturbations of the metric functions. Notice also that the perturbations of the throat radius $\delta r(l, t)$ are set to be zero in \cite{Kanti:2011yv}.

From equation (\ref{lineq}), we can see that the differential equation becomes nonhyperbolic for $q_\sigma<0$, which happens for sufficiently small wormhole ($\alpha>0.05 r_0^2$) implying that the kinetic term has a wrong sign. A similar feature was observed for the small black holes in the Lovelock theory \cite{Takahashi:2010gz}, which were found later unstable within the linear approximation against vector-type gravitational perturbations \cite{Konoplya:2017lhs}. Yet, it was pointed out in \cite{Reall:2014pwa} that the nonhyperbolicity of the perturbation equations implies that the stability of such black holes requires a more complex analysis since the initial value problem is not well posed in this case. In this sense, in order to avoid the nonhyperbolicity problem of the perturbation equation, we will consider only the configurations with sufficiently large throat.

For sufficiently large wormholes ($\alpha\leq0.05r_0^2$) the function $q_\sigma(l)$ is positive everywhere and after choosing the appropriate gauge invariant quantity function $\chi=\chi(t,l(y))$ (\ref{lineq}) can be reduced to the wavelike equation \cite{Kanti:2011yv}
\begin{equation}
\frac{\partial^2\chi}{\partial t^2}-\frac{\partial^2\chi}{\partial y^2} + V_{\rm eff}(l)\chi(t,l)= 0 \ ,
\label{wavelike}
\end{equation}
with the effective potential
\begin{equation}\label{potential}
V_{\rm eff}(l)=\frac{q_0(l)}{q_\sigma(l)}-\frac{q_1'(l)}{2 q_\sigma(l)}-\frac{q_1(l)^2}{4 q_\sigma(l)}-\frac{q_\sigma''(l)}{4q_\sigma(l)^2}+\frac{5 q_\sigma'(l)^2}{16q_\sigma(l)^3},
\end{equation}
where $y$ corresponds to the tortoise coordinate defined as
\begin{equation}\label{tortoisey}
  y= \int\sqrt{q_\sigma(l)}dl.
\end{equation}

The effective potential (\ref{potential}) is bounded on the whole range of $l$ and compatible with the behavior of other wormhole configurations, approaching constant values at both infinities \cite{Konoplya:2005et}. After imposing purely outgoing boundary conditions at both infinities $y\rightarrow\pm\infty$,

\begin{equation}
\chi  \sim e^{\pm i \omega y}, \quad y \to \pm \infty,
\end{equation}
any solution to equation (\ref{wavelike}) is a superposition of oscillations characterized by the discrete spectrum of quasinormal modes. In a similar fashion with the black hole case (see, for example, \cite{Konoplya:2011qq}), such boundary conditions describe the ``momentary'' reaction of a wormhole to the perturbation, when the source of perturbation stopped acting. In order to study stability of the wormhole perturbations, we will use the method of numerical characteristic integration of the wave-like equation described in section \ref{CI} which was used in previous chapters for the case of Einstein-Gauss-Bonnet-de Sitter black holes.

\begin{figure*}
\resizebox{\linewidth}{!}{\includegraphics*{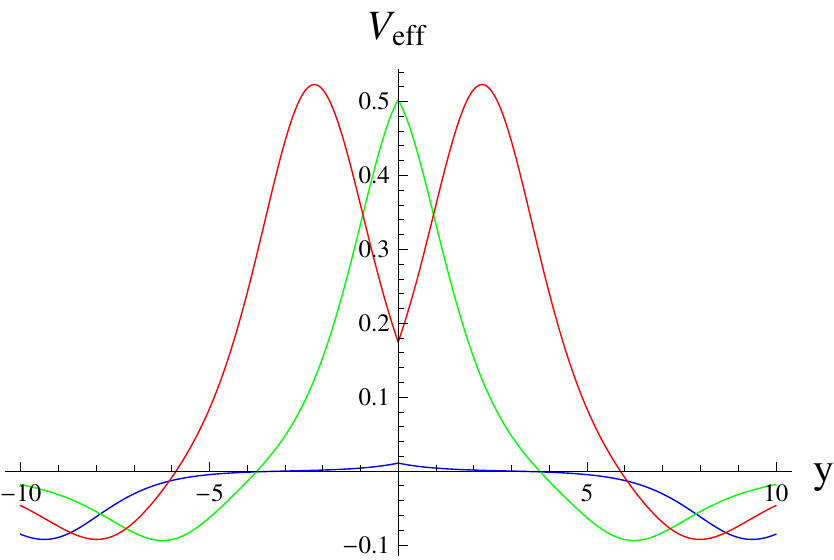}\includegraphics*{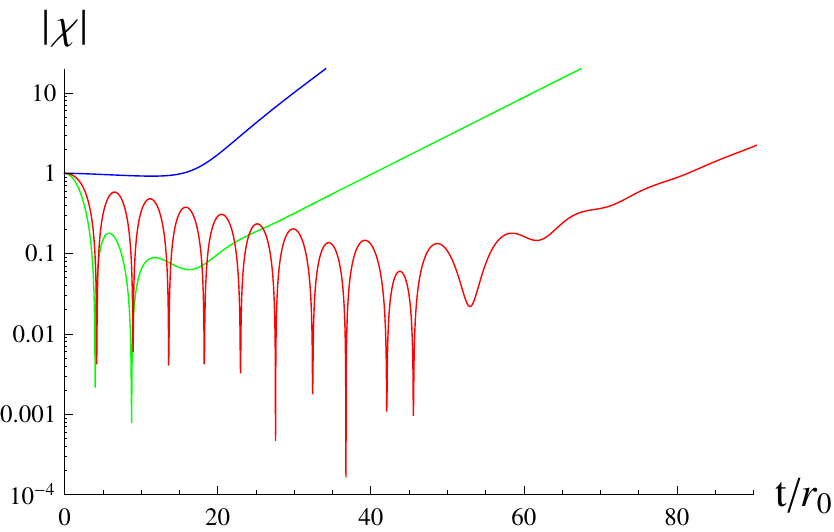}}
\caption{Effective potentials (left panel) and the time-domain profiles (right panel) for the spherically symmetric perturbations of the wormhole $\alpha=0.02r_0^2$: $f_0=1.001$ (blue), $f_0=1.1$ (green), $f_0=10$ (red).}\label{aplot}
\end{figure*}

\begin{figure*}
\resizebox{\linewidth}{!}{\includegraphics*{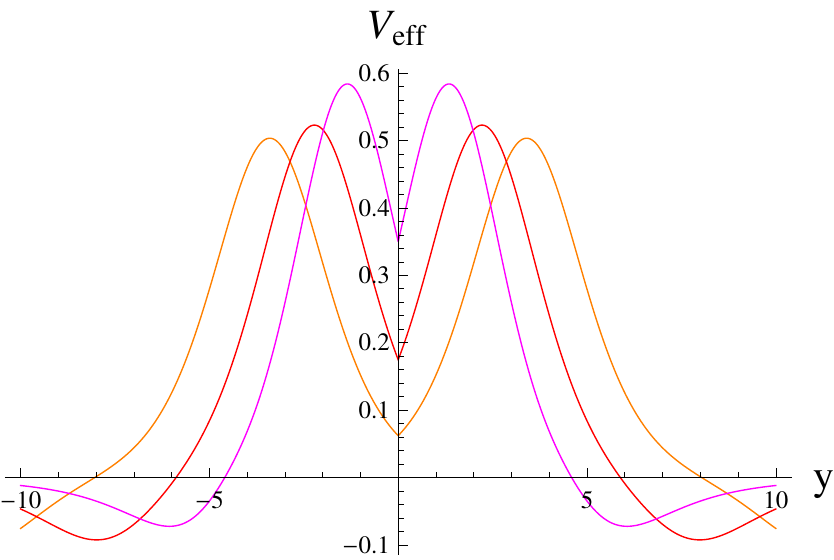}\includegraphics*{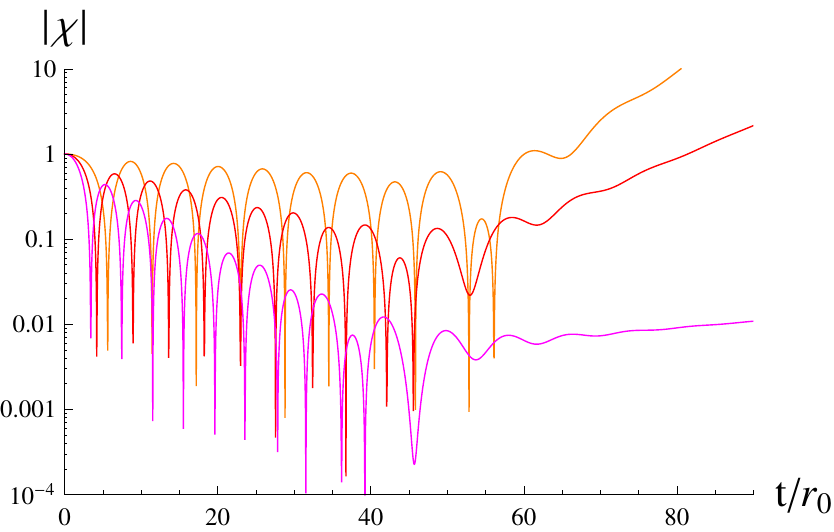}}
\caption{Effective potentials (left panel) and the time-domain profiles (right panel) for the spherically symmetric perturbations of the wormhole $f_0=10$: $\alpha=0.005r_0^2$ (orange), $\alpha=0.02r_0^2$ (red), $\alpha=0.05r_0^2$ (magenta). As $\alpha$ grows the peak becomes higher and shifts towards the throat uppering the local minimum at the throat.}\label{fplot}
\end{figure*}

\section{Instability profiles}\label{IP}
Following \cite{Bronnikov:2011if}, for the general spherically symmetric perturbations we study the time-domain profile for $\chi(t,l)$ at the throat ($l=0$).  In order to obtain the time-domain profile we use the discretization scheme proposed by Gundlach, Price, and Pullin \cite{Gundlach:1993tp} and presented in the chapter \ref{NM}.

The numerical temporal profiles were obtained for a wide range of free-parameters $(f_0,\alpha/r_0^2)$. We observed that the profiles show the typical damped oscillation for intermediate times that give way to a nonoscillating growth at late time (see figs.~\ref{aplot}~and~\ref{fplot}). In order to reach the numerical convergence necessary to our objectives and keep the error propagation under control, when integrating until very late times, we increase the accuracy of our calculations as well as diminish the step of the grid. After all, exponential growth after a long period of damped oscillations was observed in the tensor channel of higher dimensional black holes in the Einstein-Gauss-Bonnet theory \cite{Konoplya:2008ix,Cuyubamba:2016cug}.

Fig.~\ref{aplot}, shows that the higher $f_0$ is at the throat, the larger is peak of the effective potential and eventually shifts outside the throat, leading to a double peak due to symmetry. In addition, we also observe that unstable modes appear for any values of the parameter $f_0\geq 1.001$. In fig.~\ref{fplot} we show that the smaller values of $\alpha$-coupling is, the higher growth rates appear, so that when $\alpha \to 0$ it is possible that the instability growth rate may go to infinity. This means that the purely imaginary quasinormal frequency, which is responsible for the instability would increase unboundedly as $\alpha$ goes to zero. This indicates that the above mode is \emph{nonperturbative in $\alpha$}. The wormhole is therefore evidently unstable at whatever small value of $\alpha$. The same phenomenon takes place for $D>4$ dimensional asymptotically flat and anti-de Sitter black holes in various theories with higher curvature corrections
\cite{Konoplya:2008ix,Cuyubamba:2016cug,Konoplya:2017ymp,Gonzalez:2017gwa,Grozdanov:2016vgg,Grozdanov:2016fkt,Konoplya:2017zwo}.

\section{Emergence of instability}\label{DisW}
Unlike \cite{Kanti:2011jz}, we found that Kanti-Kleihaus-Kunz wormhole solutions are unstable as we have previously shown. There are some technical reasons for this discrepancy and this section will be dedicated to present the corresponding analysis.

The perturbation function considered in \cite{Kanti:2011jz} is expressed as
\begin{equation}\label{X1X}
  \delta\phi(t,l)=A(l)\chi(t,l),
\end{equation}
where $A(l)$ is a factor which diverges at the throat as $\mathcal{O}(l^{-1})$. According to \cite{Kanti:2011jz}, in order to have a finite perturbation there, the vanishing Dirichlet boundary conditions at the throat are considered ($\chi=0$ at $l=0$). Effectively this disconnects the regions of space on both sides of the throat for purely radial modes of perturbation. It is worth mentioning that the differential perturbation equation obtained in \cite{Kanti:2011jz} was found considering only the perturbations of the metric functions, $f$ and $\nu$, and the dilaton field $\phi$ (i.e. the equation (\ref{lineq}) was derived assuming that $\delta r=0$ which fixes the throat size $r_0$).

The most general radial perturbation for this wormhole must consider also the perturbation of the radial parameter $r$,
\begin{equation}
r\to r(l)+\delta r(t,l),\label{generalperturbations}
\end{equation}
where in the righthand side the function $r(l)=\sqrt{l^2+r_0^2}$ corresponds to the unperturbed radial quantity. A similar situation was considered in \cite{Bronnikov:2011if}, where various black holes supported by a non-minimal phantom scalar field were considered. In \cite{Bronnikov:2011if}, it was shown that a gauge-invariant quantity $\psi=r'\delta\phi-\phi'\delta r$ can be constructed for a general spherically symmetric wormhole solution. Therefore, It is easy to see that, in our case, the following quantity is a gauge-invariant perturbation function
\begin{equation}\label{gaugeinvariant}
\chi(t,l)= \frac{1}{A(l)}\left(\delta\phi(t,l)-\frac{r}{l}\phi'(l)\delta r(t,l)\right),
\end{equation}
which satisfies the wave-like equation (\ref{wavelike}) with the finite effective potential (\ref{potential}). When $\chi$ is finite at the throat, $\delta\phi$ does not diverge unless we fix the throat size by choosing $\delta r=0$ in~(\ref{gaugeinvariant}). Thus, the instability of the wormhole is easily understood, if one assumes perturbation of the throat's radius.

In other words, the dynamical problem of evolution of the finite wave function $\chi$ which we solved for the finite effective potential~(\ref{potential}) does not have any divergence problem. The behavior of the dilaton field at the throat, as it was shown for the other type of a scalar field in \cite{Bronnikov:2011if}, is a pure artifact of the gauge chosen in \cite{Kanti:2011yv} and, therefore, can be safely ignored. 

\chapter{Conclusions}\label{conclusions}
In the first part of this thesis, we have performed a thorough analysis of gravitational quasinormal spectrum of asymptotically de Sitter black holes in the Einstein-Gauss-Bonnet theory. Usage of the time-domain integration allowed us to take into consideration contributions of all the modes in the signal and, thereby, to judge about the (in)stability of the black hole. Gravitational perturbations can be reduced to the independent master equations for scalar, vector, and tensor types relatively the rotation group on $(D-2)$-sphere. It has been shown that the scalar channel of the gravitational perturbation are unnstable at sufficiently large values of the cosmological constant $\Lambda$. We called this instability ``the $\Lambda$-instability'', because it does not take place for asymptotically flat spacetimes. Our results have been recently confirmed within a nonlinear analysis \cite{Li:2019bqc}.
In addition, we have found that scalar and tensor channels also have instabilities for the nonzero Gauss-Bonnet coupling. This instability occurs at high multipole numbers $\ell$ and therefore was called ``the eikonal instability''. The summary of the (in)stabilities is presented in the table~\ref{tab:inst}.

\begin{table}[H]
\centering
\begin{tabular}{|c|c|c|}
  \hline
  $D$ & $\Lambda$-instability & eikonal instability \\
  \hline
  5 & scalar-type ($\ell =2$) & scalar-type\\
  \hline
  6 & scalar-type ($\ell =2$) & tensor-type\\
  \hline
  7 & scalar-type ($\ell =2$) & \\
  \hline
  8 & scalar-type ($\ell =2$) & \\
  \hline
\end{tabular}

\caption{Summary of instabilities of Einstein-Gauss-Bonnet-de Sitter black holes: each type of instability implies its parametric region. For the eikonal instability this region expands as $\ell$ increases, so that the instability region in this case corresponds to the limit $\ell\rightarrow\infty$.}\label{tab:inst}
\end{table}

It was demonstrated that the quasinormal frequencies of the scalar type of gravitational perturbations do not obey the Hod's inequality and the lowest mode in this channel of perturbation has higher damping rate than the one prescribed by the proposal. However, the other two channels, vector and tensor, have lower laying modes what, thereby, confirm the Hod's proposal. Apparently, it would be impossible to create the process of perturbation which would excite only the scalar channel and leave unperturbed the other two. If such a ``selective'' perturbation could be prepared or even theoretically modeled, the Hod's proposal would be violated.

In the second part of this thesis, we have reviewed the wormhole solution obtained by P. Kanti et al. in \cite{Kanti:2011jz} which is an important example of a wormhole supported by the string theory inspired modification of Einstein equations rather than by some exotic matter. Imposing the correct boundary conditions for the perturbation, the Kanti-Kleihaus-Kunz wormholes proved out to be unstable for whatever small values of the Gauss-Bonnet coupling constant $\alpha$. The instability develops after a long phase of damped quasinormal oscillations, which is similar to the time-domain profile of the eikonal instability observed for black holes in Einstein-Gauss-Bonnet theory. Moreover, it was recently shown in~\cite{Shinkai:2017xkx} that the fate of a perturbed spherically symmetric wormhole supported by scalar fields is either a black hole or an expanding throat depending on the total energy of the structure. This result supports our general conclusion that the Kanti-Kleihaus-Kunz wormholes are unstable.

Even being unstable, the Kanti-Kleihaus-Kunz wormhole solution is a unique and important example of a traversable wormhole supported not by an exotic matter, but by introducing the second order curvature correction and dilaton, which are inspired by string theory. This wormhole appears naturally as a result of quantum corrections to the Einsteinian theory. Therefore, in our opinion, further efforts for finding stabilizing factors could be made in the future by considering nonminimal theory with additional fields (axions, fermions, gauge fields) or other higher curvature corrections.

Nowadays, black-hole solutions are actively studied in the higher-curvature context. An analytical approximation for the four-dimensional black hole in Einstein-dilaton-Gauss-Bonnet gravity has been developed in \cite{Kokkotas:2017ymc}, quasinormal modes of such black holes have been studied for test fields \cite{Konoplya:2019hml,Zinhailo:2019rwd}. Black holes in theories with general scalar coupling to the Gauss-Bonnet term have been studied in \cite{Antoniou:2017hxj,Konoplya:2019fpy}.

Another topic of the active investigation is the eikonal instability appearing within the linear approximation in theories with higher curvature corrections. Such an instability is always accompanied by a non-hyperbolicity of the master perturbation equations \cite{Reall:2014pwa} and leads to a breakdown of the whole regime of linear perturbations. It opens a new direction of research in order to formulate the well-posed initial value problem to the perturbation equations. Thus, compact objects in higher curvature corrected gravity are important models to study, raising new important questions on gravitational theory and its testing through gravitational-wave observations. 

\begin{appendices}

\chapter{Perturbations of Einstein-Gauss-Bonnet black holes}\label{ETC}
\section{Basic formulas}
The static spherically symmetric black-hole solution within the Einstein-Gauss-Bonnet gravity is given by the line element
\begin{eqnarray}
   ds^2&=&-f(r)dt^2 + \frac{dr^2}{f(r)}+r^2{\bar \gamma}_{i j}dx^idx^j\ ,\label{eq:solutionap}\\
   f(r)&=&\kappa-r^2\psi(r) \ , \label{eq:defap}
\end{eqnarray}
where $\kappa=0,\pm 1$, $r$ is the radial coordinate, $\psi(r)$ is a real function, and $\bar{\gamma}_{ij}$ is the line-element on the $n$-sphere ($\mathcal{S}^n$). The Riemann tensor components are given by
\begin{eqnarray}
	&\ &R_{tr}{}^{tr}=-\frac{f''}{2} \ ,\quad
      R_{ti}{}^{tj}=R_{ri}{}^{rj}=-\frac{f'}{2r}\delta_{i}{}^{j} \ ,\quad \nonumber\\
      &\ &R_{ij}{}^{kl}=\left(\frac{\kappa-f}{r^2}\right)\left(\delta_{i}{}^{k}\delta_{j}{}^{l}-\delta_{i}{}^{l}\delta_{j}{}^{k}\right) \ . \label{eq:riemannap}
\end{eqnarray}
The non-vanishing Ricci tensor components are
\begin{eqnarray}
  R_t^t &=& \frac{-2f''\,f+f'^2-f\,f'^2}{4f^2}-\frac{nf'}{2r}, \\
  R_r^r &=& -\frac{f''}{2}-\frac{n\,f'}{2r}, \\
  R_i^j &=& \frac{2(\kappa-f)(n-1)-r\,f'}{r^2}\delta_i^j.
\end{eqnarray}

Using the last expressions, we can determine the elements of the modified Einstein tensor which is expressed as
\begin{equation}\label{HC:EGB:6ap}
  G_{\mu}^{\phantom{\mu}\nu}=\Lambda\,G_{(0)\mu}^{\phantom{(0)\mu}\nu}+ G_{(1)\mu}^{\phantom{(1)\mu}\nu}+\alpha\, \ G_{(2)\mu}^{\phantom{(2)\mu}\nu},
\end{equation}
where $\Lambda$ is the cosmological constant and
\begin{eqnarray}
  G_{(0)\mu}^{\phantom{(0)\mu}\nu} &=& \delta_\mu^\nu,\label{BHP:31ap} \\
  G_{(1)\mu}^{\phantom{(1)\mu}\nu} &=& R_\mu^{\phantom{\mu}\nu}-\frac{1}{2}\delta_\mu^\nu\,R,\\\label{BHP:41ap}
    G_{(2)\mu}^{\phantom{(2)\mu}\nu} &=& R_{\lambda\mu}^{\phantom{\lambda\mu}\sigma\rho}R_{\sigma\rho}^{\phantom{\sigma\rho}\lambda\nu}-
  2R_\sigma^{\phantom{\sigma}\lambda}R_{\lambda\mu}^{\phantom{\lambda\mu}\sigma\nu} -2R_\mu^{\phantom{\mu}\lambda}R_\lambda^{\phantom{\lambda}\nu}+R\,R_\mu^{\phantom{\mu}\nu}- \nonumber\\
  & & \frac{1}{4}\delta_\mu^\nu\left(R_{\lambda\sigma}^{\phantom{\lambda\sigma}\rho \delta}R_{\rho\delta}^{\phantom{\rho\delta}\lambda\sigma}
  -4R_\lambda^{\phantom{\lambda}\sigma}R_\sigma^{\phantom{\sigma}\lambda}+ R^2\right).\label{BHP:51ap}
\end{eqnarray}

In order to study the gravitational perturbation around a background solution (\ref{eq:solutionap})-(\ref{eq:defap}), we perturb the metric tensor $g_{\mu\nu}\rightarrow g_{\mu\nu}+h_{\mu\nu}$, then the variation of the modified Einstein tensor are present as follow
\begin{eqnarray}\label{delta_Gap}
	\delta{G}_t^t&=&-r^{2-n}\left(\frac{T'}{2(n-1)}\delta R_{ik}{}^{ik}+\frac{T}{r}\delta R_{ir}{}^{ir}\right)\ ,\nonumber\\
	\delta{G}_t^r&=&-\frac{T}{r^{n-1}}\delta R_{it}{}^{ir}\ ,\nonumber\\
	\delta{G}_t^i&=&r^{2-n}\left(\frac{T'}{(n-1)}\delta R_{tk}{}^{ik}+\frac{T}{r}\delta R_{tr}{}^{ir}\right)\ ,\nonumber\\
	\delta{G}_r^r&=&-r^{2-n}\left(\frac{T'}{2(n-1)}\delta R_{ik}{}^{ik}+\frac{T}{r}\delta R_{it}{}^{it}\right)\ ,\nonumber\\
	\delta{G}_r^i&=&r^{2-n}\left(\frac{T'}{(n-1)}\delta R_{rk}{}^{ik}+\frac{T}{r}\delta R_{rt}{}^{it}\right)\ ,\nonumber\\
	\delta{G}_i^j&=&-\frac{T}{r^{n-1}}\delta R_{tr}{}^{tr}\delta_i^j + \frac{T'}{(n-1)r^{n-2}}\left(\delta R_{ti}{}^{tj}+\delta R_{ri}{}^{rj}-\left(\delta R_{tk}{}^{tk}+\delta R_{rk}{}^{rk}\right)\delta_i^j\right)\nonumber\\
 &&+ \frac{T''}{2(n-1)(n-2)r^{n-3}}\left(2\delta R_{ik}{}^{jk}+\delta R_{kl}{}^{kl}\delta_i^j\right),
\end{eqnarray}
where
\begin{equation}\label{deltaRimap}
  \delta R_{\mu\nu}{}^{\rho\sigma}=\frac{1}{2}\left(R_{\mu\nu}{}^{\rho\lambda}h_\lambda{}^\sigma-R_{\mu\nu}{}^{\sigma\lambda}h_\lambda{}^\rho +\left(\nabla_\nu\nabla^\sigma\,h_\mu{}^\rho-\nabla_\mu\nabla^\sigma\,h_\nu{}^\rho\right)+\left(\nabla_\mu\nabla^\rho\, h_\nu{}^\sigma-\nabla_\nu\nabla^\rho\,h_\mu{}^\sigma\right)\right).
\end{equation}

\section{Gravitational channels}
The dynamical equations of gravitational perturbations are determined from $\delta G_{\alpha}^\beta$=0, and can be derived separately for each channel of perturbations.

\subsection*{Tensor-type perturbations}
The tensor-type perturbations are characterized as $h_{0\mu}=h_{1\mu}=0$ and $h_{ij}=r^2\phi(t,r)\mathbb{T}_{ij}$, where $\mathbb{T}_{ij}$ are symmetric tensor spherical harmonics, which satisfy
\begin{equation}\label{TTP-1ap}
  \left(\bar{\nabla}^k\bar{\nabla}_k + \left(\ell(\ell+n-1)-2\right) \right)\mathbb{T}_{ij}= 0,\qquad\mathbb{T}_{ij;k}=0,\qquad\mathbb{T}_i^{i}=0.
\end{equation}
Therefore, we can calculate the variation of the Riemann tensor
\begin{eqnarray}
	\delta R_{ti}{}^{tj}
      &=&\frac{f'}{2f}\left(\frac{{\ddot \phi}}{f'}-\frac{f\,\phi'}{2}\right) \mathbb{T}_{i}{}^{j}
      \ , \nonumber\\
      \delta R_{ri}{}^{rj}&=&-\frac{f}{2}\left(\phi''+\left(\frac{f'}{2f}+\frac{2}{r}\right)\phi'\right)\mathbb{T}_{i}{}^{j} \ , \nonumber\\
      \delta R_{il}{}^{jl}&=&
      \left( -\frac{n-2}{2} \frac{f}{r} \phi'
           +\frac{2\kappa +\left(\ell(\ell+n-1)-2\right)}{2r^2} \phi \right)\mathbb{T}_{i}{}^{j}
      \ .\label{eq:deltariemannap}
\end{eqnarray}
Substituting these equations into (\ref{delta_Gap}), we have
\begin{eqnarray}
	\delta{G}_{i}{}^{j} &=& \frac{r^{2-n}}{n-1}\left(\frac{T'}{2f}\left({\ddot\phi}-f^2\phi''\right)-\frac{T'}{2}\left(f'+\frac{2f}{r} +\frac{f\,T''}{f'}\right)\phi'+\right.\\
 &&\left.\frac{(2\kappa+\left(\ell(\ell+n-1)-2\right))T''}{2(n-2)r}\phi\right)\mathbb{T}_{i}{}^j=0\ .\nonumber
\end{eqnarray}
After introducing the wave function $\Psi(t,r)=R(r)\phi(t,r)=r\sqrt{T'}\phi(t,r)$, and using the tortoise coordinate $dr_*=dr/f(r)$ the dynamical equation for tensor-type perturbations in linear approximation takes the form,
\begin{eqnarray}
	\left(\frac{\partial^2}{\partial\,t^2}-\frac{\partial^2}{\partial\,r_*^2}+V_t(r(r_*))\right)\Psi(t,r(r_*))=0, \label{eq:potentialap}
\end{eqnarray}
where the effective potential $V_t$ for tensor-type perturbations is
\begin{eqnarray}
    V_t(r)&=&\dfrac{\ell(\ell+n-1)f(r)T^{\prime\prime}(r)}{(n-2)r\,T^{\prime}(r)}+ \dfrac{1}{R(r)}\dfrac{d^2}{dr_*^2}\left(R(r)\right),\label{BHP:11ap}\\
    T(r)&=&r^{n-1}\left(1+r^{-2}\alpha(n-1)(n-2)(\kappa-f(r))\right).\label{BHP:9ap}
\end{eqnarray}

\subsection*{Vector-type perturbations}
The the vector-type channel of black-hole perturbations generalizes the Regge-Wheeler gauge for odd perturbations. The perturbations have the form (\ref{BHP:12}) and after some calculations, the non-zero components of $\delta R_{\mu\nu}{}^{\rho\sigma}$ can be expressed as
\begin{eqnarray}
	\delta R_{tk}{}^{ij}&=&\frac{\kappa-f}{r^4}\left(\delta_k^j f_t\,\mathbb{V}^i-\delta_k^i\,f_t\,\mathbb{V}^j\right)+\frac{f'}{2r^3}\left(\delta_k^j f_t\,\mathbb{V}^i-\delta_k^i\,f_t\,\mathbb{V}^j\right)\nonumber\\
	                    &\ &-\frac{1}{2r^4}\left[-\bar{\nabla}_k\bar{\nabla}^i\,(f_t\,\mathbb{V}^j) +\bar{\nabla}_k\bar{\nabla}^j\,(f_t\,\mathbb{V}^i) +\kappa(\delta_k^j f_t\,\mathbb{V}^i-\delta_k^i\,f_t\,\mathbb{V}^j)\right]\nonumber\\
	                    &\ &-\frac{f}{2r^3}\Biggl[\delta_k^j\left((f_t\,\mathbb{V}^i)'-\partial_t((f_r\,\mathbb{V}^i))-\frac{2}{r}f_t\,\mathbb{V}^i\right)
	                               -\delta_k^i\left((f_t\,\mathbb{V}^j)'-\partial_t((f_r\,\mathbb{V}^j))-\frac{2}{r}f_t\,\mathbb{V}^j\right)\Biggr]\ ,\nonumber\\
	\delta R_{tr}{}^{ir}&=&-\frac{f'}{2r^3}f_t\,\mathbb{V}^i+\frac{f''}{2r^2}f_t\,\mathbb{V}^i\nonumber\\
	&\ &-\frac{f}{2r^3}\left((f_t\,\mathbb{V}^i)'-\partial_t((f_r\,\mathbb{V}^i))-\frac{2}{r}f_t\,\mathbb{V}^i\right)-\frac{f}{2r^2}\left((f_t\,\mathbb{V}^i)' -\partial_t((f_r\,\mathbb{V}^i))-\frac{2}{r}f_t\,\mathbb{V}^i\right)'\ ,\nonumber\\	
	\delta R_{rk}{}^{ij}&=&\frac{\kappa-f}{r^4}\left(\delta_k^j f_r\,\mathbb{V}^i-\delta_k^i\,f_r\,\mathbb{V}^j\right)+\frac{f'}{2r^3}\left(\delta_k^j f_r\,\mathbb{V}^i-\delta_k^i\,f_r\,\mathbb{V}^j\right)\nonumber\\
	                    &\ &\hspace{1cm}+ \frac{1}{2r^4}\left[\bar{\nabla}_k\bar{\nabla}^i(f_r\,\mathbb{V}^j) -\bar{\nabla}_k\bar{\nabla}^j(f_r\,\mathbb{V}^i)-\kappa(\delta_k^j\,f_r\,\mathbb{V}^i-\delta_k^i\,f_r\,\mathbb{V}^j)\right]\ ,\nonumber\\
	\delta R_{tr}{}^{ti}&=&-\frac{f'}{2r^3}f_r\,\mathbb{V}^i+\frac{f''}{2r^2}f_r\,\mathbb{V}^i-\frac{1}{2r^2f}\partial_t\left((f_t\,\mathbb{V}^i)' -\partial_t((f_r\,\mathbb{V}^i))-\frac{2}{r}f_t\,\mathbb{V}^i\right)\ , \nonumber\\
	\delta R_{ti}{}^{tj}&=&-\frac{1}{2r^2f}\left(\bar{\nabla}^j(\partial_t(f_t\,\mathbb{V}_i))+ \bar{\nabla}_i(\partial_t(f_t\,\mathbb{V}^j))\right) +\frac{f'}{4r^2}\left(\bar{\nabla}^j(f_r\,\mathbb{V}_i)+ \bar{\nabla}_i(f_r\,\mathbb{V}^j)\right)\ ,\nonumber\\
	\delta R_{ri}{}^{rj}&=&\frac{f'}{4r^2}\left(\bar{\nabla}^j(f_r\,\mathbb{V}_i)+ \bar{\nabla}_i(f_r\,\mathbb{V}^j)\right) +\frac{f}{2r^2}\left(\bar{\nabla}^j(f_r\,\mathbb{V}_i)+ \bar{\nabla}_i(f_r\,\mathbb{V}^j)\right)'\ ,\nonumber\\
	\delta R_{ij}{}^{kl}&=&\frac{f}{2r^3}\left(
	\delta_j^l\left(\bar{\nabla}^k(f_r\,\mathbb{V}_i)+ \bar{\nabla}_i(f_r\,\mathbb{V}^k)\right)-\delta_j^k\left(\bar{\nabla}^l(f_r\,\mathbb{V}_i) + \bar{\nabla}_i(f_r\,\mathbb{V}^l)\right)\right.\nonumber\\
	&\ &\left.\hspace{1cm}+\delta_i^k\left(\bar{\nabla}^l(f_r\,\mathbb{V}_j)+ \bar{\nabla}_j(f_r\,\mathbb{V}^l)\right)- \delta_i^l\left(\bar{\nabla}^k(f_r\,\mathbb{V}_j)+ \bar{\nabla}_j(f_r\,\mathbb{V}^k)\right)
	\right)\ ,
\end{eqnarray}
where $\mathbb{V}^i$ are the vector harmonic, and the real functions $f_t$ and $f_r$ both depends on $(t,r)$. Similar to the tensor-type perturbations, we need to substitute the equations into (\ref{delta_Gap}), then the non-vanishing components of the variation of Einstein tensor are
\begin{eqnarray}
	\delta {G}_t^i&=&-\frac{r^{-2-n}}{2}\left(\frac{T'}{n-1}f_t\,\bar{\nabla}_k\bar{\nabla}^k(\mathbb{V}^i)+\kappa T'f_t\,\mathbb{V}^i+f\left(r\,T\left((f_t\,\mathbb{V}^i)' -\partial_t(f_r\,\mathbb{V}^i)-\frac{2}{r}f_t\,\mathbb{V}^i\right)\right)'\right)\ ,\nonumber\\
	\delta {G}_r^i&=&-\frac{r^{-2-n}}{2}\left(\frac{T'}{n-1}f_r\bar{\nabla}_k\bar{\nabla}^k(\mathbb{V}^i)+\kappa T'f_r\,\mathbb{V}^i+\frac{r\,T}{f}\partial_t\left((f_t\,\mathbb{V}^i)'-\partial_t(f_r\,\mathbb{V}^i)-\frac{2}{r}f_t\,\mathbb{V}^i\right)\right)\ , \nonumber\\
	\delta{G}_i^j&=&-\frac{r^{-n}}{2(n-1)}\left(\frac{T'}{f}\left(\bar{\nabla}^j(\partial_t(f_t\,\mathbb{V}_i))+ \bar{\nabla}_i(\partial_t(f_t\,\mathbb{V}^j)) \right)-\left(f\,T'\left(\bar{\nabla}^j(f_r\,\mathbb{V}_i)+ \bar{\nabla}_i(f_r\,\mathbb{V}^j)\right)\right)'\right).\nonumber
\end{eqnarray}

Since the vector harmonics $\mathbb{V}_i$ satisfy $\bar{\nabla}_k\bar{\nabla}^k\mathbb{V}_i=-(\ell(\ell+n-1)-1)\mathbb{V}_i$, ($\ell>2$) for $\kappa=1$, after some algebra, the dynamical equation for the vector-type perturbations is expressed as
\begin{equation}\label{WLE-Vap}
  \ddot{\mathbb{V}}-r^2\left(\frac{f}{r^2\,T'}\right)'\left(f\,T'\,\mathbb{V}(t,r)\right)'- \frac{f}{T'}\left(f\,T'\,\mathbb{V}\right)'' +\frac{f}{r}\frac{T'}{T}\left(\frac{\ell(\ell+n-1)-1}{n-1}-\kappa\right)\mathbb{V}=0.
\end{equation}

The wave-like equation is obtained after introducing a new function $\Psi(t,r)=f\,R(r)\mathbb{V}/r$, and the tortoise coordinate. We have
\begin{eqnarray}
	\left(\frac{\partial^2}{\partial\,t^2}-\frac{\partial^2}{\partial\,r_*^2}+V_v(r(r_*))\right)\Psi(t,r(r_*))=0, \label{eq:potentialap2} \ .
\end{eqnarray}
where
\begin{equation}\label{BHP:14ap}
  V_v(r)=\frac{(\ell-1)(\ell+n)f(r)T^\prime(r)}{(n-1)r\,T(r)}+R(r)\frac{d^2}{dr_*^2}\left(\frac{1}{R(r)}\right).
\end{equation}

\subsection*{Scalar-type perturbations}
For the scalar-type channel, the metric perturbations in spherically symmetric black hole are given by
\begin{equation}
  h_{\mu\nu}=\left(
  \begin{array}{cc}
   {\large\left(\boldsymbol{\mathcal{F}_{ab}\mathbb{S}}\right)}  & {\large\left(\boldsymbol{r\,\mathcal{F}_a\,\mathbb{S}_i}\right)} \\
   {\Large{\bf Sym}} & {\large\left(\boldsymbol{2r^2\,H_L\,\bar{\gamma}_{ij}\,\mathbb{S}}\right)}
  \end{array}
  \right).
\end{equation}
where $\mathcal{F}_a$, $\mathcal{F}_{ab}$, $H_L$ are functions of $(t,r)$, ``{\bf Sym}'' represents the symmetric part of $h_{\mu\nu}$, and $\mathbb{S}$ is a scalar harmonic and $\mathbb{S}_i$ is a scalar-type vector harmonic which satisfies,
\begin{equation}\label{BHP:171ap}
  \mathbb{S}_i=-\frac{1}{\kappa_s}\bar{\nabla}_i\,\mathbb{S},
\end{equation}

After choosing an appropriate coordinate transformation to leads a gauge such that $\mathcal{F}_a=0$. Moreover, for convenience, we rewrite $\mathcal{F}_{ab}$ in the form
\begin{equation}\label{STP-1ap}
  \mathcal{F}_{tt}=f\,F_{tt},\qquad\mathcal{F}_{rr}=\frac{F_{rr}}{f},\qquad\mathcal{F}_{rt}=\dot{F}_{rt}.
\end{equation}

The variation of the Riemann tensor can be expressed as
\begin{eqnarray}
	\delta R_{ti}{}^{tj}&=&\frac{1}{2r^2}\bar{\nabla}_i\bar{\nabla}^j(F_{tt}\mathbb{S})+ \left(-\frac{1}{r}\partial_t(\dot{F}_{rt}\mathbb{S}) +\frac{1}{f}\partial^2_t(H_L\mathbb{S})-\frac{f^{'}}{2}(H_L\mathbb{S})'+\frac{f}{2r}(F_{tt}\mathbb{S})'+ \frac{f'}{2r}F_{rr}\mathbb{S}\right)\delta_{i}^{j}\ ,\nonumber\\
      \delta R_{ri}{}^{rj}&=&-\frac{1}{2r^2}\bar{\nabla}_i\bar{\nabla}^j(F_{rr}\mathbb{S}) +\left(\frac{1}{2r}(f\,F_{rr}\mathbb{S})'- \frac{f'}{2}(H_L\mathbb{S})'-\frac{2f}{r}(H_L\mathbb{S})'-(H_L\mathbb{S})''\,f\right)\delta_{i}^{j}\ ,\nonumber\\
	\delta R_{ij}{}^{kl}&=&\frac{1}{r^2}\left[\bar{\nabla}_j\bar{\nabla}^k(H_L\mathbb{S})\delta_{i}^{l}- \bar{\nabla}_i\bar{\nabla}^k(H_L\mathbb{S})\delta_{j}^{l} -\bar{\nabla}_j\bar{\nabla}^l(H_L\mathbb{S})\delta_{i}^{k} +\bar{\nabla}_i\bar{\nabla}^l(H_L\mathbb{S})\delta_{j}^{k}\right]\nonumber\\
	                    &\ &\hspace{1cm}+\left(-\frac{2\kappa}{r^2}H_L\mathbb{S}+\frac{f}{r^2}F_{rr}\mathcal{S}-\frac{2f}{r}(H_L\mathbb{S})'\right) \left(\delta_i^k\delta_j^l-\delta_i^l\delta_j^k\right)\ ,\nonumber
\end{eqnarray}
\begin{eqnarray}
	\delta R_{ti}{}^{rj}&=&-\frac{f}{2r^2}\bar{\nabla}_i\bar{\nabla}^j(\partial_t F_{rt}\,\mathbb{S})-\left(-\frac{f}{2r}\partial_t(F_{rr}\mathbb{S})-\frac{f'}{2}\partial_t(H_L\mathbb{S}) +f(\partial_t(H_L\mathbb{S}))'+ \frac{f}{r}\partial_t(H_L\mathbb{S})\right)\delta_i^j\ ,\nonumber\\
	\delta R_{ri}{}^{jk}&=&\frac{1}{2r^3}\left(\bar{\nabla}^j(F_{rr}\mathbb{S})\delta_i^k- \bar{\nabla}^k(F_{rr}\mathbb{S})\delta_i^j\right)- \frac{1}{r^2}\left(\bar{\nabla}^j(H_L\mathbb{S})'\delta_i^k- \bar{\nabla}^k(H_L\mathbb{S})'\delta_i^j\right)\nonumber\ ,\\
	\delta R_{tr}{}^{ti}&=&\frac{1}{2r^2}\bar{\nabla}^i\left(\frac{f'}{2f}(F_{rr}+F_{tt})\mathbb{S}+ (F_{tt}\mathbb{S})'- \frac{1}{r}F_{tt}\mathbb{S}- \frac{1}{f}\partial_t\left(\dot{F_{rt}}\mathbb{S}\right)\right)\ ,
	\label{Riemannap}
\end{eqnarray}
Let us substitute the above expression into equations (\ref{delta_Gap}), then $\delta G_{\mu\nu}=0$ leads to the scalar-type perturbed equations, as follow
\begin{eqnarray}
	 \delta{G}_i^j=0 &:& T'(F_{tt}-F_{rr})=2r\,T''\,H_L\ ,\label{ijap}\\
     \delta{G}_t^r=0 &:& -\ell(\ell+n-1)\dot{F}_{rt}\mathbb{S} +n\left(r(2H_L-F_{rr})\mathbb{S}+2r^2(H_L\,\mathbb{S})'- \frac{r^2f'}{f}H_L\mathbb{S}\right)^{\cdot}=0\ , \label{trap}\\
     \delta{G}_t^t=0&:& \left(-\ell(\ell+n-1)T- nr(f\,T)'\right)F_{rr}\mathbb{S}-nr\,f\,T\,(F_{rr}\mathbb{S})'+ 2r\,T'(n\kappa-\ell(\ell+n-1))H_L\,\mathbb{S}\nonumber\\
	                     &\ &\hspace{0.5cm}+\left(nT\,r^2f'+2nf(r^2T)'\right)(H_L\mathbb{S})'+2nT\,r^2f\,(H_L\mathbb{S})''=0\ ,\label{ttap}\\
     \delta{G}_r^i=0&:& rT^{'}\left(\frac{1}{r}F_{rr}\mathbb{S}-2(H_L\mathbb{S})'\right)\nonumber\\ &\ &\hspace{0.5cm}+T\left[\frac{f'}{2f}(F_{rr}\mathbb{S}+F_{tt}\mathbb{S})+ (F_{tt}\mathbb{S})'- \frac{1}{r}(F_{tt}\mathbb{S})-\frac{1}{f}\partial_t(\dot{F_{rt}}\mathbb{S})\right]=0\ , \label{riap}\\
     \delta{G}_r^r=0&:& \frac{nTr^2}{f}\left(\frac{2f}{r}\partial_t(\dot{F_{rt}}\mathbb{S})-2\partial^2_t(H_L\mathbb{S})\right)+ 2rT'(n\kappa-\ell(\ell+n-1))H_L\mathbb{S}-nr(Tf)'F_{rr}\mathbb{S}+ \nonumber\\
     &&\left(2nT'r^2f+nTr^2f'\right)(H_L\mathbb{S})' +\ell(\ell+n-1)T\,F_{tt}\mathbb{S}-nrf\,T(F_{tt}\mathbb{S})'=0\ . \label{rrap}
\end{eqnarray}

Now, we are in position to derive the master equation for the scalar-type. After defining the new function $\Psi(t,r)=\phi(t,r)\,R(r)/A(r)$, where
\begin{eqnarray}
  \phi(t,r) &=& \left(\frac{n}{\ell(\ell+n-1)}(H_L(2f-rf')-F_{rr})-2H_L\right)\mathbb{S}+2rf\left(H_L\mathbb{S}\right)' \\
  A(r) &=& -2nf+2\gamma_s+nrf^{'}\ ,
\end{eqnarray}
the wave-like equation for the dynamical scalar-type perturbations, after long calculations, can be expressed as
\begin{eqnarray}
	&&\left(\frac{\partial^2}{\partial\,t^2}-\frac{\partial^2}{\partial\,r_*^2}+V_s(r(r_*))\right)\Psi(t,r(r_*))=0, \label{eq:potentialap3} \ .\\
    &&V_s(r)=\frac{2\ell(\ell+n-1)}{nr^2\,B(r)}\frac{d}{dr_*}\left(r\,B(r)\right)+ B(r)\frac{d^2}{dr_*^2}\left(\frac{1}{B(r)}\right),
\end{eqnarray}
where
\begin{equation}\label{BHP:21ap}
  B(r)=\frac{2(\ell-1)(\ell+n)-nr^3\psi^\prime}{R(r)}T(r).
\end{equation}

\chapter{Solution of the Einstein-dilaton-Gauss-Bonnet wormhole}
\section{Basic formulas}
The wormhole metric in Einstein-dilaton-Gauss-Bonnet theory has a spherically symmetric solution which can be expressed as a diagonal matrix, as follow
\begin{equation}\label{ETC-11ap}
 g_{\mu\nu}= -e^{2\nu}dt^2+f(l)dl^2+\left(l^2+r_0^2\right)\left(d\theta^2+\sin^2\theta d\varphi^2\right),
\end{equation}
where $\nu=\nu(l)$, $f=f(l)$ and $r_0$ corresponds to the throat of the wormhole. The variable $t$ corresponds to the time coordinate and $l$ is a radial corrdinate which goes from $-\infty$ to $+\infty$. In this wormhole configuration $l=0$ is located at the throat. Furthermore, the non-vanishing Christoffel symbols could be determined,
\begin{equation}\label{ETC-2ap}
  \begin{array}{cccccc}
  \Gamma^l_{tt}&=&\frac{e^{2\nu}\nu'}{f}\qquad\qquad\qquad  & \Gamma^t_{tl}&=&\Gamma^t_{lt}=\nu', \\
  \Gamma^l_{ll}&=&\frac{f'}{2f}\qquad\qquad\qquad  & \Gamma^\theta_{l\theta}&=&\Gamma^\theta_{\theta l}=\frac{r'}{r}, \\
  \Gamma^l_{\theta\theta}&=&-\frac{r\,r'}{f}\qquad\qquad\qquad  & \Gamma^\phi_{l\phi}&=&\Gamma^\phi_{\phi l}=\frac{r'}{r}, \\
  \Gamma^l_{\phi\phi}&=&-\frac{r\,r'\sin^2\theta}{f}\qquad\qquad\qquad  & \Gamma^\phi_{\theta\phi}&=&\Gamma^\phi_{\phi\theta}=\cot\,\theta, \\
  \Gamma^\theta_{\phi\phi}&=& -\sin\theta\,\cos\theta.\qquad\qquad\qquad & &
  \end{array}
\end{equation}

Using the last expression, the non-zero Riemann components could be obtained, which are given by
\begin{equation}\label{ETC-3ap}
  \begin{array}{cccccccccc}
    R^t_{\phantom{t}ltl}&=&-R^t_{\phantom{t}llt}&=&\frac{f'\nu'}{2f}-\nu'^2-\nu'',\qquad\qquad\qquad & R^\theta_{\phantom{\theta}ll\theta}&=&-R^\theta_{\phantom{\theta}l\theta l}&=&\frac{-f'\,r'+2f\,r''}{2f\,r}, \\
    R^t_{\phantom{t}\theta t\theta}&=&-R^t_{\phantom{t}\theta\theta t}&=&-\frac{r\,r'\nu'}{f}\qquad\qquad\qquad & R^\theta_{\phantom{\theta}
    \phi\theta\phi}&=&-R^\theta_{\phantom{\theta}\phi\phi\theta}&=&\frac{\left(f-r'^2\right)\sin^2\theta}{f}, \\
    R^t_{\phantom{t}\phi t\phi}&=&-R^t_{\phantom{t}\phi\phi t}&=&-\frac{r\,r'\nu'\sin^2\theta}{f}\qquad\qquad\qquad & R^\phi_{\phantom{\phi}tt\phi}&=&-R^\phi_{\phantom{\phi}t\phi t}&=&\frac{e^{2\nu}\,r'\nu'}{f\,r}, \\
    R^l_{\phantom{l}ttl}&=&-R^l_{\phantom{l}tlt}&=&\frac{e^{2\nu}\left(f'\nu'-2f\left(\nu'^2+\nu''\right)\right)}{2f^2}\qquad\qquad\qquad & R^\phi_{\phantom{\phi}ll\phi}&=&-R^\phi_{\phantom{\phi}l\phi l}&=&\frac{-f'\,r'+2f\,r''}{2f\,r}, \\
    R^l_{\phantom{l}\theta l\theta}&=&-R^l_{\phantom{l}\theta\theta l}&=&\frac{r\left(f'\,r'-2f\,r''\right)}{2f^2}\qquad\qquad\qquad & R^\phi_{\phantom{\phi}\theta\theta\phi}&=&-R^\phi_{\phantom{\phi}\theta\phi\theta}&=&-1+\frac{r'^2}{f}, \\
    R^l_{\phantom{l}\phi l\phi}&=&-R^l_{\phantom{l}\phi\phi l}&=&\frac{r\sin^2\,\theta \left(f'\,r'-2f\,r''\right)}{2f^2}\qquad\qquad\qquad & R^\theta_{\phantom{\theta}tt\theta}&=&-R^\theta_{\phantom{\theta}t\theta t}&=&-\frac{e^{2\nu}r'\nu'}{f\,r},
  \end{array}
\end{equation}

The Ricci components will also be presented in terms of the wormholes metric functions,
\begin{equation}
  \begin{array}{ccl}
    R_{tt}&=&\frac{e^{2\nu}\left(4f\,r'\nu'+r\left(-f'\,\nu'+2f\left(\nu'^2+\nu''\right)\right)\right)}{2f^2\,r} \\
    R_{ll}&=&\frac{f'\left(2r'+r\nu'\right)-2f\left(2r''+r\left(\nu'^2+\nu''\right)\right)}{2f\,r} \\
    R_{\theta\theta}&=&\frac{2f^2+r\,f'\,r'-2f\left(r'^2+r\,r'\nu'+r\,r''\right)}{2f^2} \\
    R_{\phi\phi}&=&\frac{\left(2f^2+r\,f'\,r'-2f\left(r'^2+r\,r'\nu'+r\,r''\right)\right)\sin^2\theta}{2f^2}
  \end{array}
\end{equation}

Moreover, from the Lovelock construction, the terms $\mathcal{L}_1$ and $\mathcal{L}_2$ correspond to the Ricci scalar and the Gauss-Bonnet term respectively. Both scalar quantities in terms of the metric functions are given by
\begin{eqnarray}
  R&=&\frac{2f^2+r\,f'\left(2r'+r\,\nu'\right)-2f\left(r'^2+2r\,r'\,\nu'+r\left(2r''+r\left(\nu'^2+\nu''\right)\right)\right)}{f^2\,r^2} \\
  \mathcal{L}_2&=&\frac{4\nu'\left(-3f'\,r'^2-2f^2\nu'+f\left(f'+2r'\left(r'\nu'+2r''\right)\right)\right)+8f \left(-f+r'^2\right)\nu'')}{f^3\,r^2}
\end{eqnarray}

In the Einstein-dilaton-Gauss-Bonnet gravity, the equation of motion for the dilaton field and for the gravitational field are,
\begin{eqnarray}
\nabla^2 \phi & = & \alpha \gamma  e^{-\gamma \phi}\mathcal{L}_2,	
\label{eqsap}\\
G_{\mu\nu} & = &
\frac{1}{2}\left[\nabla_\mu \phi \nabla_\nu \phi
                 -\frac{1}{2}g_{\mu\nu}\nabla_\lambda \phi \nabla^\lambda\phi
		 \right]
\nonumber\\
& &
-\alpha e^{-\gamma \phi}
\left[	H_{\mu\nu}
  +4\left(\gamma^2\nabla^\rho \phi \nabla^\sigma \phi
           -\gamma \nabla^\rho\nabla^\sigma \phi\right)	P_{\mu\rho\nu\sigma}
		 \right],\label{eqs2ap}
\end{eqnarray}
where
\begin{eqnarray}\nonumber
H_{\mu\nu} & = & 2\left[R R_{\mu\nu} -2 R_{\mu\rho}R^\rho_\nu
                        -2 R_{\mu\rho\nu\sigma}R^{\rho\sigma}
			+R_{\mu\rho\sigma\lambda}R_\nu^{\ \rho\sigma\lambda}
		   \right]
		   -\frac{1}{2}g_{\mu\nu}\mathcal{L}_2	\ ,
\\\nonumber
P_{\mu\nu\rho\sigma} & = & R_{\mu\nu\rho\sigma}
+2 g_{\mu\sigma} R_{\rho\nu}-2 g_{\mu\rho} R_{\sigma\nu}
+2 g_{\nu\rho} R_{\sigma\mu}-2 g_{\nu\sigma} R_{\rho\mu}
+R g_{\mu\rho} g_{\sigma\nu}-R g_{\mu\sigma} g_{\rho\nu} \ .
\label{HPap}
\end{eqnarray}

The Einstein equation $G_{\mu\nu}=R_{\mu\nu}-\frac{1}{2}g_{\mu\nu}R=T_{\mu\nu}$ is modified since the tensor $T_{\mu\nu}$ has not only matter fields but also coupled geometry elements due to the dilaton. In order to obtain both sides of the equation, the corresponding left-hand-side non-vanishing elements of the modified Einstein equation are
\begin{equation}\label{ETC-6ap}
  \begin{array}{ccl}
    G_{tt}&=&\frac{e^{2\nu}\left(f^2+r\,f'\,r'-f\left(r'^2+2r\,r''\right)\right)}{f^2\,r^2}, \\
    G_{ll}&=&\frac{-f+r'\left(r'+2r\nu'\right)}{r^2}, \\
    G_{\theta\theta}&=&\frac{r\left(-f'\left(r'+r\,\nu'\right)+2f\left(r'\,\nu'+r''+r\left(\nu'^2+\nu''\right)\right)\right)}{2f^2}, \\
    G_{\phi\phi}&=&\frac{r\left(-f'\left(r'+r\nu'\right)+2f\left(r'\,\nu'+r''+r\left(\nu'^2+\nu''\right)\right)\right)\sin^2\theta}{2f^2},
  \end{array}
\end{equation}
and the corresponding right-hand-side are
\begin{equation}\label{ETC-8ap}
  \begin{array}{ccl}
    T_{tt}&=&\frac{e^{2\nu}\phi'^2}{4f}+2\alpha\gamma\frac{e^{2\nu-\gamma\phi}}{f^3\,r^2}\left(-3f'\,r'^2\phi'+2f^2\left(\gamma\phi'^2 -\phi''\right)+\right.\\
    &&\left.f\left(f'\phi'+2r'\left(2\phi'\,r''+r'\left(-\gamma\phi'^2+\phi''\right)\right)\right)\right), \\
    T_{ll}&=&\frac{4e^{-\gamma\phi}\alpha\gamma\left(f-3r'^2\right)\nu'\phi'}{f\,r^2}+\frac{1}{4}\phi'^2, \\
    T_{\theta\theta}&=&-\frac{r^2\phi'^2}{4f}+2\alpha\gamma\frac{r\,e^{-\gamma\phi}}{f^3}\left(3f'\,r'\nu'\phi'-2f\left(\nu'\phi'\,r''+\right.\right.\\ &&\left.\left.r'\left(\nu'^2\phi'+\phi'\nu''+\nu'\left(-\gamma\phi'^2+\phi''\right)\right)\right)\right), \\
    T_{\phi\phi}&=&-\frac{r^2\sin^2\theta\,\phi'^2}{4f}+2\alpha\gamma\frac{r\sin^2\theta\,e^{-\gamma\phi}}{f^3}\left(3f'\,r'\nu'\phi'- 2f\left(\nu'\phi'\,r''+\right.\right.\\
    &&\left.\left.r'\left(\nu'^2\phi'+\phi'\nu''+\nu'\left(-\gamma\phi'^2+\phi''\right)\right)\right)\right).
  \end{array}
\end{equation}
Then, after some calculations, the equation of motion for the dilaton and gravitational field are given by
\begin{eqnarray}
f' + \frac{f (r^2 f + l^2 - 2 r^2)}{l r^2}
& = &
\frac{r^2 f \phi'^2}{4 l}+2\alpha \gamma\frac{e^{-\gamma\phi}}{l r^2} \bigg\{
  2 (r^2 f - l^2)(\gamma\phi'^2-\phi'')\nonumber\\
  &&
 \left.+ \phi'\left[\frac{f'}{f}\,(r^2 f-3 l^2) +\frac{4 l r_0^2}{r^2}\right]\right\},
 \label{ode11ap} \\[1mm] && \nonumber\\
\nu' - \frac{r^2 f -l^2}{2 l r^2}
& = &
\frac{\phi'^2 r^2}{8 l}
+2\alpha \gamma\frac{e^{-\gamma\phi}}{l r^2 f}
\,\nu'\phi' (r^2 f- 3 l^2)\,,\label{ode21ap}  \\[1mm] && \nonumber
\\ \label{ode31ap}
 \nu''+\nu'^2 + \frac{\nu' (2 l f - r^2 f')}{2 r^2 f}
 & + &
  \frac{ 2 r_0^2 f - l  r^2 f'}{2 r^4 f} + \frac{\phi'^2}{4} = 2\alpha \gamma\frac{e^{-\gamma\phi}}{r^2 f}\times \\ \nonumber
&&
\left\{2 l \left[\nu' (\gamma\phi'^2-\phi'') -\phi' (\nu'^2+\nu'')\right]
+ \nu' \phi' \left(\frac{3lf'}{f}-\frac{2r_0^2}{r^2}\right)\right\},
\\ \label{ode41ap}
\phi'' + \nu'\phi' + \frac{\phi' (4 l f - r^2 f')}{2 r^2 f}
& = &
 4\alpha \gamma\frac{e^{-\gamma\phi}}{r^4 f} \times \\ \nonumber
&&
\left\{
-2 (r^2 f - l^2)(\nu'^2+\nu'')
+\nu'\left[\frac{f'}{f}\,(r^2 f-3 l^2) +\frac{4 l r_0^2}{r^2}\right]\right\}.
\end{eqnarray}
We solve these equations numerically in order to obtain the background solution of the Kanti-Kleihaus-Kunz wormhole.

\section{Wormhole perturbation}
Lets consider a spherically symmetric linear perturbation for the lowest mode $\ell=0$. After perturbing the metric parameters $\delta f$, $\delta\nu$ and the dilaton $\delta\phi$, as follow
\begin{eqnarray}\nonumber
\phi\to\phi(l)+\delta\phi(t,l),\\\label{perturbations1ap}
f\to f(l)+\delta f(t,l),\\
\nu\to\nu(l)+\delta\nu(t,l).\nonumber
\end{eqnarray}

Replacing the last equations (\ref{perturbations1ap}) into the equations of motion (\ref{ode11ap})-(\ref{ode41ap}) and then substracting to the unperturbed equations, it turns out a set of equations that govern the dynamical motion of the perturbation. From the dilaton equation (\ref{ode11ap}) is obtained

\begin{eqnarray}\label{sEDGB-Dap}
&&\delta \phi' \left(f^2 r^2 f' e^{\gamma  \phi+2 \nu}-4 f^3 r r' e^{\gamma  \phi+2 \nu}-2 f^3 r^2 \nu' e^{\gamma  \phi+2 \nu}\right)+\delta \phi  \left(24 \alpha  \gamma ^2 f e^{2 \nu} f' r'^2 \nu'-8 \alpha  \gamma ^2 f^2 e^{2 \nu} \left(f' \nu'\right.\right.\nonumber\\
&&\left.\left.+2 r' \left(2 r'' \nu'+r' \left(\nu''+\nu'^2\right)\right)\right)+16 \alpha  \gamma ^2 f^3 e^{2 \nu} \left(\nu''+\nu'^2\right)\right)+\delta f \left(72 \alpha  \gamma  e^{2 \nu} f' r'^2 \nu'\right.\nonumber\\
&&\left.-2 f e^{2 \nu} \left(f' \left(r^2 e^{\gamma  \phi} \phi'+8 \alpha  \gamma  \nu'\right)+16 \alpha  \gamma  r' \left(2 r'' \nu'+r' \left(\nu''+\nu'^2\right)\right)\right)+2 f^2 e^{2 \nu} \left(2 r r' e^{\gamma  \phi} \phi'\right.\right.\nonumber\\
&&\left.\left.+r^2 e^{\gamma  \phi} \nu' \phi'+r^2 e^{\gamma  \phi} \phi''+8 \alpha  \gamma  \nu''+8 \alpha  \gamma  \nu'^2\right)\right)+\delta \nu ' \left(-24 \alpha  \gamma  f e^{2 \nu} f' r'^2+8 \alpha  \gamma  f^2 e^{2 \nu} f'\right.\nonumber\\
&&\left.+32 \alpha  \gamma  f^2 e^{2 \nu} r'^2 \nu'+32 \alpha  \gamma  f^2 e^{2 \nu} r' r''-2 f^3 r^2 \phi' e^{\gamma  \phi+2 \nu}-32 \alpha  \gamma  f^3 e^{2 \nu} \nu'\right)+\delta f' \left(-24 \alpha  \gamma  f e^{2 \nu} r'^2 \nu'+\right.\nonumber\\
&&\left.f^2 r^2 \phi' e^{\gamma  \phi+2 \nu}+8 \alpha  \gamma  f^2 e^{2 \nu} \nu'\right)+\delta \ddot{f} \left(8 \alpha  \gamma  f^3-8 \alpha  \gamma  f^2 r'^2\right)+\delta \nu '' \left(16 \alpha  \gamma  f^2 e^{2 \nu} r'^2-16 \alpha  \gamma  f^3 e^{2 \nu}\right)\nonumber\\
&&-2 f^3 r^2 \delta \phi '' e^{\gamma  \phi+2 \nu}+2 f^4 r^2 e^{\gamma  \phi} \delta \ddot{\phi}=0
\end{eqnarray}

The following differential equations corresponds to the $\delta G_{00}$, $\delta G_{11}$ and $\delta G_{22}$ respectively,
\begin{eqnarray}
&&\delta G_{00}:\nonumber\\
&&\delta \nu \left(48 \alpha  \gamma  f f' r'^2 \phi'+8 f^2 r f' r' e^{\gamma  \phi}-16 \alpha  \gamma  f^2 \left(f' \phi'+2 r' \left(2 r'' \phi'+r' \left(\phi''-\gamma  \phi'^2\right)\right)\right)\right.\nonumber\\
&&\left.-2 f^3 \left(8 \left(r r'' e^{\gamma  \phi}-2 \alpha  \gamma  \phi''\right)+4 r'^2 e^{\gamma  \phi}+\phi'^2 \left(16 \alpha  \gamma ^2+r^2 e^{\gamma  \phi}\right)\right)+8 f^4 e^{\gamma  \phi}\right)\nonumber\\
&&+\delta \phi ' \left(24 \alpha  \gamma  f f' r'^2-8 \alpha  \gamma  f^2 f'+32 \alpha  \gamma ^2 f^2 r'^2 \phi'-32 \alpha  \gamma  f^2 r' r''-2 f^3 r^2 e^{\gamma  \phi} \phi'-32 \alpha  \gamma ^2 f^3 \phi'\right)\nonumber\\
&&+\delta \phi  \left(-24 \alpha  \gamma ^2 f f' r'^2 \phi'-8 \alpha  \gamma ^2 f^2 \left(2 r' \left(r' \left(\gamma  \phi'^2-\phi''\right)-2 r'' \phi'\right)-f' \phi'\right)-16 \alpha  \gamma ^2 f^3 \left(\phi''-\gamma  \phi'^2\right)\right)\nonumber\\
&&+\delta f \left(-72 \alpha  \gamma  f' r'^2 \phi'-8 f r f' r' e^{\gamma  \phi}+16 \alpha  \gamma  f \left(f' \phi'+2 r' \left(2 r'' \phi'+r' \left(\phi''-\gamma  \phi'^2\right)\right)\right)+8 f^2 r r'' e^{\gamma  \phi}\right.\nonumber\\
&&\left.+4 f^2 \left(r'^2 e^{\gamma  \phi}+4 \alpha  \gamma  \left(\gamma  \phi'^2-\phi''\right)\right)+f^2 r^2 e^{\gamma  \phi} \phi'^2\right)+\delta f' \left(24 \alpha  \gamma  f r'^2 \phi'+4 f^2 r r' e^{\gamma  \phi}-8 \alpha  \gamma  f^2 \phi'\right)\nonumber\\
&&-16 \alpha  \gamma  f^2 r'^2 \delta \phi ''+16 \alpha  \gamma  f^3 \delta \phi ''=0
\end{eqnarray}

\begin{eqnarray}
&&\delta G_{11}:\nonumber\\
&&\delta \phi  \left(24 \alpha  \gamma ^2 f e^{2 \nu} r'^2 \nu' \phi'-8 \alpha  \gamma ^2 f^2 e^{2 \nu} \nu' \phi'\right)+\delta f \left(2 f^2 e^{\gamma  \phi+2 \nu}+24 \alpha  \gamma  e^{2 \nu} r'^2 \nu' \phi'\right)\nonumber\\
&&+\delta \nu ' \left(-24 \alpha  \gamma  f e^{2 \nu} r'^2 \phi'-4 f^2 r r' e^{\gamma  \phi+2 \nu}+8 \alpha  \gamma  f^2 e^{2 \nu} \phi'\right)+\delta \phi ' \left(-24 \alpha  \gamma  f e^{2 \nu} r'^2 \nu'+f^2 r^2 \phi' e^{\gamma  \phi+2 \nu}\right.\nonumber\\
&&\left.+8 \alpha  \gamma  f^2 e^{2 \nu} \nu'\right)+\delta \ddot{\phi} \left(8 \alpha  \gamma  f^2 r'^2-8 \alpha  \gamma  f^3\right)=0
\end{eqnarray}

\begin{eqnarray}
&&\delta G_{22}:\nonumber\\
&&16 e^{2 \nu} \alpha  \gamma  r' \nu' \delta \phi '' f^2+\delta \phi  \left(24 e^{2 \nu} \alpha  f f' r' \nu' \phi' \gamma ^2+16 e^{2 \nu} \alpha  f^2 \left(-\nu' \phi' r''-r' \left(\phi' \nu'^2+\left(\phi''-\gamma  \phi'^2\right) \nu'\right.\right.\right.\nonumber\\
&&\left.\left.\left.+\phi' \nu''\right)\right) \gamma ^2\right)+\delta f \left(-e^{2 \nu+\gamma  \phi} \left(4 r' \nu'+4 r''+r \left(4 \nu'^2+\phi'^2+4 \nu''\right)\right) f^2+4 e^{2 \nu} \left(e^{\gamma  \phi} f' \left(r'+r \nu'\right)\right.\right.\nonumber\\
&&\left.\left.+8 \alpha  \gamma  \left(-\nu' \phi' r''-r' \left(\phi' \nu'^2+\left(\phi''-\gamma  \phi'^2\right) \nu'+\phi' \nu''\right)\right)\right) f+72 e^{2 \nu} \alpha  \gamma  f' r' \nu' \phi'\right)+\left(-2 e^{2 \nu+\gamma  \phi} r' f^2\right.\nonumber\\
&&\left.-2 e^{2 \nu+\gamma  \phi} r \nu' f^2-24 e^{2 \nu} \alpha  \gamma  r' \nu' \phi' f\right) \delta f'+\left(4 e^{2 \nu+\gamma  \phi} r' f^3+8 e^{2 \nu+\gamma  \phi} r \nu' f^3-16 e^{2 \nu} \alpha  \gamma ^2 r' \phi'^2 f^2\right.\nonumber\\
&&\left.-2 e^{2 \nu+\gamma  \phi} r f' f^2+32 e^{2 \nu} \alpha  \gamma  r' \nu' \phi' f^2+16 e^{2 \nu} \alpha  \gamma  \phi' r'' f^2+16 e^{2 \nu} \alpha  \gamma  r' \phi'' f^2-24 e^{2 \nu} \alpha  \gamma  f' r' \phi' f\right) \delta \nu '\nonumber\\
&&+\left(2 e^{2 \nu+\gamma  \phi} r \phi' f^3+16 e^{2 \nu} \alpha  \gamma  r' \nu'^2 f^2-32 e^{2 \nu} \alpha  \gamma ^2 r' \nu' \phi' f^2+16 e^{2 \nu} \alpha  \gamma  \nu' r'' f^2+16 e^{2 \nu} \alpha  \gamma  r' \nu'' f^2\right.\nonumber\\
&&\left.-24 e^{2 \nu} \alpha  \gamma  f' r' \nu' f\right) \delta \phi '+\left(4 e^{2 \nu+\gamma  \phi} r f^3+16 e^{2 \nu} \alpha  \gamma  r' \phi' f^2\right) \delta \nu ''+\left(-2 e^{\gamma  \phi} r f^3-8 \alpha  \gamma  r' \phi' f^2\right) \delta \ddot{f}\nonumber\\
&&+\left(8 \alpha  \gamma  f^2 f' r'-16 \alpha  \gamma  f^3 r''\right) \delta \ddot{\phi}=0
\end{eqnarray}

after long calculation, these set of equations can be expressed as a dynamical perturbation for the dilaton $\delta\phi$ (\ref{lineq}), where $q_1(l)$, $q_0(l)$ and $q_\sigma(l)$ have a long expression in terms the background solution of $f(l)$, $\nu(l)$ and $\phi(l)$. Once we solved the equation for the background functions, we are able to compute numerically the effective potential (\ref{potential}) for any value of the tortoise coordinate.

\end{appendices}

\addcontentsline{toc}{section}{References}

\begin{thebibliography}{100}

\bibitem{Bizon:2011gg}
Piotr Bizon and Andrzej Rostworowski.
\newblock {On weakly turbulent instability of anti-de Sitter space}.
\newblock {\em Phys. Rev. Lett.}, {\bf 107}:031102, 2011.

\bibitem{PhysRevLett.114.071102}
Nils Deppe, Allison Kolly, Andrew Frey, and Gabor Kunstatter.
\newblock Stability of anti--de sitter space in einstein-gauss-bonnet gravity.
\newblock {\em Phys. Rev. Lett.}, 114:071102, Feb 2015.

\bibitem{Dotti:2004sh}
Gustavo Dotti and Reinaldo~J. Gleiser.
\newblock {Gravitational instability of Einstein-Gauss-Bonnet black holes under
  tensor mode perturbations}.
\newblock {\em Class. Quant. Grav.}, {\bf 22}:L1, 2005.

\bibitem{Gleiser:2005ra}
Reinaldo~J. Gleiser and Gustavo Dotti.
\newblock {Linear stability of Einstein-Gauss-Bonnet static spacetimes. Part
  II: Vector and scalar perturbations}.
\newblock {\em Phys. Rev.}, {\bf D72}:124002, 2005.

\bibitem{Konoplya:2008ix}
R.~A. Konoplya and A.~Zhidenko.
\newblock {(In)stability of D-dimensional black holes in Gauss-Bonnet theory}.
\newblock {\em Phys. Rev.}, D77:104004, 2008.

\bibitem{Konoplya:2010vz}
R.~A. Konoplya and A.~Zhidenko.
\newblock {Long life of Gauss-Bonnet corrected black holes}.
\newblock {\em Phys. Rev.}, D82:084003, 2010.

\bibitem{Konoplya:2011qq}
R.~A. Konoplya and A.~Zhidenko.
\newblock {Quasinormal modes of black holes: From astrophysics to string
  theory}.
\newblock {\em Rev. Mod. Phys.}, 83:793--836, 2011.

\bibitem{Ishibashi:2011ws}
Akihiro Ishibashi and Hideo Kodama.
\newblock {Perturbations and Stability of Static Black Holes in Higher
  Dimensions}.
\newblock {\em Prog. Theor. Phys. Suppl.}, 189:165--209, 2011.

\bibitem{Kokkotas:1999bd}
Kostas~D. Kokkotas and Bernd~G. Schmidt.
\newblock {Quasinormal modes of stars and black holes}.
\newblock {\em Living Rev. Rel.}, 2:2, 1999.

\bibitem{Abbott:2016blz}
B.~P. Abbott et~al.
\newblock {Observation of Gravitational Waves from a Binary Black Hole Merger}.
\newblock {\em Phys. Rev. Lett.}, 116(6):061102, 2016.

\bibitem{TheLIGOScientific:2016src}
B.~P. Abbott et~al.
\newblock {Tests of general relativity with GW150914}.
\newblock {\em Phys. Rev. Lett.}, 116(22):221101, 2016.
\newblock [Erratum: Phys. Rev. Lett.121,no.12,129902(2018)].

\bibitem{Konoplya:2016pmh}
Roman Konoplya and Alexander Zhidenko.
\newblock {Detection of gravitational waves from black holes: Is there a window
  for alternative theories?}
\newblock {\em Phys. Lett.}, B756:350--353, 2016.

\bibitem{Hod:2006jw}
Shahar Hod.
\newblock {Universal Bound on Dynamical Relaxation Times and Black-Hole
  Quasinormal Ringing}.
\newblock {\em Phys. Rev.}, D75:064013, 2007.

\bibitem{Kanti:2011jz}
Panagiota Kanti, Burkhard Kleihaus, and Jutta Kunz.
\newblock {Wormholes in Dilatonic Einstein-Gauss-Bonnet Theory}.
\newblock {\em Phys. Rev. Lett.}, 107:271101, 2011.

\bibitem{Lobo:2005yv}
Francisco S.~N. Lobo.
\newblock {Stability of phantom wormholes}.
\newblock {\em Phys. Rev.}, D71:124022, 2005.

\bibitem{Eiroa:2003wp}
Ernesto~F. Eiroa and Gustavo~E. Romero.
\newblock {Linearized stability of charged thin shell wormholes}.
\newblock {\em Gen. Rel. Grav.}, 36:651--659, 2004.

\bibitem{Dzhunushaliev:2017syc}
Vladimir Dzhunushaliev, Vladimir Folomeev, Burkhard Kleihaus, and Jutta Kunz.
\newblock {Wormhole solutions with a complex ghost scalar field and their
  instability}.
\newblock {\em Phys. Rev.}, D97(2):024002, 2018.

\bibitem{Bronnikov:2013coa}
K.~A. Bronnikov, L.~N. Lipatova, I.~D. Novikov, and A.~A. Shatskiy.
\newblock {Example of a stable wormhole in general relativity}.
\newblock {\em Grav. Cosmol.}, 19:269--274, 2013.

\bibitem{Konoplya:2016hmd}
R.~A. Konoplya and A.~Zhidenko.
\newblock {Wormholes versus black holes: quasinormal ringing at early and late
  times}.
\newblock {\em JCAP}, 1612(12):043, 2016.

\bibitem{Gonzalez:2008wd}
J.~A. Gonzalez, F.~S. Guzman, and O.~Sarbach.
\newblock {Instability of wormholes supported by a ghost scalar field. I.
  Linear stability analysis}.
\newblock {\em Class. Quant. Grav.}, 26:015010, 2009.

\bibitem{Bronnikov:2012ch}
K.~A. Bronnikov, R.~A. Konoplya, and A.~Zhidenko.
\newblock {Instabilities of wormholes and regular black holes supported by a
  phantom scalar field}.
\newblock {\em Phys. Rev.}, D86:024028, 2012.

\bibitem{Kodama:2003kk}
Hideo Kodama and Akihiro Ishibashi.
\newblock {Master equations for perturbations of generalized static black holes
  with charge in higher dimensions}.
\newblock {\em Prog. Theor. Phys.}, 111:29--73, 2004.

\bibitem{Gundlach:1993tp}
Carsten Gundlach, Richard~H. Price, and Jorge Pullin.
\newblock {Late time behavior of stellar collapse and explosions: 1. Linearized
  perturbations}.
\newblock {\em Phys. Rev.}, D49:883--889, 1994.

\bibitem{Cuyubamba:2016cug}
M.~A. Cuyubamba, R.~A. Konoplya, and A.~Zhidenko.
\newblock {Quasinormal modes and a new instability of Einstein-Gauss-Bonnet
  black holes in the de Sitter world}.
\newblock {\em Phys. Rev.}, D93(10):104053, 2016.

\bibitem{Cuyubamba:2018jdl}
M.~A. Cuyubamba, R.~A. Konoplya, and A.~Zhidenko.
\newblock {No stable wormholes in Einstein-dilaton-Gauss-Bonnet theory}.
\newblock {\em Phys. Rev.}, D98(4):044040, 2018.

\bibitem{EinsteinGR}
A~Einstein.
\newblock Die feldgleichungen der gravitation.
\newblock {\em Sitzungsber. Preuss. Akad. Wiss.}, 6:88 -- 92, 08 2006.

\bibitem{Vermeil:1917au}
Vermeil H.
\newblock Notiz über das mittlere krümmungsmaf einer n-fach ausgedehnten
  riemannschen mannigfaltigkeit.
\newblock {\em Nachr. Ges. Wiss. Göttingen, Math. Phys. Klasse}, 334-344,
  1917.

\bibitem{Weisberg:1981mt}
J.~M. Weisberg, J.~H. Taylor, and L.~A. Fowler.
\newblock {GRAVITATIONAL WAVES FROM AN ORBITING PULSAR}.
\newblock {\em Sci. Am.}, 245:66--74, 1981.

\bibitem{Akiyama:2019cqa}
Kazunori Akiyama et~al.
\newblock {First M87 Event Horizon Telescope Results. I. The Shadow of the
  Supermassive Black Hole}.
\newblock {\em Astrophys. J.}, 875(1):L1, 2019.

\bibitem{Schwarzschild:1916uq}
Karl Schwarzschild.
\newblock {On the gravitational field of a mass point according to Einstein's
  theory}.
\newblock {\em Sitzungsber. Preuss. Akad. Wiss. Berlin (Math. Phys.)},
  1916:189--196, 1916.

\bibitem{PhysRevLett.14.57}
Roger Penrose.
\newblock Gravitational collapse and space-time singularities.
\newblock {\em Phys. Rev. Lett.}, 14:57--59, Jan 1965.

\bibitem{ReissnerQ}
H.~Reissner.
\newblock Über die eigengravitation des elektrischen feldes nach der
  einsteinschen theorie.
\newblock {\em Annalen der Physik}, 355(9):106--120, 1916.

\bibitem{NordstromQ}
G.~{Nordström}.
\newblock {On the Energy of the Gravitation field in Einstein's Theory}.
\newblock {\em Koninklijke Nederlandse Akademie van Wetenschappen Proceedings
  Series B Physical Sciences}, 20:1238--1245, 1918.

\bibitem{PhysRevLett.11.237}
Roy~P. Kerr.
\newblock Gravitational field of a spinning mass as an example of algebraically
  special metrics.
\newblock {\em Phys. Rev. Lett.}, 11:237--238, Sep 1963.

\bibitem{BoyerLinquist}
Robert~H. Boyer and Richard~W. Lindquist.
\newblock Maximal analytic extension of the kerr metric.
\newblock {\em Journal of Mathematical Physics}, 8(2):265--281, 1967.

\bibitem{NewmanQJ}
E.~T. Newman, E.~Couch, K.~Chinnapared, A.~Exton, A.~Prakash, and R.~Torrence.
\newblock Metric of a rotating, charged mass.
\newblock {\em Journal of Mathematical Physics}, 6(6):918--919, 1965.

\bibitem{Zwiebach:1985uq}
Barton Zwiebach.
\newblock {Curvature Squared Terms and String Theories}.
\newblock {\em Phys. Lett.}, 156B:315--317, 1985.

\bibitem{Lovelock:1971}
D.~Lovelock.
\newblock {The Einstein Tensor and Its Generalizations}.
\newblock {\em J. Math. Phys.}, 12:498, 1971.

\bibitem{PhysRevLett.25.1596}
Demetrios Christodoulou.
\newblock Reversible and irreversible transformations in black-hole physics.
\newblock {\em Phys. Rev. Lett.}, 25:1596--1597, Nov 1970.

\bibitem{PhysRevLett.26.1344}
S.~W. Hawking.
\newblock Gravitational radiation from colliding black holes.
\newblock {\em Phys. Rev. Lett.}, 26:1344--1346, May 1971.

\bibitem{PhysRevD.7.2333}
Jacob~D. Bekenstein.
\newblock Black holes and entropy.
\newblock {\em Phys. Rev. D}, 7:2333--2346, Apr 1973.

\bibitem{Bardeen1973}
J.~M. Bardeen, B.~Carter, and S.~W. Hawking.
\newblock The four laws of black hole mechanics.
\newblock {\em Communications in Mathematical Physics}, 31(2):161--170, Jun
  1973.

\bibitem{Sarkar:2012wy}
Sudipta Sarkar and Swastik Bhattacharya.
\newblock {Issue of zeroth law for Killing horizons in Lanczos-Lovelock
  gravity}.
\newblock {\em Phys. Rev.}, D87(4):044023, 2013.

\bibitem{PhysRevD.38.2434}
Robert~C. Myers and Jonathan~Z. Simon.
\newblock Black-hole thermodynamics in lovelock gravity.
\newblock {\em Phys. Rev. D}, 38:2434--2444, Oct 1988.

\bibitem{Jacobson:1993xs}
Ted Jacobson and Robert~C. Myers.
\newblock {Black hole entropy and higher curvature interactions}.
\newblock {\em Phys. Rev. Lett.}, 70:3684--3687, 1993.

\bibitem{Kastor:2010gq}
David Kastor, Sourya Ray, and Jennie Traschen.
\newblock {Smarr Formula and an Extended First Law for Lovelock Gravity}.
\newblock {\em Class. Quant. Grav.}, 27:235014, 2010.

\bibitem{Akbar:2008vz}
M.~Akbar.
\newblock {Generalized Second Law of Thermodynamics in Extended Theories of
  Gravity}.
\newblock {\em Int. J. Theor. Phys.}, 48:2665--2671, 2009.

\bibitem{Sarkar:2010xp}
Sudipta Sarkar and Aron~C. Wall.
\newblock {Second Law Violations in Lovelock Gravity for Black Hole Mergers}.
\newblock {\em Phys. Rev.}, D83:124048, 2011.

\bibitem{Kolekar:2012tq}
Sanved Kolekar, T.~Padmanabhan, and Sudipta Sarkar.
\newblock {Entropy Increase during Physical Processes for Black Holes in
  Lanczos-Lovelock Gravity}.
\newblock {\em Phys. Rev.}, D86:021501, 2012.

\bibitem{Torii_2006}
Takashi Torii.
\newblock Black holes in higher curvature theory and third law of
  thermodynamics.
\newblock {\em Journal of Physics: Conference Series}, 31:175--176, mar 2006.

\bibitem{PhysRevLett.116.241103}
B.~P. Abbott et~al.
\newblock Gw151226: Observation of gravitational waves from a 22-solar-mass
  binary black hole coalescence.
\newblock {\em Phys. Rev. Lett.}, 116:241103, Jun 2016.

\bibitem{LIGOScientific:2018mvr}
B.~P. Abbott et~al.
\newblock {GWTC-1: A Gravitational-Wave Transient Catalog of Compact Binary
  Mergers Observed by LIGO and Virgo during the First and Second Observing
  Runs}.
\newblock 2018.

\bibitem{Takahashi:2010ye}
Tomohiro Takahashi and Jiro Soda.
\newblock {Master Equations for Gravitational Perturbations of Static Lovelock
  Black Holes in Higher Dimensions}.
\newblock {\em Prog. Theor. Phys.}, 124:911--924, 2010.

\bibitem{Iyer_1989}
B~R Iyer, S~Iyer, and C~V Vishveshwara.
\newblock Scalar waves in the boulware-deser black-hole background.
\newblock {\em Classical and Quantum Gravity}, 6(11):1627--1639, nov 1989.

\bibitem{Konoplya:2004xx}
Roman Konoplya.
\newblock {Quasinormal modes of the charged black hole in Gauss-Bonnet
  gravity}.
\newblock {\em Phys. Rev.}, D71:024038, 2005.

\bibitem{Abdalla:2005hu}
E.~Abdalla, R.~A. Konoplya, and C.~Molina.
\newblock {Scalar field evolution in Gauss-Bonnet black holes}.
\newblock {\em Phys. Rev.}, D72:084006, 2005.

\bibitem{Zhidenko:2008fp}
Alexander Zhidenko.
\newblock {Quasinormal modes of brane-localized standard model fields in
  Gauss-Bonnet theory}.
\newblock {\em Phys. Rev.}, D78:024007, 2008.

\bibitem{Gonzalez:2017gwa}
P.~A. González, R.~A. Konoplya, and Yerko Vásquez.
\newblock {Quasinormal modes of a scalar field in the Einstein-Gauss-Bonnet-AdS
  black hole background: Perturbative and nonperturbative branches}.
\newblock {\em Phys. Rev.}, D95(12):124012, 2017.

\bibitem{Dotti:2005sq}
Gustavo Dotti and Reinaldo~J. Gleiser.
\newblock {Linear stability of Einstein-Gauss-Bonnet static spacetimes. Part I.
  Tensor perturbations}.
\newblock {\em Phys. Rev.}, D72:044018, 2005.

\bibitem{Konoplya:2017ymp}
R.~A. Konoplya and A.~Zhidenko.
\newblock {Eikonal instability of Gauss-Bonnet (anti)-de Sitter black holes}.
\newblock {\em Phys. Rev.}, D95(10):104005, 2017.

\bibitem{Takahashi:2010gz}
Tomohiro Takahashi and Jiro Soda.
\newblock {Catastrophic Instability of Small Lovelock Black Holes}.
\newblock {\em Prog. Theor. Phys.}, 124:711--729, 2010.

\bibitem{Yoshida:2015vua}
Daiske Yoshida and Jiro Soda.
\newblock {Quasinormal modes of black holes in Lovelock gravity}.
\newblock {\em Phys. Rev.}, D93(4):044024, 2016.

\bibitem{Takahashi:2011qda}
Tomohiro Takahashi.
\newblock {Fatal Effects of Charges on Stability of Black Holes in Lovelock
  Theory}.
\newblock {\em Prog. Theor. Phys.}, 125:1289--1310, 2011.

\bibitem{Takahashi:2012np}
Tomohiro Takahashi.
\newblock {Instability of Charged Lovelock Black Holes: Vector Perturbations
  and Scalar Perturbations}.
\newblock {\em PTEP}, 2013:013E02, 2013.

\bibitem{Gannouji:2013eka}
Radouane Gannouji and Naresh Dadhich.
\newblock {Stability and existence analysis of static black holes in pure
  Lovelock theories}.
\newblock {\em Class. Quant. Grav.}, 31:165016, 2014.

\bibitem{Konoplya:2008yy}
R.~A. Konoplya, Keiju Murata, Jiro Soda, and A.~Zhidenko.
\newblock {Looking at the Gregory-Laflamme instability through quasi-normal
  modes}.
\newblock {\em Phys. Rev.}, D78:084012, 2008.

\bibitem{Konoplya:2003ii}
R.~A. Konoplya.
\newblock {Quasinormal behavior of the d-dimensional Schwarzschild black hole
  and higher order WKB approach}.
\newblock {\em Phys. Rev.}, D68:024018, 2003.

\bibitem{Zhidenko:2003wq}
A.~Zhidenko.
\newblock {Quasinormal modes of Schwarzschild de Sitter black holes}.
\newblock {\em Class. Quant. Grav.}, 21:273--280, 2004.

\bibitem{Horowitz:1999jd}
Gary~T. Horowitz and Veronika~E. Hubeny.
\newblock {Quasinormal modes of AdS black holes and the approach to thermal
  equilibrium}.
\newblock {\em Phys. Rev.}, D62:024027, 2000.

\bibitem{Konoplya:2003dd}
R.~A. Konoplya.
\newblock {Gravitational quasinormal radiation of higher dimensional black
  holes}.
\newblock {\em Phys. Rev.}, D68:124017, 2003.

\bibitem{Konoplya:2007jv}
R.~A. Konoplya and A.~Zhidenko.
\newblock {Stability of multidimensional black holes: Complete numerical
  analysis}.
\newblock {\em Nucl. Phys.}, B777:182--202, 2007.

\bibitem{PhysRev.48.73}
A.~Einstein and N.~Rosen.
\newblock The particle problem in the general theory of relativity.
\newblock {\em Phys. Rev.}, 48:73--77, Jul 1935.

\bibitem{MISNER1957525}
Charles~W Misner and John~A Wheeler.
\newblock Classical physics as geometry.
\newblock {\em Annals of Physics}, 2(6):525 -- 603, 1957.

\bibitem{PhysRevLett.33.442}
Douglas~M. Eardley.
\newblock Death of white holes in the early universe.
\newblock {\em Phys. Rev. Lett.}, 33:442--444, Aug 1974.

\bibitem{Ellis:1973yv}
H.~G. Ellis.
\newblock {Ether flow through a drainhole - a particle model in general
  relativity}.
\newblock {\em J. Math. Phys.}, 14:104--118, 1973.

\bibitem{Bronnikov:1973fh}
K.~A. Bronnikov.
\newblock {Scalar-tensor theory and scalar charge}.
\newblock {\em Acta Phys. Polon.}, B4:251--266, 1973.

\bibitem{Morris:1988cz}
M.~S. Morris and K.~S. Thorne.
\newblock {Wormholes in space-time and their use for interstellar travel: A
  tool for teaching general relativity}.
\newblock {\em Am. J. Phys.}, 56:395--412, 1988.

\bibitem{Bronnikov:2005gm}
K.~A. Bronnikov and J.~C. Fabris.
\newblock {Regular phantom black holes}.
\newblock {\em Phys. Rev. Lett.}, 96:251101, 2006.

\bibitem{Bolokhov:2012kn}
S.~V. Bolokhov, K.~A. Bronnikov, and M.~V. Skvortsova.
\newblock {Magnetic black universes and wormholes with a phantom scalar}.
\newblock {\em Class. Quant. Grav.}, 29:245006, 2012.

\bibitem{Kanti:2011yv}
Panagiota Kanti, Burkhard Kleihaus, and Jutta Kunz.
\newblock {Stable Lorentzian Wormholes in Dilatonic Einstein-Gauss-Bonnet
  Theory}.
\newblock {\em Phys. Rev.}, D85:044007, 2012.

\bibitem{Blazquez-Salcedo:2016yka}
Jose~Luis Blázquez-Salcedo et~al.
\newblock {Black holes in Einstein-Gauss-Bonnet-dilaton theory}.
\newblock {\em IAU Symp.}, 324:265--272, 2016.

\bibitem{Cunha:2016wzk}
Pedro V.~P. Cunha, Carlos A.~R. Herdeiro, Burkhard Kleihaus, Jutta Kunz, and
  Eugen Radu.
\newblock {Shadows of Einstein-dilaton-Gauss-Bonnet black holes}.
\newblock {\em Phys. Lett.}, B768:373--379, 2017.

\bibitem{METSAEV1987385}
R.R. Metsaev and A.A. Tseytlin.
\newblock Order $\alpha$ (two-loop) equivalence of the string equations of
  motion and the $\sigma$-model weyl invariance conditions: Dependence on the
  dilaton and the antisymmetric tensor.
\newblock {\em Nuclear Physics B}, 293:385 -- 419, 1987.

\bibitem{GROSS198741}
David~J. Gross and John~H. Sloan.
\newblock The quartic effective action for the heterotic string.
\newblock {\em Nuclear Physics B}, 291:41 -- 89, 1987.

\bibitem{Kanti:1995vq}
P.~Kanti, N.~E. Mavromatos, J.~Rizos, K.~Tamvakis, and E.~Winstanley.
\newblock {Dilatonic black holes in higher curvature string gravity}.
\newblock {\em Phys. Rev.}, D54:5049--5058, 1996.

\bibitem{Torii:1996yi}
Takashi Torii, Hiroki Yajima, and Kei-ichi Maeda.
\newblock {Dilatonic black holes with Gauss-Bonnet term}.
\newblock {\em Phys. Rev.}, D55:739--753, 1997.

\bibitem{Pani:2009wy}
Paolo Pani and Vitor Cardoso.
\newblock {Are black holes in alternative theories serious astrophysical
  candidates? The Case for Einstein-Dilaton-Gauss-Bonnet black holes}.
\newblock {\em Phys. Rev.}, D79:084031, 2009.

\bibitem{Guo:2008hf}
Zong-Kuan Guo, Nobuyoshi Ohta, and Takashi Torii.
\newblock {Black Holes in the Dilatonic Einstein-Gauss-Bonnet Theory in Various
  Dimensions. I. Asymptotically Flat Black Holes}.
\newblock {\em Prog. Theor. Phys.}, 120:581--607, 2008.

\bibitem{Guo:2008eq}
Zong-Kuan Guo, Nobuyoshi Ohta, and Takashi Torii.
\newblock {Black Holes in the Dilatonic Einstein-Gauss-Bonnet Theory in Various
  Dimensions II. Asymptotically AdS Topological Black Holes}.
\newblock {\em Prog. Theor. Phys.}, 121:253--273, 2009.

\bibitem{Konoplya:2017lhs}
R.~A. Konoplya and A.~Zhidenko.
\newblock {The portrait of eikonal instability in Lovelock theories}.
\newblock {\em JCAP}, 1705(05):050, 2017.

\bibitem{Reall:2014pwa}
Harvey Reall, Norihiro Tanahashi, and Benson Way.
\newblock {Causality and Hyperbolicity of Lovelock Theories}.
\newblock {\em Class. Quant. Grav.}, 31:205005, 2014.

\bibitem{Konoplya:2005et}
R.~A. Konoplya and C.~Molina.
\newblock {The Ringing wormholes}.
\newblock {\em Phys. Rev.}, D71:124009, 2005.

\bibitem{Bronnikov:2011if}
K.~A. Bronnikov, J.~C. Fabris, and A.~Zhidenko.
\newblock {On the stability of scalar-vacuum space-times}.
\newblock {\em Eur. Phys. J.}, C71:1791, 2011.

\bibitem{Grozdanov:2016vgg}
Sašo Grozdanov, Nikolaos Kaplis, and Andrei~O. Starinets.
\newblock {From strong to weak coupling in holographic models of
  thermalization}.
\newblock {\em JHEP}, 07:151, 2016.

\bibitem{Grozdanov:2016fkt}
Sašo Grozdanov and Andrei~O. Starinets.
\newblock {Second-order transport, quasinormal modes and zero-viscosity limit
  in the Gauss-Bonnet holographic fluid}.
\newblock {\em JHEP}, 03:166, 2017.

\bibitem{Konoplya:2017zwo}
R.~A. Konoplya and A.~Zhidenko.
\newblock {Quasinormal modes of Gauss-Bonnet-AdS black holes: towards
  holographic description of finite coupling}.
\newblock {\em JHEP}, 09:139, 2017.

\bibitem{Li:2019bqc}
Peng-Cheng Li, Cheng-Yong Zhang, and Bin Chen.
\newblock {The Fate of Instability of de Sitter Black Holes at Large $D$}.
\newblock [arXiv:1909.02685], 2019.

\bibitem{Shinkai:2017xkx}
Hisa-Aki Shinkai and Takashi Torii.
\newblock {Nonlinear Dynamics in the Einstein-Gauss-Bonnet gravity}.
\newblock {\em Phys. Rev.}, D96(4):044009, 2017.

\bibitem{Kokkotas:2017ymc}
K.~D. Kokkotas, R.~A. Konoplya, and A.~Zhidenko.
\newblock {Analytical approximation for the Einstein-dilaton-Gauss-Bonnet black
  hole metric}.
\newblock {\em Phys. Rev.}, D96(6):064004, 2017.

\bibitem{Konoplya:2019hml}
R.~A. Konoplya, A.~F. Zinhailo, and Z.~Stuchlík.
\newblock {Quasinormal modes, scattering, and Hawking radiation in the vicinity
  of an Einstein-dilaton-Gauss-Bonnet black hole}.
\newblock {\em Phys. Rev.}, D99(12):124042, 2019.

\bibitem{Zinhailo:2019rwd}
A.~F. Zinhailo.
\newblock {Quasinormal modes of Dirac field in the
  Einstein-dilaton-Gauss-Bonnet and Einstein-Weyl gravities}.
\newblock [arXiv:1909.12664], 2019.

\bibitem{Antoniou:2017hxj}
G.~Antoniou, A.~Bakopoulos, and P.~Kanti.
\newblock {Black-Hole Solutions with Scalar Hair in
  Einstein-Scalar-Gauss-Bonnet Theories}.
\newblock {\em Phys. Rev.}, D97(8):084037, 2018.

\bibitem{Konoplya:2019fpy}
Roman~A. Konoplya, Thomas Pappas, and Alexander Zhidenko.
\newblock {Einstein--scalar--Gauss--Bonnet black holes: Analytical
  approximation for the metric and applications to calculations of shadows}.
\newblock [arXiv:1907.10112], 2019.

\end{thebibliography}

\end{document}